\documentclass[traditabstract]{aa} 
\usepackage{graphicx}
\usepackage{txfonts}
\usepackage{wasysym}
\newcommand{\halley}{{\em Halley}}
\newcommand{\houtman}{{\em Houtman}}
\newcommand{\keplere}{{\em KeplerE}}
\newcommand{\variants}{{\em Variants}}

\newcommand{\classis}{{\em Classis}}
\newcommand{\secunda}{{\em Secunda Classis}}
\newcommand{\tertia}{{\em Tertia Classis}}
\newcommand{\aliter}{{\em Aliter}}
\newcommand{\gtap}{\mathrel{\hbox{\rlap{\lower.55ex \hbox {$\sim$}}
                   \kern-.3em \raise.4ex \hbox{$>$}}}}
\newcommand{\ltap}{\mathrel{\hbox{\rlap{\lower.55ex \hbox {$\sim$}}
                   \kern-.3em \raise.4ex \hbox{$<$}}}}

\begin{document}
  \title{Early star catalogues of the southern sky\thanks{The full Tables \houtman,
      \classis, \aliter\ and \halley\ (see Tables \ref{t:cathout},
      \ref{t:catclass}, \ref{t:cathal}) are
      available in electronic form only at the CDS via anonymous ftp
      to cdsarc.u-strasbg.fr (130.79.128.5) or via
      http://cdsweb.u-strasbg.fr/cgi-bin/qcat?J/A+A/}}
  \subtitle{De Houtman,  Kepler (Second and Third Classes), and Halley}

  \author{Frank Verbunt\inst{1,2} \and Robert H. van Gent\inst{3,4}}

  \institute{Astronomical Institute, Utrecht University, PO Box 80\,000,
    3508 TA Utrecht, The Netherlands; \email{f.w.m.verbunt@uu.nl}
  \and SRON Netherlands Institute for Space Research, Sorbonnelaan 2,
  3584 CA Utrecht, The Netherlands
  \and until Jan 2010: URU-Explokart, Faculty of Geosciences, Utrecht University, PO Box 80\,115,
    3508 TC Utrecht, The Netherlands
  \and Institute for the History and Foundations of Science,  PO Box 80\,000,
    3508 TA Utrecht, The Netherlands; \email{r.h.vangent@uu.nl} }

  \date{Received February 27, 2011 / Accepted April 5, 2011}

  \abstract{De Houtman in 1603, Kepler in 1627 and Halley in 1679
    published the earliest modern catalogues of the southern sky. We
    provide machine-readable versions of these catalogues, make some
    comparisons between them, and briefly discuss their accuracy on
    the basis of comparison with data from the modern {\em Hipparcos
      Catalogue}. We also compare our results for De Houtman with
    those by Knobel (1917) finding good overall agreement. About half
    of the $\sim$200 new stars (with respect to Ptolemaios) added by
    De Houtman are in twelve new constellations, half in old
    constellations like Centaurus, Lupus and Argo.  The right
    ascensions and declinations given by De Houtman have error
    distributions with widths of about 40\arcmin, the longitudes and
    latitudes given by Kepler have error distributions with widths of
    about 45\arcmin. Halley improves on this by more than an order of
    magnitude to widths of about 3\arcmin, and all entries in his
    catalogue can be identified. The measurement errors of Halley are
    due to a systematic deviation of his sextant (increasing with
    angle to 2\arcmin\ at 60\degr) and random errors
    of 0\farcm7. The position errors in the catalogue of Halley
    are dominated by the position errors in the reference stars, which
    he took from Brahe.}


  \maketitle

\section{Introduction}

The star catalogues of Tycho Brahe -- available first in manuscript
and then in print (Brahe 1598, 1602) -- set a new standard of accuracy
for positional astronomy.  Brahe made his measurements at the island
of Hven, at a latitude of 55\degr52\arcmin, and his catalogue
accordingly was limited to declinations
$\delta\mathrm{(1601)}\gtap-30\degr$.  Stars further south were known
from Ptolemaios, who lived in Alexandria, at a latitude of
31\degr12\arcmin. An area around the southern equatorial pole,
$\delta(150)\ltap-53\degr$, was wholly unknown to ancient Greek
astronomers and Brahe's European contemporaries alike.  In the course
of the 17th century four catalogues were published to remedy this
situation, by De Houtman (1603), Kepler (1627, two catalogues), and
Halley (1679).

\begin{table}
\caption{Constellations in the catalogue of De Houtman. \label{t:houtman}}
\begin{tabular}{rc@{\hspace{0.2cm}}c@{\hspace{0.2cm}}rrll}
C & C$_\mathrm{K}$ & & N & F & constellation \\
\hline
1 & 52 & Phe & 13 &  1 & {\bf Den voghel Fenicx}\\
2 & 49 & CrA & 16 & 14 & De Zuyder Croon \\
3 & 38 & Eri  &   7 & 30 & Het Zuyder eynde van den Nyli \\
4 & 62 & Hyi & 16 & 37 & {\bf De Waterslang}  \\
5 & 60 & Dor & 4 & 53 & {\bf Den Dorado} \\
6 & (40) & Col & 11 & 57 & De Duyve \\
7 & 42  & Arg & 56  & 68 & Argo Navis, het Schip\\
8 & 46 & Cen & 48 & 124 & Centaurus  \\
9 &  (46)  & Cru & 5 & 172 & De Cruzeiro \\
10 & 56 & Mus & 4 & 177 & {\bf De Vlieghe}\\
11 & 59 & Vol & 5 & 181 & {\bf De Vliegende Visch}\\
12 & 57 & Cha & 9 & 186 & {\bf Het Chameljoen} \\
13 & 47 & Lup & 29 & 195 & Lupus, den Wolf\\
14 & 58 & TrA & 4 & 224 & {\bf Den Zuyder Trianghel}\\
15 & 55 & Aps & 9 & 228 & {\bf De Paradijs Voghel} \\
16 & 48 & Ara & 12 & 237 & Het Outaer\\
17 & 31 & Sco & 8$^a$ & 249 & De steert van Scorpio\\
18 & 54 & Pav & 19 & 257 & {\bf De Pauw} \\
19 & 53 & Ind & 11 & 276 & {\bf De Indiaen}\\
20 & 51 & Gru & 12 & 288 & {\bf Den Reygher}\\
21 & 61 & Tuc & 6 & 299 & {\bf Den Indiaenschen Exster,}\\
 & & all & 304 & & $\backslash$ {\bf op Indies Lang ghenaemt} \\
\end{tabular}
\tablefoot{For each constellation the columns give the sequence number $C$,
  the sequence number $C_\mathrm{K}$ of the corresponding
  constellation in Kepler, the
  abbreviation we use, the number of stars in the constellation $N$,
  the sequence number of the first star in the constellation $F$, and
  the Dutch name as given by De Houtman. (The English translation
  of the constellation names is given by Knobel 1917). Brackets (~) around
$C_K$ indicate that the stars from the constellation in \houtman\ are 
listed in a different constellation in \secunda. The twelve new
constelllations are indicated with boldface.} 

\vspace*{0.2cm}

\noindent$^a$ A ninth star is decribed (`the extreme one of the tail of
Scorpio'),
but no position is given; we omit it from \houtman.
\end{table}

In 1603 Frederick de Houtman published a catalogue of `{\em many fixed
  stars, located around the south pole, never seen before this time}',
as he phrases it (in Dutch). De Houtman had sailed in 1595 with the
first voyage by Dutch merchants from Amsterdam to Java and Sumatra,
which is known in Dutch history as {\em de Eerste Schipvaert} (the
first sailing of ships) to the Far East. De Houtman returned to
Amsterdan in 1597, left in 1598 on a second voyage by competing
merchants from Middelburg, was taken prisoner by the Sultan of Atjeh,
was released in 1601, and returned to Amsterdam in July 1602. (Further
details are given by Dekker 1987). In the preface to the catalogue, de
Houtman states that it is based on his own 
measurements during these voyages, upon which he improved while at
Atjeh (Northern Sumatra).  De Houtman distributed his stars over 21
constellations, among which 12 new ones (see Table\,\ref{t:houtman}).

These 12 new constellations had been delineated already in 1597 by
Plancius based on positions of southern stars brought to Amsterdam
with the returning {\em Eerste Schipvaert}. They were first shown on a
globe made by Hondius (Van der Krogt 1993, p.152 sqq.).  The
measurements of the stars in the twelve new constelllations were
ascribed on this globe to Pieter Dirksz Keyser -- who died during the
voyage. There has been some debate whether the catalogue by De Houtman
is independent from the work by Keyser. We will return to this
question in Section\,\ref{s:houtclas}.  De Houtman's catalogue was
little known, perhaps not suprisingly as he published it as an
appendix to a dictionary of the Malaysian and Madagaskar
languages\ldots However, his star positions were used on 
celestial globes made by Blaeu in 1603 and later (Dekker 1987).
 
\begin{table}
\caption{Constellations in the {\em Secunda Classis} 
    of Kepler\label{t:classis2}}
\begin{tabular}{r@{\hspace{0.2cm}}c@{\hspace{0.2cm}}rrrll}
C$_\mathrm{K}$ &      & $N$ & K\phantom{m} & OT & constellation & Fig.\\
\hline
 4 & Cep &  1 & 1005 & 12 & Cepheus   & I-C.6 \\
 5 & Boo &  1 & 1006 & 29 & Bootes    & I-C.7 \\
 7 & Her &  3 & 1007 & 29 & Hercules  & I-C.9 \\
 9 & Cyg &  1 & 1010 & 28 & Cygnus    & I-C.11 \\
11 & Per &  1 & 1011 & 34 & Perseus   & I-C.15 \\
13 & Oph & 26 & 1012 & 38 & Ophiuchus &  \ref{f:ophiuchus} \\
14 & Ser & 13 & 1038 & 14 & Serpens   & \ref{f:ophiuchus}  \\
21 & And &  3 & 1051 & 24 & Andromeda & I-C.27 \\
24 & Ari &  2 & 1054 & 22 & Aries     & I-C.30,I-C.45 \\ 
25 & Tau & 16 & 1056 & 44 & Taurus    & \ref{f:taurus} \\
26 & Gem &  1 & 1072 & 30 & Gemini    & I-C.31 \\
27 & Cnc &  2 & 1073 & 16 & Cancer    & I-C.35 \\
29 & Vir &  2 & 1075 & 40 & Virgo     & I-C.37 \\
30 & Lib &  2 & 1077 & 19 & Libra & I-C.38 \\
31 & Sco & 17 & 1079 & 11 & Scorpius  & \ref{f:scorpius} \\
32 & Sgr & 17 & 1096 & 30 & Sagittarius& \ref{f:sagittarius} \\
34 & Aqr &  4 & 1113 & 42 & Aquarius   & I-C.43 \\
35 & Psc &  6 & 1117 & 37 & Pisces     & I-C.44 \\
36 & Cet &  4 & 1123 & 22 & Cetus      & I-C.45 \\
38 & Eri & 20 & 1127 & 20 & Eridanus   & \ref{f:eridanusc}\\
40 & CMa & 16 & 1147 & 14 & Canis Maior& \ref{f:colomba} \\
42 & Arg & 42 & 1163 & 12 & Argo       & \ref{f:argo} \\
43 & Hya &  9 & 1205 & 25 & Hydra      & \ref{f:hydra} \\
46 & Cen & 33 & 1214 &  5 & Centaurus  & \ref{f:centaurus} \\
47 & Lup & 19 & 1247 &  1 & Lupus      & \ref{f:lupus} \\
48 & Ara &  7 & 1266 &  1 & Ara        & \ref{f:ara} \\
49 & CrA & 13 & 1273 & 1 & Corona Australis & \ref{f:coraust}\\
50 & PsA & 17 & 1286 & 1 & Piscis Austrinus & \ref{f:pisaustr} \\
\multicolumn{2}{c}{all} & 298
\end{tabular}
\tablefoot{For each constellation the columns give
  the sequence number C$_\mathrm{K}$ (in \keplere\ for
  C$_\mathrm{K}<46$), its abbreviation, the number $N$ of 
  stars added in {\em Secunda Classis}  to 
  \keplere, and for the first star in the constellation the 
  sequence number in the whole catalogue K and the
  sequence number within the constellation OT. Note that
  the numbers C$_\mathrm{K}$ ($\geq46$), K and OT are
  continued from \keplere.  
  We also give the Figure in which the new stars are depicted,\
  where prefix `I-' refers to Paper\,I.}
\end{table}

A new edition of Brahe's star catalogue was edited in 1627 by Kepler,
in his {\em Tabulae Rudolphinae}. To this catalogue Kepler appended
two more, with stars which he decribed as belonging to the {\em
  Secunda classis} and {\em Tertia classis} (second and third class),
respectively. He describes these classes in the beginning of each
catalogue as follows (we translate from the Latin; words between [\,]
are added).

\begin{quotation} {\bf [The second class]} includes those fixed
  [stars] from the old catalogue of Hipparchos, retrieved and emended
  by Ptolemaios, that Tycho omitted.  It is convenient to call them
  semi-Tychonian: indeed having sought them out from the [Greek]
  manuscript of Ptolemaios, and also using the [Latin] version of
  [George of] Trebizond published by Schreckenfuchs 76 years ago in
  T\"ubingen [actually Basel], I have converted them to the [end of the] year 1600,
  with the addition to the positions of the longitude recorded by
  Ptolemaios of such an angle as Tycho added to some other nearby
  bright [star]; and having added or subtracted to the latitude as
  much of the angle as at any place the obliquity under Ptolemaios is
  believed to have been greater; in any case in such a way that it
  would have the ratio of the adjacent round number.

  Furthermore I have considered it expedient
  to follow this Greek text of  Ptolemaios more closely, than among others
  the Prutentic, Copernican, and Alfonsine [texts], which appear to
  have followed the Arabic version of the Almagest; as in this way
  I would offer the opportunity to compare the versions among themselves,
  since it is uncertain whether the Arabs have corrected anything
  in these Ptolemaic [numbers], or whether all diversity among the versions
  originates from the  inaccuracy of the copyists. There are a few,
  to which I have put my hand myself, in the book on the star in
  Serpentarius [i.e.\ SN1604], as well as others, which I have rendered
 in old  character, to notify the reader of this.
\end{quotation}

\begin{quotation} {\bf The third class} of fixed stars[:] comprising
  twelve celestial images, which can not be seen at all in our
  moderate northern zone. In his Uranometria Joh.\ Bayer reports that
  these have been observed by Amerigo Vespucci, Andreas Corsali, and
  Pedro de Medina, the first among Europeans, and declares that
  they were for the first time corrected to astronomical standard by
  Pieter Dicksz. [Keyser]. Jacobus Bartsch from Lausitz, a diligent
  young man, famous for some time now for his great merits concerning
  the celestial globe, assembled these same [constellations] into
  numbers and a map from the last tables and manuscripts of Johann
  Bayer himself (a splendid little collection of Christian
  constellations extracted from the Uranographia of Schiller, the
  publication of which is forthcoming in accordance with the last will
  of the author); and he has promised that he will subsequently
  publish the most perfect maps, by producing a one-an-a-half foot
  globe with the ancient images, as more conform with the version 
  of Tycho.
\end{quotation}

\begin{table}
\caption{Constellations in the {\em Tertia
    Classis} of Kepler\label{t:classis3}}
\begin{tabular}{r@{\hspace{0.2cm}}c@{\hspace{0.2cm}}rrll}
 C$_\mathrm{K}$ & & $N$ & K\phantom{m} & constellation & Fig. \\
\hline
 51 & Gru & 13 & 1303 &  Grus & \ref{f:grus}      \\
 52 & Phe & 15 & 1316 &  Phoenix & \ref{f:phoenix}   \\
 53 & Ind & 12 & 1331 &  Indus   &\ref{f:indus}   \\
 54 & Pav & 23 & 1343 &  Pavo & \ref{f:pavo}      \\
55 & Aps & 11 & 1366 &  Apus, Avis Indica & \ref{f:apus}      \\
56 & Mus &  4 & 1377 &  Apis, Musca   & \ref{f:musca}   \\
57 & Cha & 10 & 1381 &  Chamaeleon & \ref{f:chamaeleon} \\
58 & TrA &  5 & 1391 &  Triangulum Aus.  & \ref{f:triangaus}\\
59 & Vol &  7 & 1396 &  Piscis Volans, Passer &  \ref{f:volans}  \\
60 & Dor &  7 & 1403 &  Dorado, Xiphias & \ref{f:hydrus}     \\
61 & Tuc &  8 & 1410 &  Toucan, Anser Americanus & \ref{f:tucana}    \\
62 & Hyi & 21 & 1418 &  Hydrus & \ref{f:hydrus}    \\
\multicolumn{2}{c}{all} & 136 \\
\end{tabular}
\tablefoot{Columns as in Table\,\ref{t:classis2}, without the OT}
\end{table}
 
Uranographia of Schiller refers to his celestial atlas, the
{\em Coelum Stellatum Christianum} of 1627.
Kepler thus put the stars known from Ptolemaios, but with
positions more-or-less improved by comparison with improved
positions of nearby stars, and corrected for the different value
of the obliquity used by Ptolemaios, in his second class, together
with some new stars measured by himself. In the third class he
puts the stars of the new constellations.
Kepler states that the positions of the third class are
based on those of Petrus Theodorus, i.e.\ Pieter Dirksz. [Keyser],
through the intermediary Bartsch.

To our knowledge, no analysis has ever been published of the
\secunda\ and/or \tertia.

The third catalogue of the stars in the southern sky was published by
Edmond Halley (1679), on the basis of the measurements he made during
a year (roughly Feb 1677 to Feb 1678) on the island of St.\ Helena,
latitude $-$15\degr57\arcmin. As Halley explains in his introduction
to the catalogue, persistent bad weather prevented him from making
planetary observations to determine the obliquity, and therefore he
decided to determine the positions of the southern stars from the
angular distance to stars from the catalogue of Brahe.  In doing so he
kept the ecliptic positions as given by Brahe.  The bad weather also
forced Halley to forego observations of the faint stars in Piscis
Austrinus and the stars in Indus, in favor of more important brighter
stars in other constellations.

Halley was 20 years and a bachelor student when he left; upon his
return, with a recommendation from king Charles\,II, he was granted a
master degree. For further details we refer to the biography of Halley
writen by Cook (1998).  An analysis of the star catalogue of Halley
has been published by Baily (1843).

In this paper, we describe machine-readable versions of the
star catalogues of De Houtman, Kepler (second and third classes)
and Halley. The machine-readable tables give our identifications
with stars from the  (modern) Hipparcos catalogue, and on the
basis of these the accuracy of the positions and magnitudes
tabulated in the old catalogues. For the catalogue of De Houtman
we compare our identifications with those made by Knobel (1917).

\begin{table}
\caption{Constellations in \halley. \label{t:halley}}
\begin{tabular}{r@{\hspace{0.2cm}}c@{\hspace{0.2cm}}rrll}
 C & & $N$ & E\phantom{0} & constellation & Fig. \\
\hline
 1 & Sco & 29 & 1 & Scorpius & \ref{e:scorpius} \\
 2 & Sgr & 21 & 30 & Sagittarius & \ref{e:sagittarius}\\
 3 & Eri & 30 & 51 & Eridanus & \ref{e:eridanus} \\
 4 & CMa &  5 & 81 & Canis Maior & \ref{e:canismaior} \\
 5 & PsA &  1 & 86 & Piscis Austrinus & \ref{e:pisaustr} \\
 6 & Col &  10 & 87 & Colomba Noachi & \ref{e:colomba} \\
 7 & Arg & 46 & 97 & Argo Navis & \ref{e:argo} \\
 8 & RCa & 12 & 143 & Robur Carolinum & \ref{e:robur} \\
 9 & Hya & 5 & 155 & Hydra & \ref{e:hydra} \\
10 & Cen & 35 & 160 & Centaurus & \ref{e:centaurus} \\
11 & Lup & 23 & 195 & Lupus & \ref{e:lupus} \\
12 & Ara & 9 & 218 & Ara, Thuribulum & \ref{e:ara} \\
13 & CrA & 12 & 227 & Corona Australis & \ref{e:coraustr} \\
14 & Gru & 13 & 239 & Grus & \ref{e:grus} \\
15 & Phe & 13 & 252 & Phoenix & \ref{e:phoenix} \\
16 & Pav & 14 & 265 & Pavo & \ref{e:pavo} \\
17 & Aps & 11 & 279 & Apus Avis, Inidica & \ref{e:apus} \\
18 & Mus & 4 & 290 & Musca Apis & \ref{e:musca} \\
19 & Cha & 10 & 294 & Chamaeleon & \ref{e:chamaeleon} \\
20 & TrA & 5 & 304 & Triangulum Australe & \ref{e:triaustr} \\
21 & Vol & 8 & 309 & Piscis Volans & \ref{e:volans} \\
22 & Dor & 6 & 317 & Dorado, Xiphias & \ref{e:dorado} \\
23 & Tuc & 9 & 323 & Toucan, Anser Americanus & \ref{e:tucana} \\
24 & Hyi & 10 & 332 & Hydrus & \ref{e:hydrus} \\
\multicolumn{2}{c}{all} & 341 \\
\end{tabular}
\tablefoot{For each constellation the columns give the sequence number $C$,
  the abbreviation we use, the number of stars in the constellation $N$,
  the sequence number of the first star in the constellation $E$, the
  constellation name and the Figure where it is shown. Due to bad
  weather, Halley observed only 1 star in Piscis Austrinus, and none in
Indus.}
\end{table}

In the following we refer to (our machine-readable versions of) the
star catalogues of De Houtman (1603) as \houtman; and of Halley (1679)
as \halley. An F or E number is the sequence number of a star in
\houtman, and \halley, respectively.  (These letters stand for
Frederick and Edmond, respectively, and are used to avoid confusion
beweeen \houtman, \halley, and the catalogue of Hevelius 1690.)  Thus
F\,273 is the 273rd entry in \houtman, and E\,55 the 55th entry in
\halley. Our emended, machine-readable versions of Kepler's (1627)
edition of the star catalogue of Brahe (Verbunt \&\ Van Gent 2010a,
hereafter: Paper\,I) and of his Second and Third Classes are referred
to as \keplere, \secunda\ and \tertia, respectively. A K-number refers
to an entry in these catalogues, where we continue the numbering:
K\,1004 is the last star in \keplere, K\,1005 and K\,1302 the first
and last entry in \secunda, and K\,1303 the first entry in \tertia.
The sequence number within a constellation is indicated by a number
following the abbreviated name of the constellation: thus Phe\,3 in
the third star in Phoenix, in the catalogue under discussion.

We follow Hevelius (1690) in continuing the sequence numbering 
within each constelllation between \keplere\ and \secunda,
indicated with OT (Ordo Tychonis) in the Tables (see also
Verbunt \&\ Van Gent 2010b). Thus Cep\,11
is the last star of Cepheus in \keplere, to which Cep\,12 is added in
\secunda.

\section{Description of the catalogues\label{s:desc}}

All catalogues are organized by constellation.

\subsection{\houtman\ (1603)\label{s:houtdesc}}

The star catalogue of De Houtman (1603) 
contains 21 constellations, of which twelve are new, 8 are from
Ptolemaios, and one, Cruzeiro, is the Southern Cross, split off from
Centaurus. Table\,\ref{t:houtman} lists the
constellations and the number of stars in them.  Each constellation
starts with a statement of how many stars it contains, followed by
a table with sequence number within the constellation, a short
description of the star followed by five columns of integer positive
numbers.The first two numbers give the right ascension in degrees
($G$) and minutes ($M$), the next two the declination in degrees ($G$)
and minutes ($M$), and the fifth the magnitude of the star. Zeros are
indicated either by an empty slot, or explicitly with 0.  The
coordinates then are
$$ \alpha = G +M/60;\qquad \delta = - (G+M/60) $$
i.e.\ the minus-sign of the declination is not given in the tables
but implicitly assumed. 

Interestingly, \houtman\ gives the coordinates in the equatorial system.
The equinox of the catalogue is not given. We assume it is the same as
that of \keplere, i.e.\ AD 1601.0, JD\,2\,305\,824.

\subsection{\secunda, \tertia, Kepler (1627)\label{s:kepdesc}}

The star catalogue given as the \secunda\ (second class) by Kepler
(1627)  starts with stars to be added
to the 46 constellations listed in \keplere, and ends with four
constellations that were listed by Ptolemaios but which are too far
south to be in \keplere.  Table\,\ref{t:classis2} lists the
constellations and the numbers of stars added to them with respect to
\keplere. For each entry a Latin description of the star is followed
by the ecliptic longitude and latitude, and the magnitude.  The
longitude is given in integer degrees ($G$), integer minutes ($M$) and
zodiacal sign ($Z$). For one entry, K\,1059, $\frac{1}{2}$ is indicated
behind the integer $M$. The longitude in decimal degrees thus is
$$ \lambda=(Z-1)*30+G+M/60$$
The latitude is given in integer degrees ($G$), integer minutes ($M$)
and B (for borealis, north) or A (australis, south). Often the B or A
is not given explicitly, but implicitly taken to be the same as for
the previous entry. The latitude in decimal degrees thus is:
$$ \beta = \pm(G+M/60); \qquad  +/- \mathrm{for ~B/A} $$

Kepler gives a number of alternative positions in the Secunda Classis,
in text between the catalogue entries, and sometimes in the
entry lines themselves, usuallly as {\em  aliter} (alternatively), in
two cases (K\,1066, K\,1067) as {\em Schreckenf} (Schreckenfuchs). 
In the entry lines, the number for which a variant is given is 
marked by an explicit degree or arcminute sign.

The star catalogue given as the \tertia\ (third class) by Kepler
(1627) has the same layout as the second class. The constellations in
it are the twelve new constellations delineated by Plancius.
Table\,\ref{t:classis3} lists the constellations in this catalogue and
the number of stars in each of them.

Since \secunda\ and \tertia\ immediately follow the star catalogue
of Brahe in Kepler (1627), we assume that they have the same equinox
as \keplere.

\subsection{\halley (1679)\label{s:haldesc}}

The {\em catalogue of the southern stars, or supplement to the
  Tychonian catalogue} of Halley (1679) is given in six columns. The
first column contains the descriptions of the stars. The second and
third column are headed {\em unde observata} (from where observed) and
{\em Distantia observata} (observed distance), and contain 
the names of two reference stars for each entry and 
the angular distance of the entry to each of them, in
degrees (gr.), minutes (m.) and seconds (s.), all in integers.  For
most entries two reference stars with angular distance are
given, for some entries three or four reference stars; 
occasionally just one reference star is given, together with 
an indication that  a longitude or latitude of the entry was used
as given by Brahe or Kepler. The
fourth column has two lines per entry, the first one giving the
ecliptic longitude in zodiacal sign, degrees and minutes, and the
second one the ecliptic latitude in degrees and minutes. An $s$ (for
semi) may follow the minute indicating an extra 0.5 (both for
longitude and latitude). The latitudes are followed by $b$ (for
borealis) if the star has a northern latitude; for the other entries a
southern latitude is implictly assumed.  The fifth column gives the
longitude and latitude in the same notation (without the $s$'s and
$b$'s) {\em E Catal. vetusto} (from the old catalogue). Halley
explains in his introduction that with old catalogue he refers to
Clavius, {\em Commentaries on Sacrobosco's De Sphaera} and Bartsch's
excerpt of the Rudolfine Tables by Kepler.  The sixth column gives the
magnitude as an integer.

Halley notes in the catalogue that due to bad weather he did
not observe the faint stars in Piscis Austrinus, nor the
stars in Indus. The last six stars of Lupus (E\,212 - E\,217) he
observed {\em inter navigandum}  (while sailing), and for these
no angles to reference stars are given, but only the rough positions.

The equinox of the catalogue is given by Halley as {\em Annum
  Completum 1677}, old style. We assume that this corresponds
to 1678 Jan 1, JD\,2333948.

\section{Identification procedure\label{sec:id}}

The procedure that we follow for the identification of the stars in
the different catalogues is {\em mutatis mutandis} the same as the
procedure we followed for the catalogue of Brahe, and we refer to
Paper\,I for details.
Briefly, we select all stars for the {\em Hipparcos Catalogue} with a
Johnson visual magnitude brighter than 6.0, we correct their
equatorial position for proper motion between the Hipparcos epoch
1991.25 and the equinox of the catalogue, and precess the resulting
equatorial coordinates to the equinox of the catalogue. For \secunda,
\tertia, and \halley, we convert the equatorial coordinates to
ecliptic coordinates, using the obliquity appropriate for the
catalogue equinox.

\begin{table}
\caption{Certain identifications with objects not in the
 Hipparcos catalogue\label{t:nonhip}}
\begin{tabular}{ll}
entry & identification \\ 
\hline
K\,1009 & globular cluster M\,13 \\
K\,1036 & supernova 1604 (Kepler) \\
K\,1230 & globular cluster $\omega$\,Cen \\
K\,1408 & Large Magellanic Cloud \\
K\,1433 & Small Magellanic Cloud \\
F\,46 & Small Magellanic Cloud \\
E\,20 & open cluster NGC\,6231 \\
E\,29 & open cluster M\,7 \\
E\,146 & highly variable star $\eta$\,Car \\
E\,180 & globular cluster $\omega$\,Cen \\
\end{tabular}
\end{table}

For each entry in the old catalogue we then find the nearest -- in
terms of angular separation -- counterpart in the Hipparcos
Catalogue. Often, this star is designated a secure counterpart by us,
and given an identification flag\,1. This flag is also used for
certain identifications with objects not in the {\em Hipparcos
  Catalogue} (see Table\,\ref{t:nonhip}).  Occasionally, a star
brighter than the nearest star is considered by us to be the secure
counterpart, and we flag such an identification with 2. This may
happen when the brighter star is at a marginally larger angular
distance, or when it is part of a recognizable pattern.  Examples are
F\,7 in Phoenix (near the center in Fig.\,\ref{f:phoenix}), and many
stars in the constellation Corona Australis, both in \houtman\ and in
\secunda\ (Fig.\,\ref{f:coraust}).  Identifications we
consider plausible but uncertain are flagged 3, and cases where
several equally plausible counterparts are found, are flagged 4. An
unidentified catalogue entry is flagged 5. As we will see, all repeat
entries in \secunda\ except one are repeating entries from \keplere\
-- with slightly different position, i.e.\ they are repeat
entries only if \secunda\ and \keplere\ are considered as one
catalogue. We mark these, and the only repeat entry within \secunda,
with a 6.  \tertia\ and \houtman\ do not contain repeat entries.

\begin{figure}
\includegraphics[angle=270,width=\columnwidth]{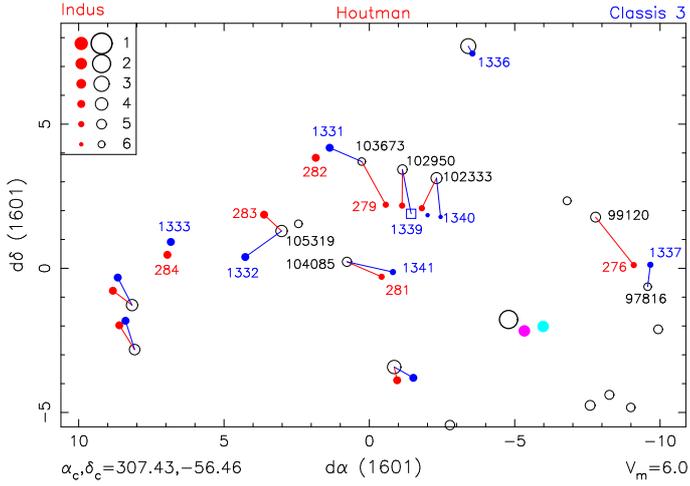}

\caption{Comparison of our identifications of entries in Indus in
 \houtman\ and \secunda.\label{f:annotindus}
  In this projected image of the constellation Indus red
  stars and numbers refer to \houtman, blue to \secunda\ and black to
  the {\em Hipparcos Catalogue}. The axes give the distances in right
  ascension and declination to the center of the constellation, roughly
  in degrees (for details of the projection used, see Sect.\,\ref{s:figures}).}
\end{figure}

\begin{figure}
\includegraphics[angle=270,width=\columnwidth]{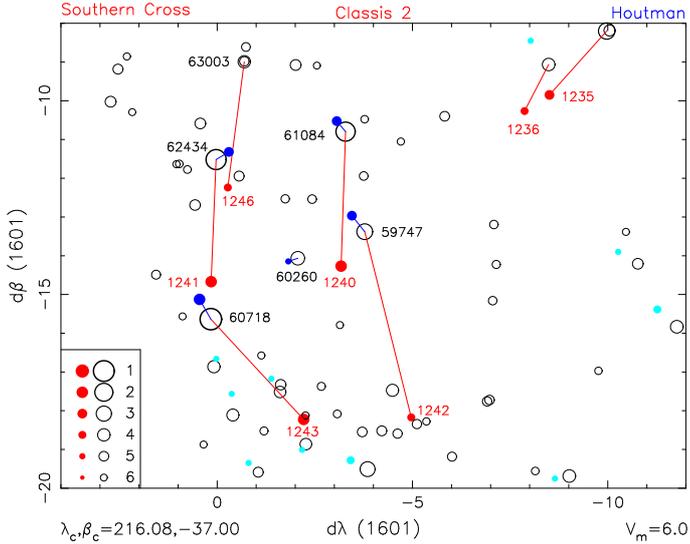}

\caption{Comparison of our identifications of entries in Crux in
 \secunda\ and \houtman.\label{f:annotcrux}
 In this projected image of the constellation Crux red
  stars and numbers refer to \secunda, blue to \houtman\ and black to
  the {\em Hipparcos Catalogue}. The axes give the distances in ecliptic
  longitude and latitude to the center of the constellation, roughly
  in degrees (for details of the projection used, see Sect.\,\ref{s:figures}).}
\end{figure}

When no good Hipparcos star is found as identification, we investigate
the possibility of extended sources such as star clusters, or extinct
sources such as (super)novae. In a number of cases this leads to a good
identification, as listed in Table\,\ref{t:nonhip}. When a bright star
is near or in a star cluster, the choice for single star or extended
object is ambiguous. In two such cases, E\,20 and E\,29 we list the 
Hipparcos star in the catalogue, and the cluster in Table\,\ref{t:nonhip}.

In finalizing the identifications we have also compared our initial
results for \secunda\ and \tertia\ with those for \houtman; this
occasionally made us change an initial identification.
In crowded constellations we have limited a first search for
counterparts to Hipparcos stars with $V<5.0$; this limits
confusion by fainter and therefore less likely counterparts.
In particular in Argo, many identifications flagged 2,  i.e.\ the
counterpart is not the nearest star with $V<6.0$, are the
nearest Hipparcos star  with $V<5.0$.

It may be argued that the majority of stars in \secunda\ that
are taken from the catalogue of Ptolemaios should properly be
identified from that catalogue. We have in fact applied our
identification procedure independently both to the \secunda\ and to
the catalogue of Ptolemaios, for which we use the edition by Toomer
(1998). In almost all cases, the identifications are the same for
corresponding entries in both catalogues; where they are not, they
reflect the uncertainty in choosing between several plausible 
possibilities. We discuss this further in Sect.\,\ref{s:ptolclas}.

To illustrate some of the problems encountered in trying to identify
the catalogue entries, and our solutions of them, we discuss two of
the more ambiguous cases.  The first of these is the constellation
Indus, illustrated in Fig.\,\ref{f:annotindus}. First, we note that
with the sole exception of K\,1336 ($\alpha$\,Ind) every entry in
\tertia\ is matched with an entry in \houtman. Thus, one may expect
the same identification for each matched pair, and mostly this is
indeed the case. An exception is the pair F\,276/K\,1337, where in the
case of F\,276 we choose the further but brighter HIP\,99120 as
counterpart and in the case of K\,1337 the closer but fainter
HIP\,97816. We might have chosen to identify both F\,274 and K\,1339
with the same star, be it the fainter HIP\,97816 or the brighter
HIP\,99120.  Another exception involves HIP\,102333, HIP\,102950 and
HIP\,103673, three stars in a row. It is tempting to identify these
with the three entries in a row about 1\fdg5 to the south, and this we
have done for the catalogue of De Houtman. This implies that F\,282 is
unidentified.  In the case of \tertia\ we have chosen to identify
HIP\,103673 with K\,1331, and this implies that K\,1338 -- the star
between K\,1339 and K\,1340 -- remains unidentified.  We have further
chosen to identify F\,283/K\,1332 with HIP\,105319.  One could
consider to identify K\,1332 and K\,1333 with HIP\,104085 and
HIP\,105319 respectively, and leave K\,1341 unidentified. We have
chosen to identify K\,1332 and K\,1341 with HIP\,105319 and
HIP\,104085, respectively, and leave K\,1333 unidentified. The
matching entries F\,281, F\,283 and F\,284 suggest that this is the
right choice.  These examples show that a certain arbitrariness can
not be avoided. Knobel (1917) made the same choices in identifying
the entries in Indus in \houtman\ as we do.

As the second example we consider the Southern Cross, which in
\secunda\ is part of Centaurus (Fig.\,\ref{f:annotcrux}). The
identification of the stars in \houtman\ is straightforward, the
identification of the stars in \secunda\ far from straightforward.
If we identify the bright entries K\,1240 and K\,1241 with
HIP\,61084 and HIP\,62434, respectively, identification of
K\,1242 and K\,1246 with HIP\,59747 and HIP\,63003 becomes
plausible. What about K\,1243? From its brightness, identification
with HIP\,60718 appears best, but from the direction of the 
offsets of the other identifications HIP\,60260 may be possible as
well.  The positions as given by De Houtman for these stars are a
marked improvement on (Kepler and through him on) Ptolemaios.

\begin{table*}
\caption{First lines from the machine-readable table \houtman}. \label{t:cathout}
\begin{tabular}{rr@{\quad}r@{\quad}rr@{\quad}rr@{\quad}rcrrrrrrr}
F & C &  & & $G$ & $M$ & $G$\phantom{0} & $M$ 
& $V_F$ & HIP\phantom{P} & I & K & $V$ & 
$\Delta\alpha$ & $\Delta\delta$ & $\Delta$ \\\hline
\\
  1 & 1 &=Phe &  1 & 347 & 20  & 40 & 50 & 5& 116231&  1& 1 & 4.4 &  29.6 &  49.1 &  54.1 \\
  2 & 1 &=Phe &  2 & 348 & 35  & 44 &15  &5 &116389 & 1 &3 & 4.7  &$-$15.5 &  $-$33.7 &  35.5\\ 
  3 & 1 &=Phe &  3 & 349 &45  & 42 &45  &4   &   0      &0 &3 & 0.0  & 0.0 &   0.0 &   0.0\\
  4 & 1 &=Phe &  4 & 357 &10  & 47 &45  &4   & 765 & 1& 1 & 3.9   & 3.0  & $-$12.0 &  12.2\\ 
  5 & 1 &=Phe &  5  &  1   &00  & 44 &05  &2   &2081&  1 &1 & 2.4  & 34.7  & $-$24.0 &  34.6 \\
  6 & 1 &=Phe &  6  & 1   &30  & 45 &34  &4   &2072 & 1 &1&  3.9   & 5.1  & $-$20.0  & 20.3 \\
  7 & 1 &=Phe &  7  & 2   & 40  &50 &15  &4  & 2472 & 2 &1 & 4.8  & 18.3  & $-$46.1 &  47.6\\ 
\end{tabular}
\tablefoot{For explanation of the columns see Sect.\,\ref{s:houtmachine}.}
\end{table*}

\begin{table*}
\caption{First lines from the machine-readable table \classis}. \label{t:catclass}
\begin{tabular}{cc@{\quad}r@{\quad}rr@{\quad}r@{\quad}lr@{\quad}r@{\quad}ccrrrrrrrr}
K & $C_\mathrm{K}$ &  & & $Z$ & $G$ & $M$ & $G$ & $M$ & $S$ 
& $V_K$ & HIP\phantom{P} & I & $V$ & 
$\Delta\lambda$ & $\Delta\beta$ & $\Delta$ & P & I$_\mathrm{P}$\\\hline
\\
1005 &  4 &=Cep &12 & 1& 04& 30.  &64& 00 &B & 5 &107259& 1 & 4.2 &
$-$15.4 &  10.9 &  12.9 & 86 & 1\\
1006  & 5 &=Boo &29 & 7& 28& 45.  &45& 45 &B & 4 & 75312 &6  &5.0
&165.0  & 65.6  & 131.5  & 91 & 1    \\
1007  & 7 &=Her &29 & 9& 01& 20.  &55& 55 &B & 5 & 83313 &1  &5.3
&64.5    &1.9 &  36.2 & 131 & 1\\
1008  & 7 &=Her &30 & 9& 02& 30.  &58& 15 &B & 5 & 83838 &1  &5.4
&92.9   &15.7  & 51.2 & 132 & 1\\
1009  & 7 &=Her &31 & 8& 24& 39.  &57& 30 &B & 5  &    0      &1 & 5.8
&$-$40.9   &23.8 &  32.3 & 0 & 3 \\
1010  & 9 &=Cyg &28 &12& 03& 00. &  63& 20 &B & 9 &101138& 2 & 4.9
&$-$147.3 &  43.6 &  78.5 & 175 & 1\\
\end{tabular}
\tablefoot{For explanation of the columns see
  Sect.\,\ref{s:kepmachine}. The format of the file with alternative
  positions,  \aliter, is identical}
\end{table*}

\begin{table*}
\caption{First lines from the machine-readable table \halley}. \label{t:cathal}
\begin{tabular}{c@{\quad}c@{~}r@{\quad}rr@{\quad}r@{\quad}rr@{\quad}r@{\quad}c
r@{\quad}r@{\quad}r@{\quad}r@{\quad}r@{\quad}r@{\quad}r@{\quad}rr@{\quad}r@{\quad}rrrr}
$E$ & $C$ &  & & $Z$ & $G$ & $M$ & $G$ & $M$ & $V_E$ & 
HIP\phantom{P} & $G$ & $M$ & $S$ &HIP\phantom{P} & $G$ & $M$ & $S$ &
HIP\phantom{P} & I & $V$ & 
$\Delta\lambda$ & $\Delta\beta$ & $\Delta$ \\\hline
\\
  1 & 1 &=Sco & 1 & 8& 28 &41.0 & 01 & 5.0& 2.5 & 65474& 39& 28 & 0 &
  79593& 16 &15 &20 & 78820 &1 & 2.6 &  0.7 &  $-$2.1 & 2.2   \\
  2 & 1 &=Sco & 2 & 8& 28 & 4.0  &$-$01& 54.0& 2.5 & 65474& 38& 42& 20 & 
 79593& 19& 13 &10 & 78401& 1 & 2.3 & 0.6 &  $-$2.6 & 2.7     \\
  3 & 1 &=Sco & 3 & 8& 28 &26.0 &$-$05& 23.0& 3.0 & 65474& 39 & 9& 30 &
 79593& 22& 42 &10 & 78265 &1 & 2.9  & 0.8 &  $-$3.0 & 3.2    \\
  4 & 1 &=Sco & 4 & 8& 28 &38.5 &$-$08& 28.0& 3.5 & 65474& 39 &39 & 0 & 
77070& 35 &44 & 0 & 78104 &1 & 3.9 &   0.8 &  $-$5.5 &   5.5  \\
  5 & 1 &=Sco & 5 & 9& 00 & 8.0 & 01 &42.0 &4.0 & 65474& 40& 57 &40 & 
79593& 15& 46& 40 & 79374 & 1 & 4.0 &   0.9 &  $-$1.5 &   1.8  \\
  6 & 1 &=Sco & 6 & 8& 29 &11.0 & 00 &16.0 &5.0 & 65474& 39 &53& 20 &
 79593& 17 & 4& 20 & 78933 & 1 & 3.9 &  $-$0.5 &  $-$0.3 &   0.6    
\end{tabular}
\tablefoot{For explanation of the columns see Sect.\,\ref{s:halmachine}.}
\end{table*}

\subsection{Identifications of \houtman\ by Knobel} 

Knobel (1917) identifies stars from \houtman\ with entries
in the {\em Uranometria Argentina} (Gould 1879). We use the
machine-readable version prepared by Pilcher (2010)  which takes into
account the (very few) corrections later made by Gould and others
to the 1879 version, and which adds among others the HD number
of the stars. We convert this to the Hipparcos number,
for comparison with our identification.
(The {\em Hipparcos Catalogue} provides the HD number
for each of its entries.)
When the identification by Knobel is identical to ours we flag it K=1,
when it is different from ours, we flag it 3. (Flag 2, when we choose
from two possible identifications the other one than he, does not occur.)
Flag 0 indicates that Knobel has no identification.

\section{The machine-readable catalogues}

\subsection{The star catalogue of De Houtman\label{s:houtmachine}}

The machine-readable table \houtman\ contains the following information
(see Table\,\ref{t:cathout}). The first column gives the sequence
number F. The second and third column give the sequence number of the
constellation $C$ and the abbreviation of the constellation name.  The
fourth column gives the sequence number within the constellation.
Columns 5 and 6 give the right ascension in degrees ($G$) and minutes
($M$), and columns 7 and 8 the declination in degrees ($G$) and
minutes ($M$). The negative declination (which all entries
have) is implicit. Column 9 gives the magnitude
according to De Houtman.

Columns 10-16 provide additional information from our analysis, viz.\
the Hipparcos number of our identification HIP, the flag I indicating
the quality of the identification, the flag K which compares our
identification with that by Knobel, the visual (Johnson) magnitude $V$
given in the {\em Hipparcos Catalogue} for our identification, the
differences in right ascension $\Delta\alpha$ and declination
$\Delta\delta$ in minutes, and the angle $\Delta$ between the catalogue entry
and our Hipparcos identification, in arcminutes (\arcmin).
If the catalogue entry for minutes $M$ as given by De Houtman is
$M_\mathrm{F}$ and the value computed from the position and proper
motion in the {\em Hipparcos Catalogue} $M_\mathrm{HIP}$, then columns
14 and 15 give $M_\mathrm{HIP}-M_\mathrm{F}$.

\subsection{\secunda\ and \tertia\ from Kepler; variants
to \secunda\label{s:kepmachine}}

The machine-readable table \classis\ combines the \secunda\ and
\tertia. As noted above, Kepler gives alternative positions for some
entries in \secunda. Some of these alternative positions occur in
\keplere\ or its variants that we collected in \variants\ in
Paper\,I. Such is the case for the alternative positions given for UMa
39, 40, 41 and 56 at the very beginning of Secunda Classis, and for
Tau\,30 (Electra in the Pleiades). Variants that do not occur in these
earlier catalogues we collect in \aliter, with the exception of those
for K\,1139 and K\,1140, which we use to emend the main catalogue
entry (see Sect.\,\ref{s:classisem}).

The files \classis\ and \aliter\ contain the following information
(see Table\,\ref{t:catclass}). The first column gives the sequence
number K. The second and third column give the sequence number of the
constellation $C_\mathrm{K}$ and the abbreviation of the constellation name.  The
fourth column gives the sequence number within the constellation.
Columns 5 - 7 give the longitude in zodiacal sign ($Z$), degrees ($G$) and minutes
($M$), and columns 8 - 10 the latitude in degrees ($G$),
minutes ($M$) and sign $S$.  Column 11 gives the magnitude
according to Kepler. A tabulated magnitude 8 indicates that Kepler
does not give a magnitude; a 9 indicates a magnitude given by
Kepler as ne(bulous).

Columns 12-18 provide additional information from our analysis, viz.\
the Hipparcos number of our identification HIP, the flag I indicating
the quality of the identification, the visual (Johnson) magnitude $V$
given in the {\em Hipparcos Catalogue} for our identification, the
differences in longitude $\Delta\lambda$ and latitude $\Delta\beta$ in
minutes, and the angle $\Delta$ between the catalogue entry and our
Hipparcos identification, in arcminutes (\arcmin).  If the catalogue
entry for minutes $M$ as given by Kepler is $M_\mathrm{K}$ and the
value computed from the position and proper motion in the {\em
  Hipparcos Catalogue} $M_\mathrm{HIP}$, then columns 16 and 17 give
$M_\mathrm{HIP}-M_\mathrm{K}$.

Columns 19 and 20 indicate the connection with the catalogue of
Ptolemaios. Column 19 gives the sequence number $P$ of the entry in
the catalogue of Ptolemaios which corresponds to the entry in \classis,
in most cases based on a common Hipparcos identification, in which
case column 20 contains a 1. In some cases the correspondence is based
on positional coicidence (after taking precession into account), but the
identification is different: these have a 2 in column 20. This
includes cases where one or both of the corresponding stars is
unidentified.  A zero in column 19 indicates that no corresponding
entry is found, in these cases colum 20 has a 3, indicating a
genuinely new entry.

\subsection{The catalogue of Halley \label{s:halmachine}}

The machine-readable table \halley\ uses one line for each 2-line
entry in the catalogue as printed by Halley. Column 1 gives the
sequence number for the entry in the catalogue as a whole $E$ (for
Edmond), columns 2 and 3 the sequence number $C$ and abbreviation of
the constellation, and column 4 the sequence number of the entry
within the constellation. Columns 5-7 give the longitude in integers
$Z$ and $G$ and real $M$, such that the longitude in decimal
degrees is
$$ \lambda=(Z-1)*30+G+M/60$$.
Columns 8-9 give the latitude with integer $G$ and real $M$ 
such that the latitude in decimal degrees is
$$ \beta = \pm(|G|+M/60); \qquad  +/- \mathrm{for ~}G>0/G<0 $$
i.e.\ the sign of $G$ gives the sign of $\beta$.  
Column 10 gives the magnitude according to Halley.

Columns 11-14 give the Hipparcos number HIP of the first reference
star and the angular distance of the entry to it in degrees $G$,
arcminutes $M$ and arcseconds $S$, and columns 15-18 the same for the
second reference star.  In place of the distance to the second
reference star a longitude or latitude from Brahe or Kepler is used
in five cases; we indicate these with a negative value for HIP: 
$-$1 when a longitude from Brahe is used
(for E\,12), $-$2 latitude from Brahe (for E\,9), $-$3 longitude from
Kepler (for E\,11, E\,16 and E\,17).  Columns 15-20 give the results
from our analysis, viz. the Hipparcos number of our identification, a
flag $I$ indicating the quality of the identification, as explained in
Sect.\,\ref{sec:id}, the magnitude of the Hipparcos entry of column
15, the differences $\Delta\lambda\equiv\lambda_\mathrm{HIP}-\lambda$
in minutes as tabulated and
$\Delta\beta\equiv\beta_\mathrm{HIP}-\beta$ in arcminutes between the
longitude $\lambda_\mathrm{HIP}$ and latitude $\beta_\mathrm{HIP}$ as
derived from the Hipparcos catalogue and the values $\lambda$ and
$\beta$ derived from the values in \halley, and the distance $\Delta$ in
arcminutes between the position derived from the Hipparcos catalogue
and the position according to Halley.

The last four lines of the catalogue give additional angles to
reference stars, with sequence number of star, sequence number and
abbreviation of constellation, sequence number in constellation, and
Hipparcos identification with flag, in columns 1--4 and 19, 20; and
the Hipparcos number of the reference stars and the associated angles,
in column 11--14 and 15-18. 

\begin{figure}
\includegraphics[angle=270,width=0.95\columnwidth]{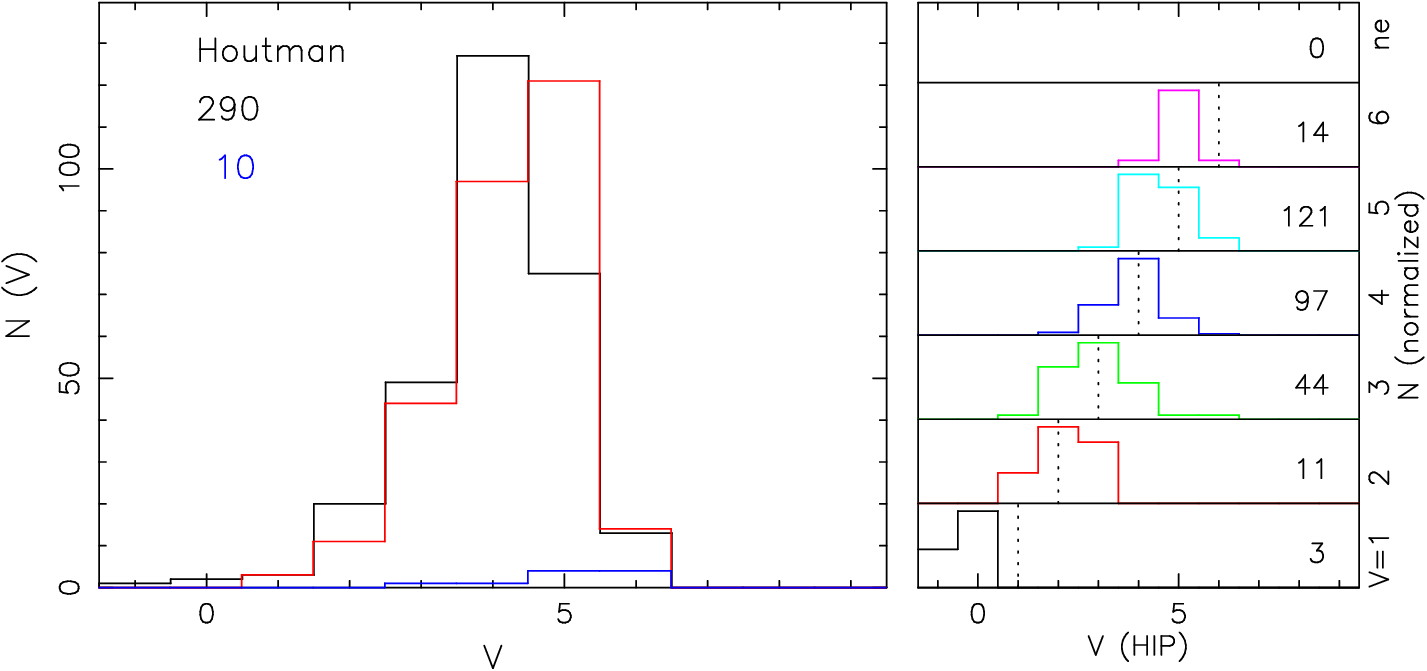}

\centerline{
\includegraphics[angle=270,width=0.475\columnwidth]{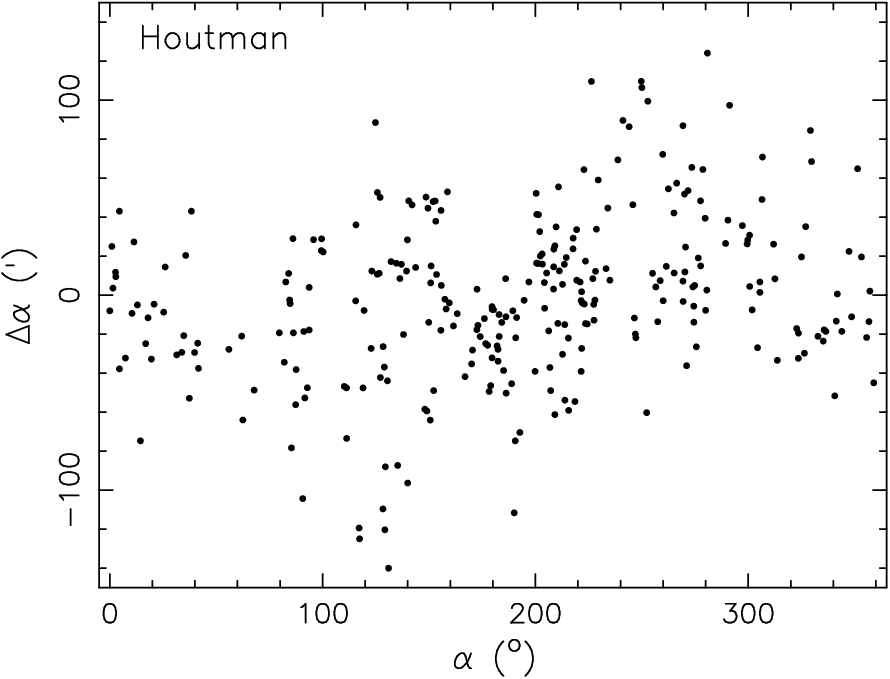}
\includegraphics[angle=270,width=0.475\columnwidth]{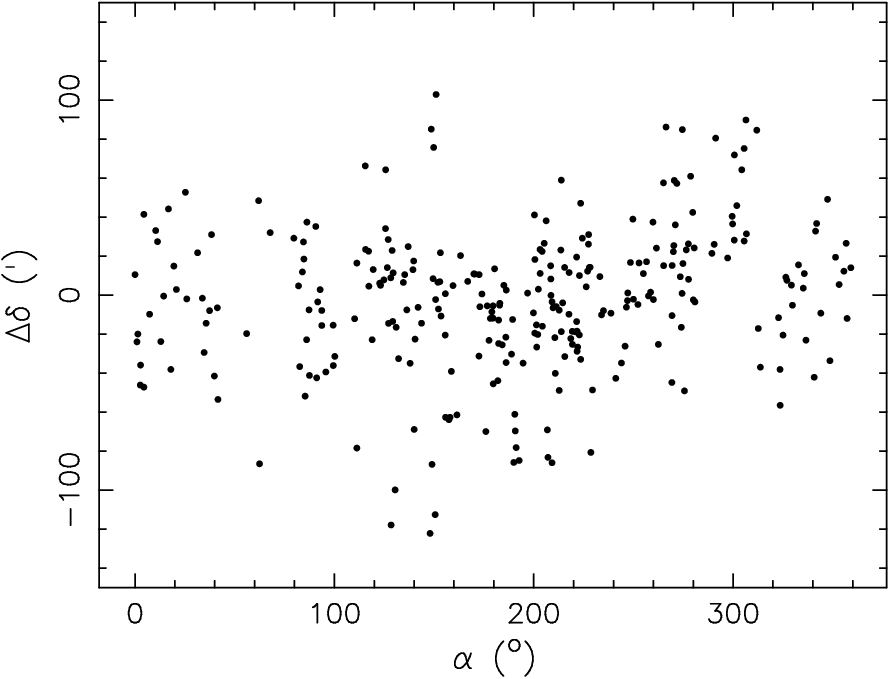}}

\centerline{\includegraphics[angle=270,width=0.95\columnwidth]{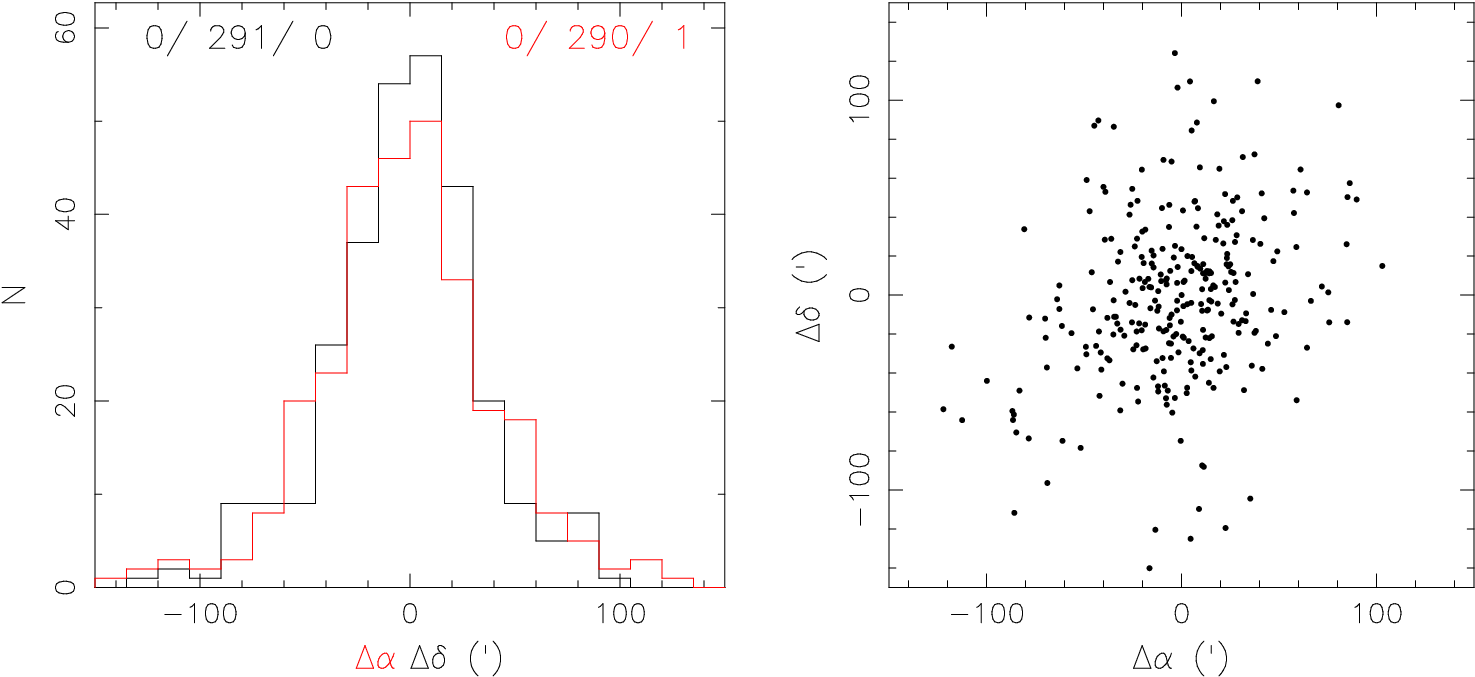}}

\includegraphics[angle=270,width=0.95\columnwidth]{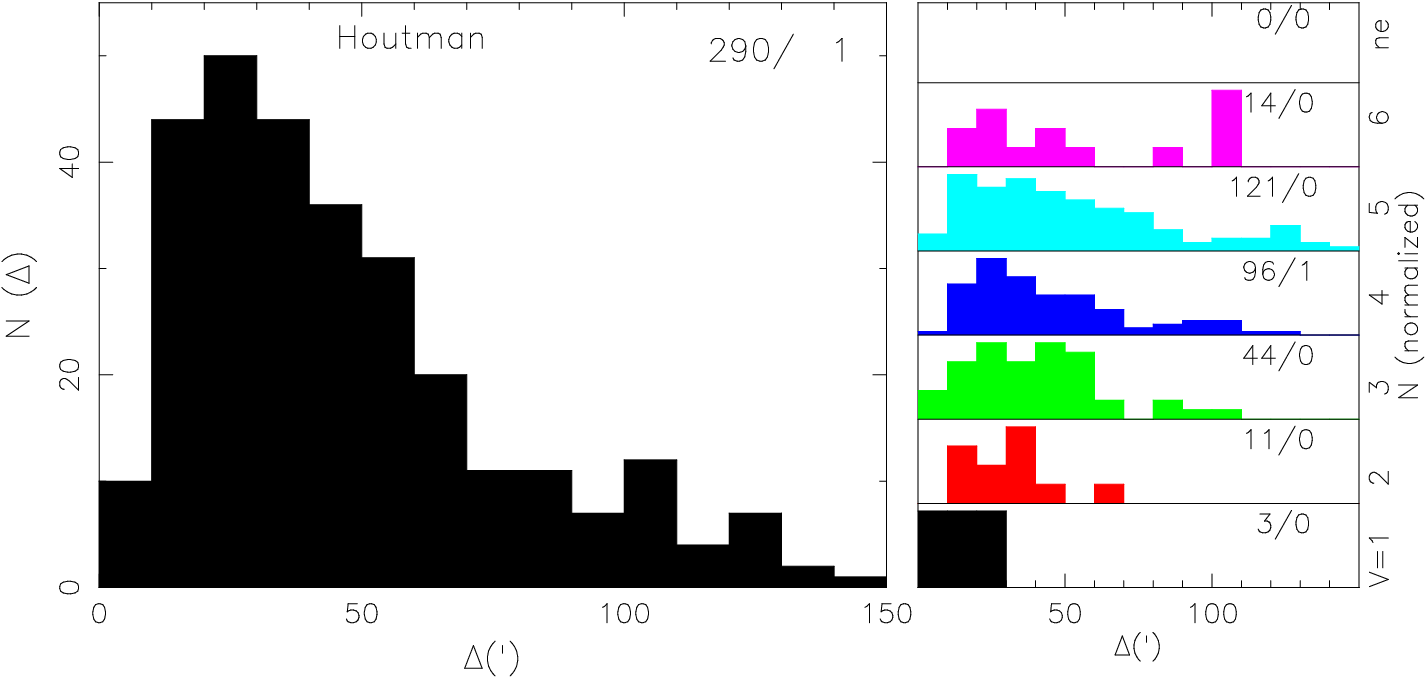}
\caption{Magnitude and position errors and their correlations in
  \houtman. \label{f:acchout}
 The top frame shows the distributions of magnitudes
  according to \houtman\ for securely identified stars (identification
  flags 1,2; red) and for not securely identified stars (flags 3-5;
  blue), and of the magnitudes in the {\em Hipparcos Catalogue} for
  all securely identified stars (black) in the large frame, and for each
  magnitude according to \houtman\ separately in the small frames.
  The middle frames show the errors $\Delta\alpha$ in right ascension
  and $\Delta\delta$ in declination as a function of right ascension
  and of one another, and histograms of the position errors for the
  securely identified stars separately for $\Delta\alpha$ (red) and
  $\Delta\delta$ (black). The numbers indicate stars within the frame
 (middle) and outside the frame to the left or right.
 The lower frame shows the distributions of position error $\Delta$,
 for all securely identified sources in the large frame, and as a
 function of magnitude in \houtman\ in the small frames. The numbers
 indicate stars within/outside each frame.}
\end{figure}

\section{Analysis and discussion}

\begin{table}
\caption{Frequency of identification flags in the four catalogues.
\label{t:knobel}}
\begin{tabular}{l|rrrrr|@{\hspace{0.5cm}}|r|r|@{\hspace{0.5cm}}|r}
I\verb+\+K & 0 & 1 & 2 & 3 & all & I$_\mathrm{S}$ & I$_\mathrm{T}$ & I$_\mathrm{E}$ \\
\hline 
1 &   3 & 185 &   0 &  19 & 207 & 162 & 82& 336 \\
2 &   0 &  76 &   0 &   8 &  84  & 71 & 48 & 5 \\
3 &   0 &   4 &   0 &   4 &   8 & 17 & 2 & 0 \\
4 &   0 &   2 &   0 &   0 &   2 & 0& 1 & 0 \\
5 &   2 &   0 &   0 &   1 &   3 & 9 & 3 & 0 \\
6 &  0 & 0 & 0 &        0 &  0  & 39 &0& 0 \\
all &  5 & 267 & 0 &  32 & 304 & 298&136 & 341
\end{tabular}
\tablefoot{The distribution of the identification flags is given
  for \houtman\ (I), \secunda\ (I$_\mathrm{S}$), \tertia\
  (I$_\mathrm{T}$), and \halley\  (I$_\mathrm{E}$). For \houtman\ we 
  further give the distribution of the flags K comparing
  identifications by Knobel (1917) with ours. (The meaning of the
  flags is explained in Sect.\,\ref{sec:id}.)}
\end{table}

\subsection{The accuracy of \houtman\label{s:acchout}}

Figure\,\ref{f:acchout} and Table\,\ref{t:knobel} display some results
of our analysis of the star catalogue of De Houtman (1603). The catalogue
consists of 304 entries, of which we claim to have identified 291 with
certainty, one of which (F\,46) with the Large Magellanic Cloud. In
ten cases we suggest plausible or possible identifications, and in
three cases we do not find a counterpart. Comparison with
\classis\ suggests that the first of these unidentified entries,
F\,3 (near $-$9.9,6.5 in Fig.\,\ref{f:phoenix}), corresponds
to K\,1319 (near $-$8.8,1.0), but obtained a wrong position due to a computational
or scribal error. The other two, in Indus, were discussed in 
Sect.\,\ref{sec:id} (see Fig.\,\ref{f:annotindus}).

As shown in Table\,\ref{t:knobel}, our identifications agree on the
whole with those by Knobel (1917). We identify three of the five stars
that Knobel (1917) left unidentified.  In 27 cases we do not agree
with the identification given by Knobel (1917) and give another
identification; in 4 cases we prefer an alternative but consider his
identification possible as well; and in the case of F\,3 comparison
with \classis\ (see discussion in the previous paragraph and in
Sect.\,\ref{s:notes}) leads us to reject his identification.

\begin{figure}
\centerline{\includegraphics[angle=270,width=0.85\columnwidth]{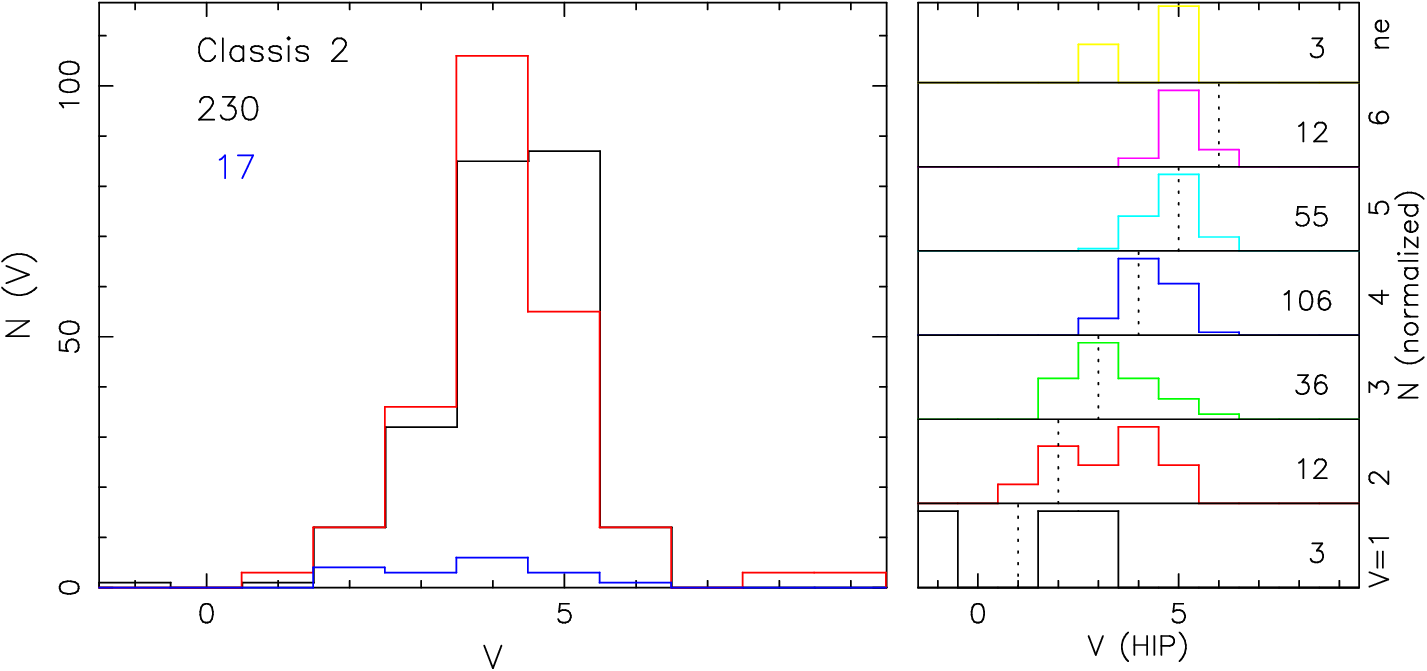}}

\centerline{\includegraphics[angle=270,width=0.85\columnwidth]{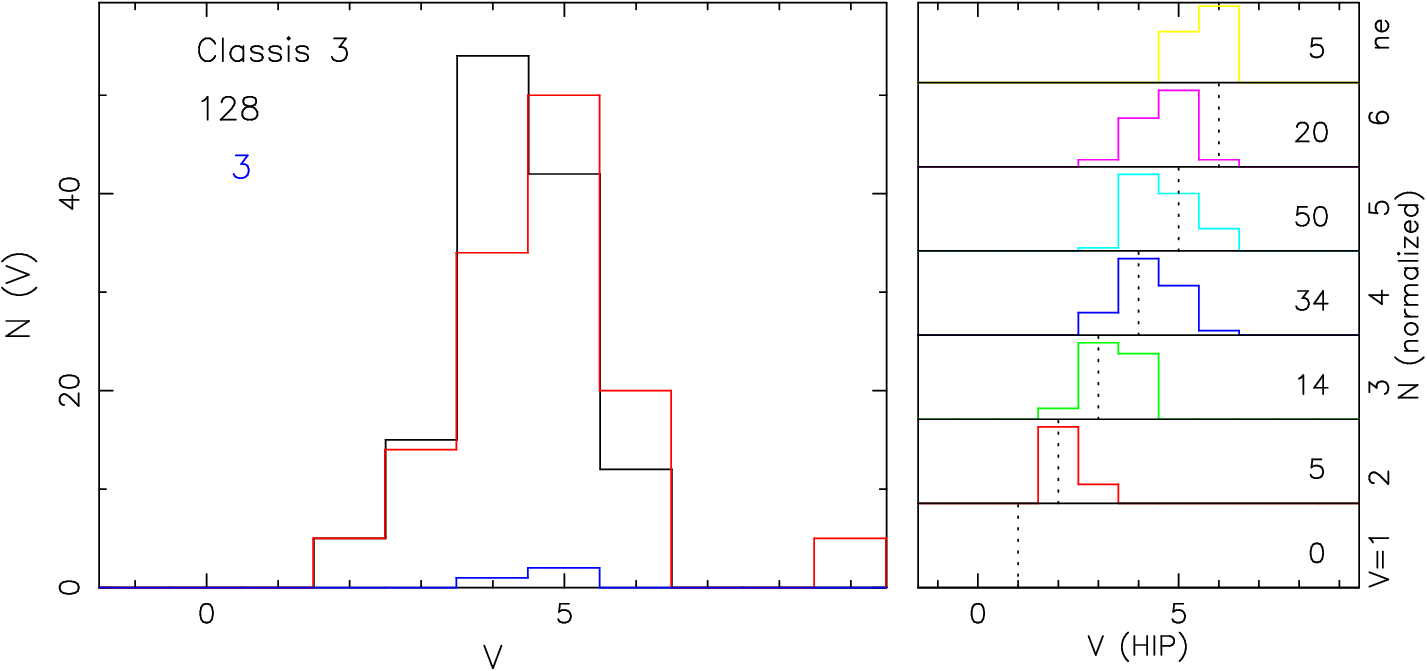}}

\centerline{
\includegraphics[angle=270,width=0.45\columnwidth]{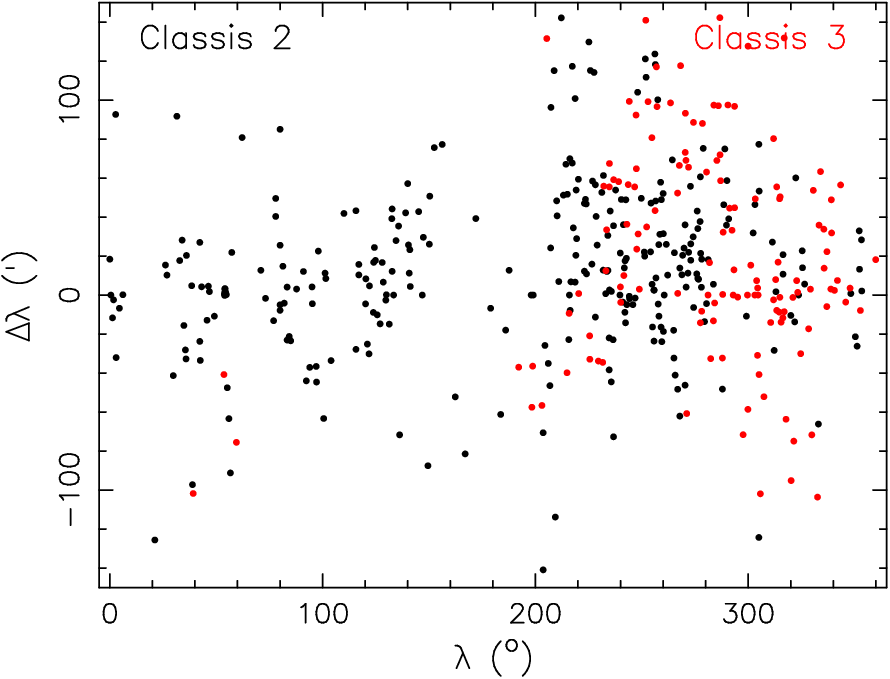}
\includegraphics[angle=270,width=0.45\columnwidth]{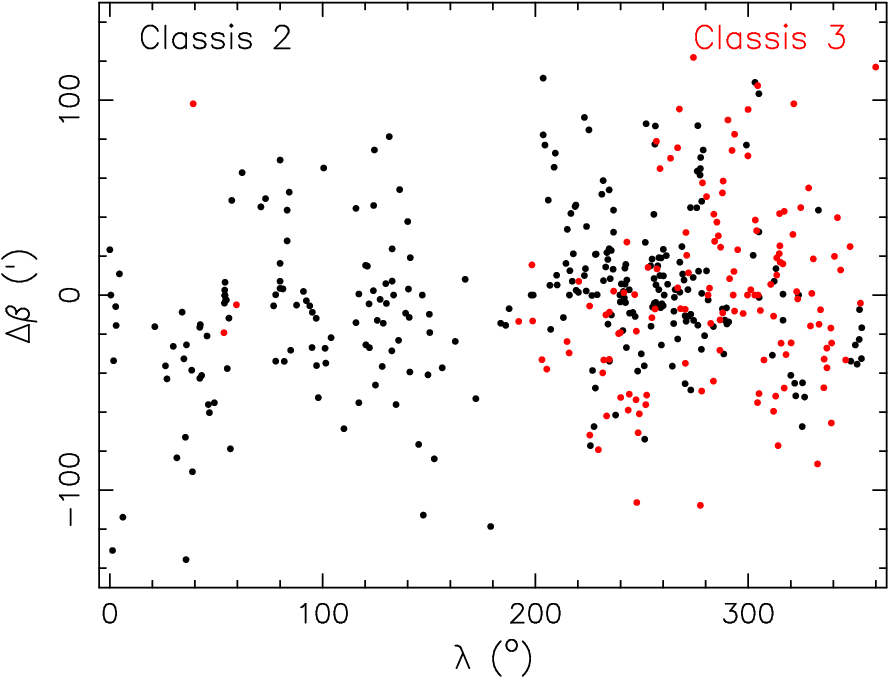}}

\centerline{\includegraphics[angle=270,width=0.9\columnwidth]{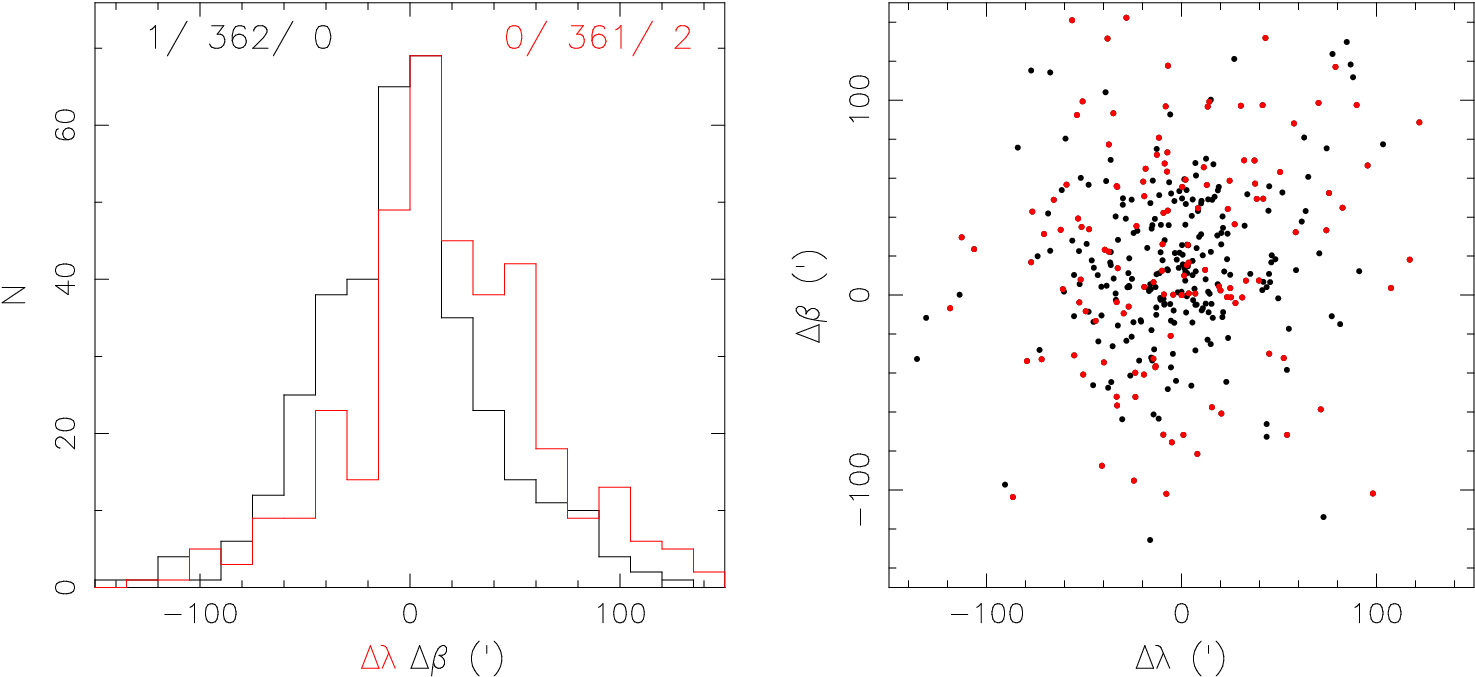}}

\includegraphics[angle=270,width=0.9\columnwidth]{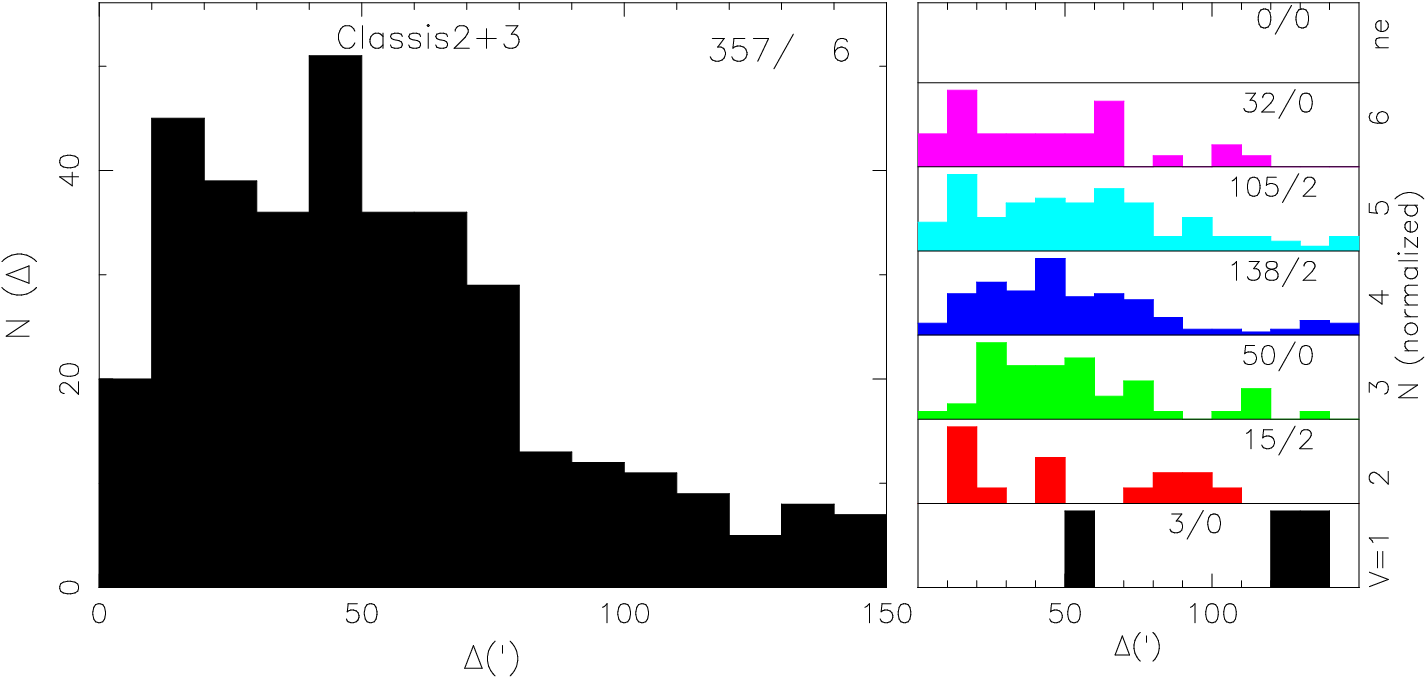}
\caption{Magnitude and position errors and their correlations in
  \secunda\ and \tertia. \label{f:accclas}
  As Fig.\,\ref{f:acchout}, but with position errrors in the
  ecliptic system. In the middle correlation frames black/red dots
  indicate stars from \secunda\ and \tertia, respectively}
\end{figure}

The magnitude distributions of the 290 entries securely identified
with a star in the {\em Hipparcos Catalogue} show a good correlation
between the magnitudes as assigned by De Houtman and those of the
counterparts in the {\em Hipparcos Catalogue}. The brightest category
in \houtman\ is magnitude 1: all three stars in this category
($\alpha$\,Eri = Achernar, $\alpha$\,Car = Canopus, and $\beta$\,Cen)
have $V_\mathrm{HIP}<0.5$. The faintest category in \houtman\ has
magnitude 6; most stars in this category have $V_\mathrm{HIP}=5$.

The position errors $\Delta\alpha$ and $\Delta\delta$ are a mixture of
random measurement errors and occasionaly large systematic errors, as
illustrated by the Figures in Appendix\,\ref{s:figures}, e.g.\
Fig.\,\ref{f:coraust}. As a result, the error distributions are not
gaussians. Separate fits of gaussians to the central peak ($N>$10 in
error bins of 15\arcmin, as shown in Figure\,\ref{f:acchout}) and the
whole distribution out to errors of 150\arcmin, we obtain
$\sigma\sim25$\arcmin\ and $\sigma\sim40$\arcmin, respectively.  Since
gaussians are not acceptable fits, these numbers should be considered as
rough indicators only of the position accuracy.  The errors in right
ascension and in declination do not show clear correlations with right
ascension or with one another.  The distribution of the total position
error $\Delta$ peaks near 25\arcmin, and shows a broad tail towards
$\sim100$\arcmin, with larger errors for fainter magnitudes.

\begin{table}
\caption{Stars in \secunda\ repeating a star in \keplere.
\label{t:doubles}}
\begin{tabular}{ccrrccr}
\multicolumn{3}{c}{\secunda} & & \multicolumn{3}{c}{\keplere} \\
K & Con & $\Delta_S$(\arcmin) & HIP & K & Con & $\Delta_K$(\arcmin)  \\ 
\hline 
1006 & Boo &  131.5 &  75312 & 143 & Boo &    2.8 \\
1012 & Oph &   10.4 &  78727 & 685 & Lib &    3.2 \\
1013 & Oph &   34.0 &  78207 & 684 & Lib &    2.6 \\
1014 & Oph &   19.5 &  77853 & 683 & Lib &    3.3 \\
1015 & Oph &   18.6 &  80628 & 348 & Oph &    3.9 \\
1023 & Oph &   25.1 &  86284 & 336 & Oph &   46.4 \\
1024 & Oph &   25.1 &  84893 & 341 & Oph &   55.7 \\
1027 & Oph &    5.1 &  84970 & 355 & Oph &  131.6 \\
1028 & Oph &   10.3 &  85340 & 356 & Oph &   86.7 \\
1029 & Oph &   22.0 &  85755 & 357 & Oph &   97.5 \\
1032 & Oph &   33.4 &  88149 & 349 & Oph &    3.3 \\
1033 & Oph &   23.2 &  88192 & 350 & Oph &    2.7 \\
1034 & Oph &   53.4 &  88601 & 352 & Oph &    4.9 \\
1035 & Oph &   41.1 &  88290 & 351 & Oph &    2.7 \\
1037 & Oph &   64.3 &  88771 & 342 & Oph &    4.4 \\
1038 & Ser &   15.9 &  77257 & 343 & Oph &    2.6 \\
1040 & Ser &   17.5 &  84880 & 344 & Oph &    5.9 \\
1041 & Ser &   24.3 &  86263 & 345 & Oph &    4.7 \\
1042 & Ser &   21.2 &  86565 & 346 & Oph &    2.8 \\
1054 & Ari &   31.6 &  13061 & 504 & Ari &    3.8 \\
1055 & Ari &   32.6 &  12828 & 823 & Cet &   13.3 \\
1057 & Tau &    4.1 &  17499 & 534 & Tau &   37.1 \\
1058 & Tau &    2.8 &  17702 & 536 & Tau &    1.5 \\
1059 & Tau &    3.3 &  17608 & 535 & Tau &    8.6 \\
1062 & Tau &    2.5 &  17847 & 537 & Tau &    2.5 \\
1063 & Tau &   56.3 &  16852 & 918 & Eri &    0.6 \\
1064 & Tau &   47.0 &  23497 & 522 & Tau &   35.8 \\
1068 & Tau &   69.7 &  26640 & 326 & Aur &    1.3 \\
1069 & Tau &   34.2 &  27468 & 327 & Aur &    0.9 \\
1070 & Tau &   36.1 &  27830 & 324 & Aur &    2.1 \\
1071 & Tau &   56.9 &  28237 & 325 & Aur &    2.0 \\
1072 & Gem &    6.0 &  29696 & 323 & Aur &    1.4 \\
1127 & Eri &   53.3 &  20507 & 917 & Eri &    2.2 \\
1128 & Eri &  120.1 &  19849 & 916 & Eri &    0.7 \\
1129 & Eri &  123.6 &  13701 & 911 & Eri &    0.6 \\
1148 & CMa &   38.5 &  31592 & 938 & CMA &   18.8 \\
1205 & Hya &   78.3 &  42313 & 980 & Hyd &    1.4 \\
1206 & Hya &   14.9 &  42799 & 963 & Hyd &    1.3 \\
\multicolumn{3}{c}{\secunda} & & \multicolumn{3}{c}{\tertia} \\
1296 & PsA &   41.0 & 108085 & 1303 & Gru & 11.0 
\end{tabular}
\tablefoot{For each repeat star in \secunda, the table gives
  the constellation and position error $\Delta_s$ to its identification HIP, 
  and the sequence number and constellation of the corresponding 
  entry in \keplere\ with its position error $\Delta_K$. In one
  case the corresponding entry is in \tertia}
\end{table}

\subsection{The accuracy of \secunda\ and \tertia}

Figure\,\ref{f:accclas} and Table\,\ref{t:knobel} display some results
of our analysis of the second and third classes of Kepler (1627).  The
Table shows that \secunda\ contains 39 repeat entries; all except one
correspond to entries from \keplere\ (Table\,\ref{t:doubles}). K\,1296 repeats
(or foreshadows) K\,1303 in \tertia. With few exceptions, the repeat
entries in \secunda\ have rather less accurate positions than the
corresponding entries in \keplere. Among the exceptions are some
southern stars in Ophiuchus (Fig.\,\ref{f:ophiuchus}).

The three securely identified entries (flags 1, 2) with no magnitude
or a magnitude `nebulous' in the upper small frame for \secunda\ in
Fig.\,\ref{f:accclas} are K\,1010 in Cygnus (possibly indicating a
combination of two stars, see Sect.\,\ref{s:notes}), K\,1099 in
Sagittarius (the close binary $\nu^1,\nu^2$\,Sgr), and K\,1093 a
single star for which it is not clear why it would be marked nebulous.
For three new stars of the Pleiades \secunda\ gives no magnitude. The
five securely identified (flags 1,2) stars with a magnitude `nebulous'
in the frame for \tertia\ are K\,1324 and K\,1325 in Phoenix
(Fig.\,\ref{f:phoenix}), K\,1339 in Indus (Fig.\,\ref{f:annotindus}),
and K\,1349 and K\,1351 in Pavo (Fig.\,\ref{f:pavo}), for none of
which there is an obvious reason for the label nebulous.

Separate fits of gaussians to the central peak of the distributions of
errors $\Delta\lambda$ in longitude and $\Delta\beta$ in latitude
($N>$10 in error bins of 15\arcmin, as shown in
Figure\,\ref{f:accclas}) and the whole distribution out to errors of
150\arcmin, give $\sigma\sim35$\arcmin and $\sigma\sim45$\arcmin,
respectively.  Separate fits for \secunda\ and \tertia\ show that the
errors in \secunda\ are on average about 5\arcmin\ smaller; those in
\tertia\ are larger by up to 10\arcmin. However, as the errors are the
sum of random and systematic errors, not well described by gaussian
distributions, these numbers must be considered as rough indicators of
the position accuracy only.  The errors in longitude and in latitude
do not show clear correlations with right ascension or with one
another.  The distribution of the total position error $\Delta$ has a
broad peak from 20 to 70\arcmin, and a broad tail towards larger
error. The importance of systematic errors causes the errors for
bright stars to be similar to those of fainter stars.

\subsection{Comparison between \classis\ and the star catalogue of
  Ptolemaios\label{s:ptolclas}}

To compare the stars in \classis\ with those in the star catalogue of
Ptolemaios we use the edition of the latter by Toomer (1998).  For
this we convert the identifications given by Toomer to Hipparcos
numbers; and precess the coordinates of the entries in \classis\ to
the epoch of the catalogue of Ptolemaios. In doing so, we assume that the
positions in the star catalogue of Ptolemaios, after subtraction from the
longitude of 2\degr40\arcmin, correspond to the epoch of $-$128 (= 129
BC), as stated by Ptolemaios in the Almagest (Toomer 1998, p.333).
\footnote{The longitudes thus found for the catalogue of
  Ptolemaios are more accurate than the tabulated longitudes for the
  epoch of Ptolemaios, 137 AD, which are systematically too low by
  about a degree. This suggests that Ptolemaios did not make his own
  measurements, but corrected the positions measured by Hipparchos
  with his estimate for the change due to precession, i.e.\
  2\degr40\arcmin; the correct value being $\simeq$3\degr40\arcmin.  An
  ingenious analysis by Duke (2003) shows that Ptolemaios did indeed
  copy most of his star positions from Hipparchos.}
In some cases the correspondence with Ptolemaios is explicit in the
description of the entry in \classis. For example, K\,1006 is
described 10 Ptol.\ in \classis, and K\,1040-1042 as 13-15 Ptol.  In
the other cases we determine corresponding pairs by their having the
same counterpart from the Hipparcos Catalogue, and graphically by
similar positions in the constelllation.

\begin{figure}
\centerline{\includegraphics[angle=270,width=\columnwidth]{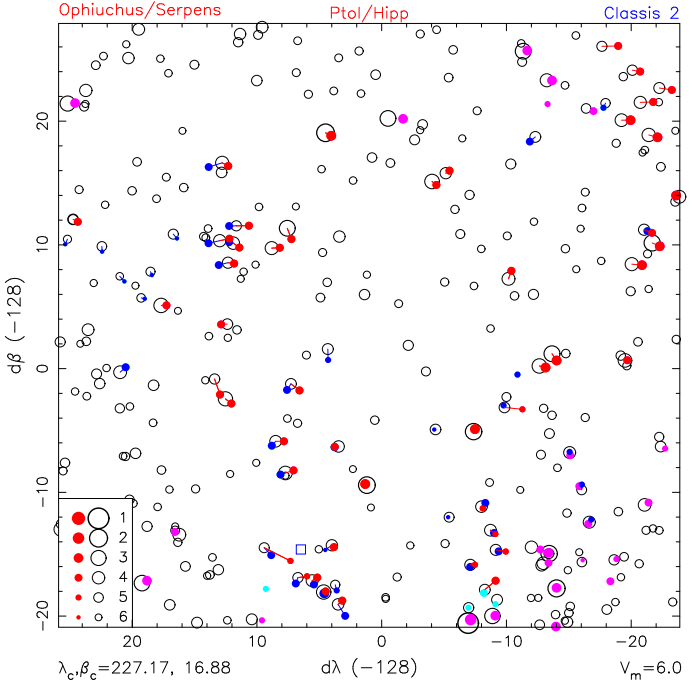}}

\caption{The stars of Ophiuchus and Serpens in the star catalogue of
  Ptolemaios (red; other stars from this catalogue in purple) and in
  \secunda\ after precession of the coordinates to the epoch of
  Ptolemaios/Hipparchos (dark blue; other stars in \classis\ in light
  blue). Open black circles are stars from the modern Hipparcos
  Catalogue. Short solid lines indicate identifications.  
  In many cases stars in \classis\ are matched by an entry in
  the star catalogue of Ptolemaios, occasionally by an entry in a 
  different constellation. New stars in \classis, not matched by an
  entry in Ptolemaios' catalogue are also seen, including the
  supernova of 1604, indicated as a blue square near 6,$-$14.
  \label{f:ptolclas}}
\end{figure}

An example of the graphical comparison is given in
Fig.\,\ref{f:ptolclas}, in which we show the constellations of
Ophiuchus and Serpens, the constellations to which Kepler added most
new stars.  Short solid lines connect the catalogue entries with the
center of their Hipparcos identifications in this Figure, enabling
for many \classis\ entries  a straightforward determination whether they
have a corresponding entry in the star catalogue of Ptolemaios.
(Note that in some cases the entry from the Hipparcos
catalogue is almost hidden behind the \classis\ entry in the Figure,
{\em viz.} K\,1043 and K\,1019, near 3.8,$-$6.3 and $-$9.1,$-$13.4
in Fig.\,\ref{f:ptolclas}.)

Among the entries in \classis\ that are not matched in the star
catalogue of Ptolemaios we note the globular clusters M\,13
in Hercules (K\,1009) and $\omega$\,Cen, SN\,1604 in Ophiuchus (K\,1036), 
and four new stars in the Pleiades (K\,1056 and K\,1058-1060),

\subsection{Comparison between \houtman\ and \classis\label{s:houtclas}}

In Table\,\ref{t:matches} we compare the numbers of entries in those
constellations that are listed both in \houtman\ and in \classis.  and
the numbers of matching pairs, i.e.\ entries in \houtman\ and
\classis\ that are identified with the same star in the {\em Hipparcos
  Catalogue}. We see from the Table that 99 of the 112 entries in the twelve 
new constelllations in \houtman\ are matched by a star in \classis,
all of them in \tertia. To this may be added 8 matches not
with the same Hipparcos identification, viz.\ F\,3 (see
Sect.\,\ref{s:notes}) in Phoenix, F\,46 (the Small Magellanic Cloud),
and three stars each in Pavo and Indus whose pattern matches
between \houtman\  and \classis, even though identified with
different Hipparcos stars (Figs.\,\ref{f:annotindus}, \ref{f:pavo}).
Thus the new constellations as listed in \houtman\ contain only five
entries not found in \classis. Conversely, \tertia\ contains 29 stars
not in \houtman.

\begin{table}
\caption{Numbers of matching stars in \houtman\ and \tertia\ and
\secunda.
\label{t:matches}}
\begin{tabular}{crrr|crrr}
Con & H\phantom{\,} & 3C & M & Con & H\phantom{\,} & 2C & M \\
\hline 
Phe & 13 & 15 & 12+1 & CrA & 16 & 13 & 11 \\
Hyi & 16 & 21 & 12+1 & Eri & 7 & 20 & 1 \\
Dor & 4 & 7 & 3 & Col & 11 & 16 & 9+1/1 \\
Mus & 4 & 4 & 4 & Arg & 56 & 42 & 22 \\
Vol & 5 & 7 & 5 & Cen & 53 & 33 & 27/1 \\
Cha & 9 & 10 & 9 & Lup & 29 & 19 & 14 \\
TrA & 4 & 5 & 4 & Ara & 12 & 7 & 6/1 \\
Aps & 9 & 11 & 7/2 & Sco & 8 & 17 & 8 \\
Pav & 19 & 23 & 16+3 \\
Ind & 11 & 12 & 8+3 \\
Gru & 12 & 13 & 12 \\
Tuc & 6 & 8 & 5 \\
\hline
tot & 112 & 136 & 99+8 & tot & 192 & 167 & 101+1
\end{tabular}
\tablefoot{For each constellation the numbers of stars in \houtman\
and in \secunda\ (2C) or \tertia\ (3C) is given as well as the number of
matching pairs. Most numbers indicates matches based on identifiction
with the same star in the {\em Hipparcos Catalogue}.
A match of a star in \houtman\ with a star in a
different constellation in \classis\ is indicated behind a
slash; Colomba and Canis Maior are treated as one constellation.
The numbers behind a + sign indicate pairs that we consider
matches in addition, even though identified with different Hipparcos
stars, or unidentified.}
\end{table}

In comparing constelllations from Ptolemaios in \houtman\ with
\classis, we see that 101 entries in \houtman\ are identified with the
same star in the {\em Hipparcos Catalogue} as a star in \classis, all
in \secunda.  The pair F\,63--K\,1153 in Colomba / Canis Maior
may also be considered a match even though they are identified
with different Hipparcos stars (near 3,$-$0.5 in
Fig.\,\ref{f:colomba}). Thus there are 90 stars in these
constellations in \houtman\ not matched in \classis.
Four of these, the northernmost stars in Centaurus, are actually
in \keplere.

Knobel (1917) gives a table in which the numbers of stars new with
respect to Ptolemaios are given, for each constellation separately.
For the twelve new constellations, these numbers are virtually
identical to the total number of stars in them; of the stars in
\houtman\ but not in \tertia, one star in Hydrus and one in Grus are
considered by Knobel to be present in the catalogue of Ptolemaios (we
find one in Grus, F\,287=P\,1022, but none in Hydrus).  For the old
constellations, Knobel's numbers are very close to the numbers of
stars in each constellation in \houtman\ minus the numbers of matches
with \secunda. This is understandable, as \secunda\ is mainly composed
of stars from the catalogue of Ptolemaios (see previous Section). One
star in Corona Austrinus, two in Centaurus/Crux (not counting the four
matches with \keplere), and one in Lupus, not matched in \secunda, are
considered by Knobel to be matched with stars from the catalogue of
Ptolemaios.

\begin{figure}
\centerline{\includegraphics[angle=270,width=0.85\columnwidth]{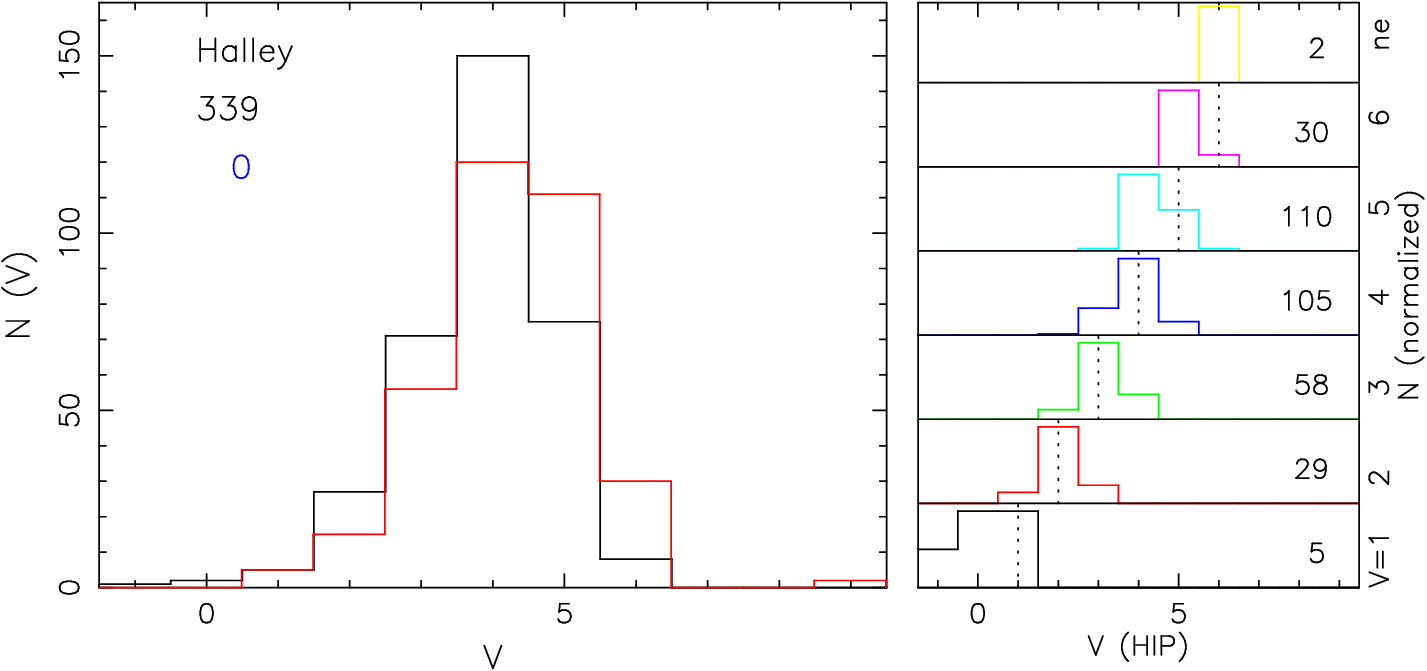}}

\centerline{
\includegraphics[angle=270,width=0.45\columnwidth]{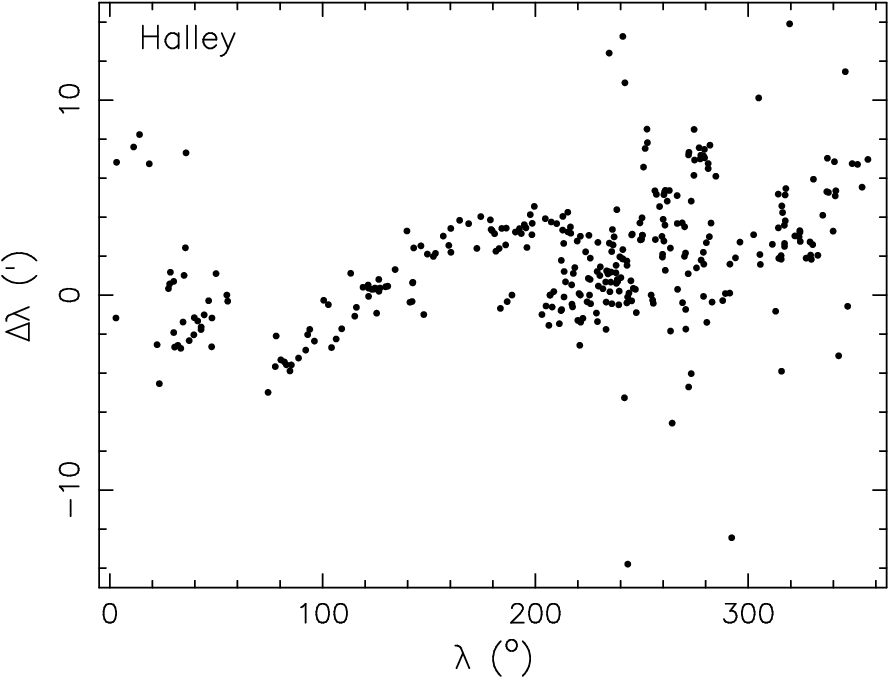}
\includegraphics[angle=270,width=0.45\columnwidth]{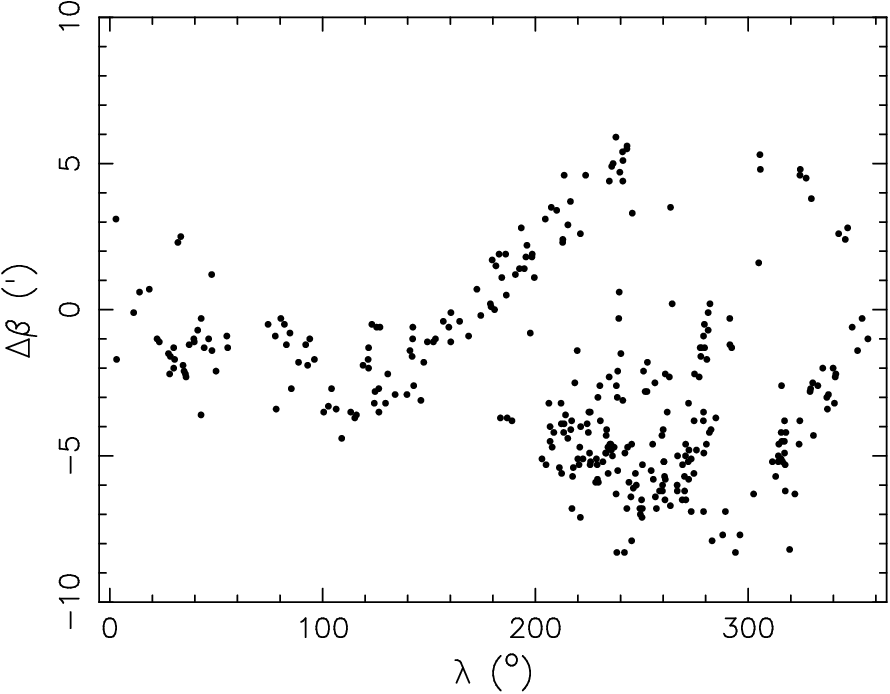}}

\centerline{\includegraphics[angle=270,width=0.9\columnwidth]{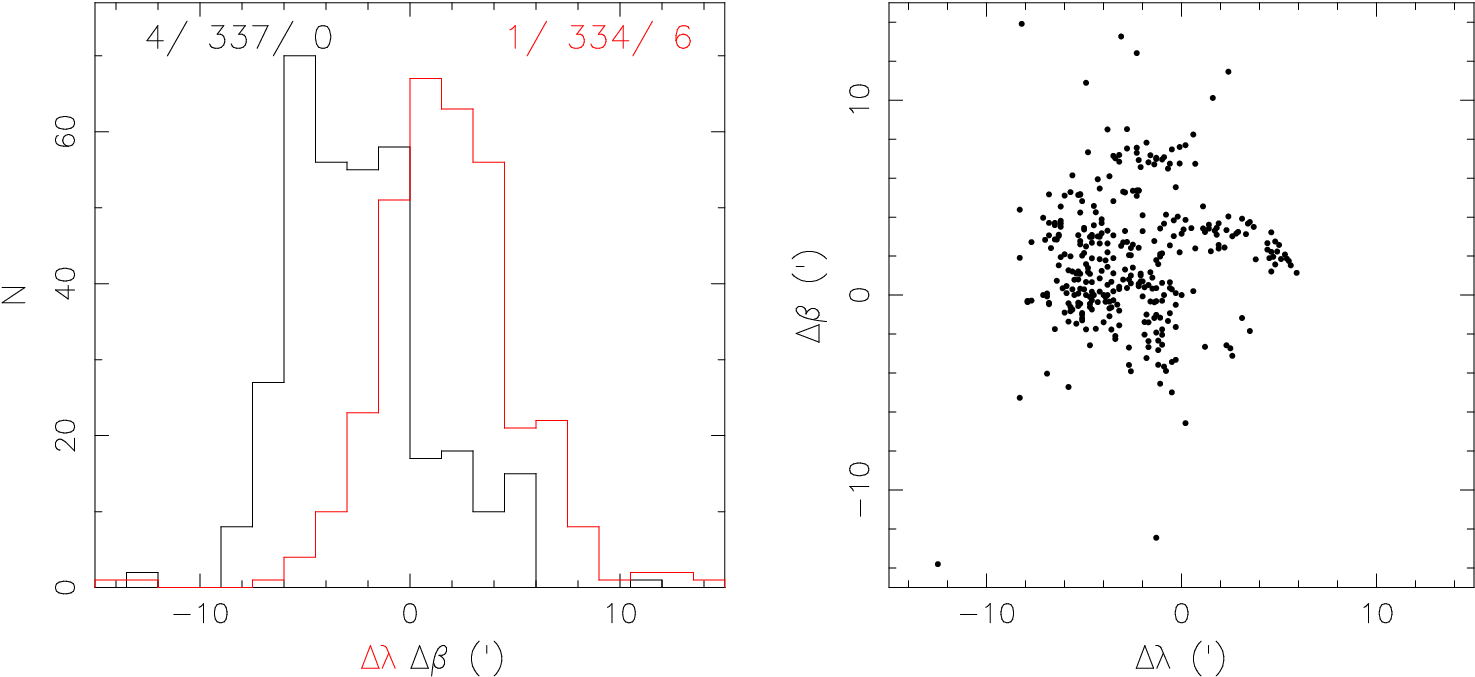}}

\includegraphics[angle=270,width=0.9\columnwidth]{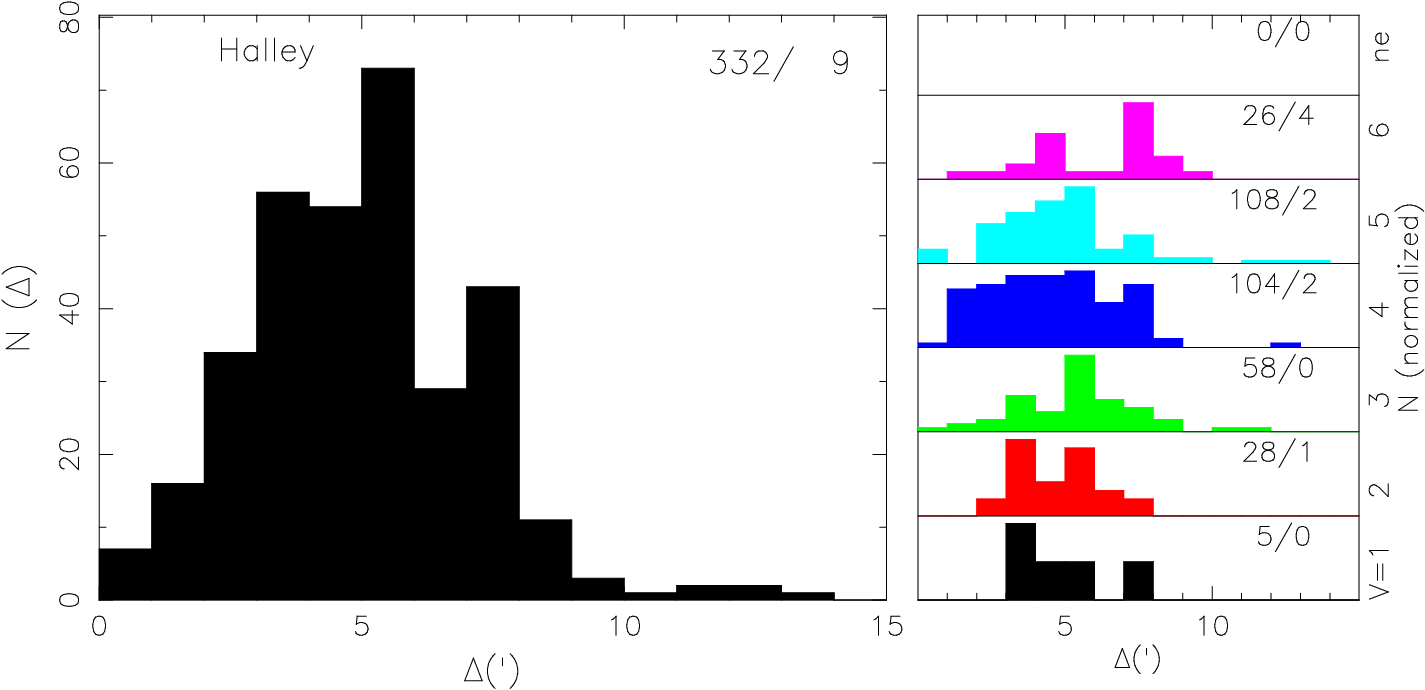}
\caption{Magnitude and position errors and their correlations in
  \halley. \label{f:acchall}
  As Fig.\,\ref{f:acchout}, but with position errrors in the
  ecliptic system.}
\end{figure}

\begin{figure}
\includegraphics[angle=270,width=\columnwidth]{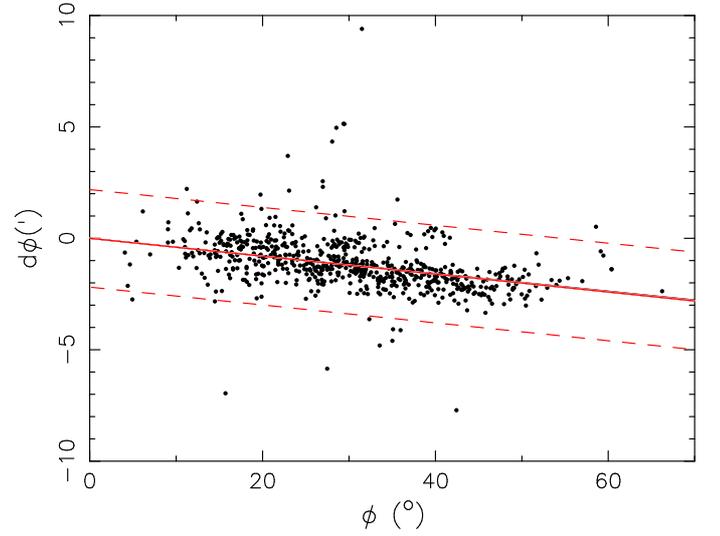}

\caption{For each angle $\phi$ between an entry in \halley\ and a
reference star, the figures shows the difference $d\phi\equiv\phi_E-\phi$ between
the catalogue value $\phi_E$ and the modern value $\phi$, as a function
of $\phi$.\label{f:sextant} The red line shows a linear fit, forced to
go through (0,0), in which points deviating more than 3-sigma have
been excluded. The 3-sigma range is indicated with the dashed lines.}
\end{figure}

Our analysis of \secunda, \tertia, and \houtman\ leads us to agree
with the conclusions of Stein (1917) that the data for the twelve new
constellations were mostly obtained during the {\em Eerste
  Schipvaert}, to which Keyser and De Houtman both contributed, and
that the data for the new stars in the Ptolemaean constellations were
obtained by De Houtman during his stay in Sumatra. Indeed, this agrees
with the statement of De Houtman in his introduction to the star
catalogue, except that he does not acknowledge a contribution by Keyser.

Knobel (1917) correctly notes that the twelve new constellations can not
have been observed by De Houtman from Sumatra, but his conclusion
that De Houtman merely plagiarized Keyser is not warranted.
The fact that \secunda\ has none of the stars added by \houtman\ to
the Ptolemaean constellations, indicates that Kepler in 1627 used
only data from the {\em Eerste Schipvaert}. The large similarity
between the lists of stars in \tertia\ with those for the twelve new
constellations in \houtman\ then indicates a connection between the two
{\em viz.} the observations obtained during the {\em Eerste
  Schipvaert}.

Dekker (1987) shows significant differences between the depictions of
the twelve new constellations by Hondius on globes in 1598 and 1601
(and therefore based on data from the {\em Eerste Schipvaert}) and
those by Blaeu on his globe of 1603 (based on data from De Houtman).
Dekker concludes that Keyser and De Houtman recorded their
observational data separately. The large overlap in the star lists
lead us to suggest an alternative explanation, {\em viz.} independent
reductions by Plancius and Blaeu of a shared set of observational data.
The presence in \tertia\ of 29 stars not in \houtman\ then shows that de 
Houtman did not use all stars from the shared set, possibly because he
omitted the stars that he had not measured himself.

\subsection{The accuracy of \halley}

Figure\,\ref{f:acchall} and Table\,\ref{t:knobel} display some results
of our analysis of the star catalogue of Halley (1679), and illustrate
its remarkable accuracy.  Separate fits of gaussians to the
distributions of errors $\Delta\lambda$ in longitude and $\Delta\beta$
in latitude as shown in Figure\,\ref{f:acchall} give
$\sigma\sim3$\arcmin\ for both, with offsets of 2\arcmin\ for
$\Delta\lambda$ and $-$2.6\arcmin\  for $\delta\beta$.
These numbers justify what Halley writes in the
introduction to the catalogue (we translate from the Latin):
\begin{quotation} Then there is a rumour that a certain Dutchman
Frederick [de] Houtman has made an effort on these stars on the island of
Sumatra, and that Willem Blaeu [used] his observations to correct the
celestial globe which he [Blaeu] published. Which instruments he used is not
known to me, but from a comparison made of his globe with our
catalogue, it is sufficiently and abundantly clear that this observer
was little practised in this arena.
\end{quotation}
This is somewhat harsh on De Houtman, who probably used an ordinary
mariner's cross-staff, whereas Halley used an astronomical sextant
with telescopic sights, made specifically for his observations in
St.\,Helena, probably in the Ordnance Office (Cook, 1998, pp.\,38).

Fig.\,\ref{f:acchall} shows that the correlations of the errors in
longitude $\Delta\lambda$ and latitude $\Delta\beta$ with longitude
$\lambda$ clearly are not random. (The largest errors, of three
stars in Lupus, observed only from the ship while sailing, are
outside the frames.) The position errors of the stars in
\halley\ are the result of his own measurement errors and of the
errors in the positions of the reference stars that he used, from the
catalogue of Brahe. Our identifications of the reference stars are
given in Table\,\ref{t:halleyrs}.  The angular distances given in the
catalogue of each entry to the reference stars allow us to test the
measurements errors of Halley, by comparing them with the values
derived from the {\em Hipparcos Catalogue}, taking into account the
proper motion between 1991.25 and 1678.0.  The result is shown in
Figure\,\ref{f:sextant}.  The sextant shows a systematic error,
increasing roughly linearly to about 2\arcmin\ at the largest angles of
about 60\degr, on which errors with a standard deviation
$\sigma\simeq$0\farcm7 are superposed.

Table\,\ref{t:halleyrs} shows that both the systematic errors and the
random errors in the angles with the reference stars are small with
respect to the position errors of many reference stars themselves,
both of some primary references (e.g.\ $\sigma$\,Sgr and Fomalhaut)
and of secondary references derived from them. These errors explain
both the range and the systematics seen in the error correlation plots
in Fig.\,\ref{f:acchall}. Even so, it may be concluded that Halley
achieved a stunning accuracy in his catalogue.

A problem of identification with earlier catalogues is mentioned by
Halley following his list of the stars in Argo. He remarks that
Ptolemaios assigns a magnitude 2 to the 31st star in his list for the
constellation of Argo, but that there is now little or no remanant of
this star. Halley qualifies this by noting that in particular the
southern stars in Argo have such large positional errors that
identification with the stars in his catalogue cannot be be made
safely. In fact, it appears to us that Halley's entry E\,118 is a good
match for Arg\,31 of Ptolemaios: both entries have magnitude 2 in the
old catalogues, and both are identified with HIP\,44816
($\lambda$\,Vel), a $V$=2.2 star.

\begin{acknowledgements}
  We thank Jetze Touber and Rienk Vermij for help in translating the
  Latin texts heading \secunda\ and \tertia.  This research has made
  use of the SIMBAD database, operated at CDS, Strasbourg, France, and
  was supported by the Netherlands Organisation for Scientific
  Research under grant 614.000.425.

\end{acknowledgements}

\begin{appendix}
\section{Annotations and Emendations}

\subsection{Emendations to Knobel}

\noindent Vel\,55 is identified by Knobel as 215=x\,Vel from 
the  Uranometria Argentina; we emend  to 225=x\,Vel

\noindent Pav\,4. Knobel has 32=$\nu$\,Pav; we emend to 33=$\nu$\,Pav

\subsection{Annotations and emendations to De Houtman}

\noindent F\,273 and F\,274 have $\alpha$=330\degr36\arcmin and
209\degr45\arcmin\ in the catalogue, which puts both well outside
the constellation. Since the descriptions {\em het hert} (the heart) and
{\em Een onder dese} (one below this) puts F\,272 in the
body and F\,274 below it, we emend to $\alpha$=300\degr36\arcmin
and 299\degr45\arcmin. The emended positions are near 6.5,$-$3.3
in Fig.\,\ref{f:pavo}.

\subsection{Emendations to Classis\label{s:classisem}}

\noindent K\,1064 has $\lambda$=\taurus\ 11\,00 in the catalogue, which
puts it well outside the constellation. We emend to \gemini\ 11\,00,
in accordance with the description {\em Trium supra australe cornu
  praeced.} (the leading of the three above the southern horn).

\noindent K\,1139 \&\ K\,1140 have zodiacal sign \aries\ for their
longitude in \classis, but Kepler notes \taurus\ as an alternative.
This corresponds to their location in the star catalogue of
Ptolemaios, and we make the emendation. Kepler also gives an
alternative latitude $\beta$=53\degr\ A for K\,1139, but since this
does not fit its description as being north of K\,1140 we do not make
this emendation.

\subsection{Annotations and emendations to Halley\label{s:annothal}}

\noindent E\,14 and E\,15 have zodiacal sign \scorpio\ in the catalogue;
we emend to \sagittarius.

\noindent E\,40. The angle to reference star HIP\,97649 ({\em lucida Aquilae}) is
given as the angle to reference star HIP\,86032 ({\em caput Ophiuchi}) and
v.v.\ in the catalogue. We emend the star names.

\noindent E\,157 has $\lambda=$\libra\,3\degr59\arcmin; we emend to
\libra\,8\degr59\arcmin.

\noindent E\,193 = $\beta$\,Cen has an uncharacteristically large
position error $\Delta$=44\farcm1. It is used as a reference star for the four stars in
Musca: these stars have position errors between 5\farcm2 and
5\farcm9, implying that Halley used a more accurate position for
E\,193. The table has $\lambda$ of E\,193 as \virgo 18 18.5, Halley
probably used \virgo 19 18.5, which gives $\Delta$=5\farcm9.
We do not emend this, as it may not have been obvious to a
contemporary of Halley.

\noindent E\,300. The angle to reference star HIP\,30438 ({\em Canopus}) is given
as the angle to reference star HIP\,52727 ({\em clarior in summ. Roboris
Carolini}) and v.v.\ in the catalogue. We emend the star names.

The descriptions of the reference stars given by Halley are listed in
Table\,\ref{t:halleyrs}, together with our identifiation of them, and
with the error $\Delta$ in their position.

\begin{table}
\caption{Reference stars of Halley and their position error.
 \label{t:halleyrs}}
\begin{tabular}{lcrr}
name in Halley & Bayer & HIP & $\Delta$(\arcmin)\\
\hline
Spica Virginis & $\alpha$\,Vir & 65474 & 2.6\\
Yed Ophiuchi & $\delta$\,Oph & 79593 & 1.0\\ 
lucida colli Serp. & $\alpha$\,Ser & 77070 & 2.7\\
Lanx Austrina & $\alpha^2$\,Lib & 72622 & 3.0\\
genus Ophiuchi & $\eta$\,Oph & 84012 & 3.7\\
humerus (sin.) \sagittarius & $\sigma$\,Sgr  & 92855 & 7.7 \\
Cor Scorp. & $\alpha$\,Sco & 80763 & 4.2 \\
oculus Pavonis & $\alpha$\,Pav & 100751 & (6.9)  \\
lucida Aquilae & $\alpha$\,Aql & 97649 & 2.5 \\
caput Ophiuchi & $\alpha$\,Oph & 86032 & 2.7\\
sequent. capit. \sagittarius & $\pi$\,Sgr & 94141 & 2.5 \\
5tus spond. caud. & $\theta$\,Sco & 86228 & (5.8) \\
lucida Arietis & $\alpha$\,Ari & 9884 & 0.8 \\
Aldebaran & $\alpha$\,Tau & 21421 & 1.3\\
(Lux) Mandibula Ceti & $\alpha$\,Cet & 14135 & 2.1\\
Aust. Cadae Ceti & $\beta$\,Cet & 3419 & 2.0\\
Regel Orionis & $\beta$\,Ori&  24436 & 1.8 \\
Achernar & $\alpha$\,Eri & 7588 & (7.6)\\
Fomalhaut & $\alpha$\,PsA & 113368 & 5.9 \\
Cor Hydrae & $\alpha$\,Hya & 46390 & 1.6\\
Sirius & $\alpha$\,CMa & 32349 & 0.9\\
Marchab Pegasi & $\alpha$\,Peg & 113963 & 1.3\\
Procyon & $\alpha$\,CMi & 37279 & 1.5 \\
sect.trans.(Navis) & $\lambda$\,Vel & 44816 & (3.1)\\
Canobus & $\alpha$\,Car& 30438 & (3.5)\\
Lucida in Tran. & $\zeta$\,Pup& 39429 & (3.2)\\
Cauda Leonis & $\beta$\,Leo & 57632 & 2.0 \\
Cor Leonis & $\alpha$\,Leo & 49669 & 0.6 \\
Lanx Borea & $\beta$\,Lib& 74785 & 2.5 \\
pes (dext.) Cent. & $\alpha$\,Cen & 71683 & (4.7) \\
Pes Crucis & $\alpha^1$\,Cru & 60718 & (5.8)\\
Aust.(Part.)arc.\sagittarius & $\epsilon$\,Sgr & 90185 & (5.4)\\
Caput Phoenicis & $\alpha$\,Phe & 2081 & (5.6)\\
 lucida ad rad. Roboris Carolini & $\beta$\,Car & 45238 & (5.2) \\
genus sinistr. Cent. & $\beta$\,Cen & 68702 & (44.1)$^a$\\
clarior.in summ. Roboris Carolini & $\mu$\,Vel& 52727 &(3.2) \\
ult.Erid.Ptol. & $\theta^1$\,Eri& 13847 & (6.8) \\
ala gruis & $\alpha$\,Gru & 109268 & (5.8)\
\end{tabular}
\tablefoot{The position errors are taken from \keplere, those
in brackets refer to stars not in \keplere, and are taken from
\halley. Small variations in the descriptions given by Halley are
ignored; words between brackets are occasionally omitted by him.}

\vspace*{0.2cm}

$^a$see annotation for E\,193 in Sect.\,\ref{s:annothal}.
\end{table}

\end{appendix}

\begin{appendix}
\section{Notes on individual identifications \label{s:notes}}

\subsection{De Houtman}

\begin{figure}
\centerline{\includegraphics[angle=270,width=\columnwidth]{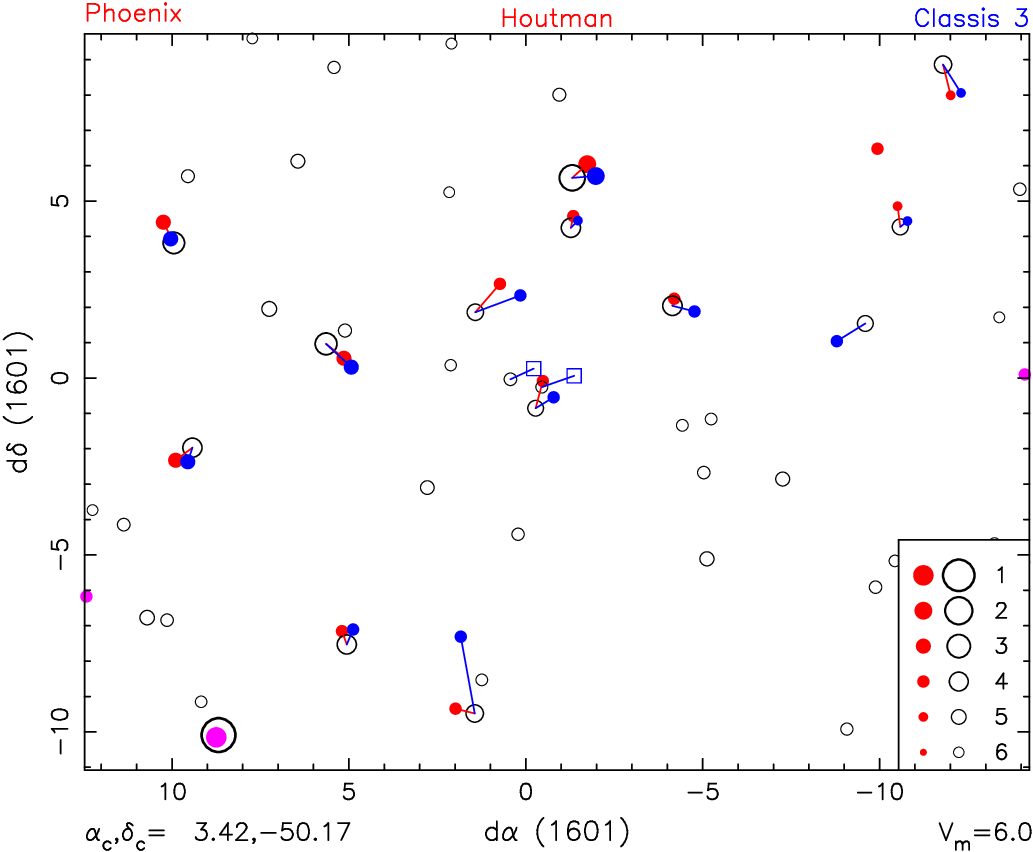}}
\caption{Phoenix
 \label{f:phoenix}}
\end{figure}

\noindent F\,2, F\,3, are near $-$10.5,4.9 and $-$9.9,6.5, respectively
in Fig.\,\ref{f:phoenix}. We identify F\,2 with its nearest Hipparcos
star, HIP116389 and leave F\,3 unidentified; Knobel identifies F\,3
with HIP\,116389 and F\,2 with HIP\,116602 near $-$9.6,1.5. 
Comparison with \tertia\ leads us to think that F\,3 corresponded 
to HIP\,116602, but that its position in \houtman\ is corrupted.

\noindent F\,7, near $-$0.5,$-$0.1 in Fig.\,\ref{f:phoenix} is closest
to HIP\,2383 ($V$=5.7) but we identify it with the brighter HIP\,2472
below it.

\noindent F\,14-29, Corona Australis. The pattern of this
constellation appears shifted as a whole, and we identify the stars
accordingly, which in a number of cases leads us to prefer an
Hipparcos counterpart at larger angular distance to the nearest
Hipparcos star (Fig.\,\ref{f:coraust}). In all cases, Knobel agrees.
For F\,19, near 0.5,$-$3.9 he prefers HIP\,93049, which however has
$V=6.3$.

\begin{figure}
\centerline{\includegraphics[angle=270,width=\columnwidth]{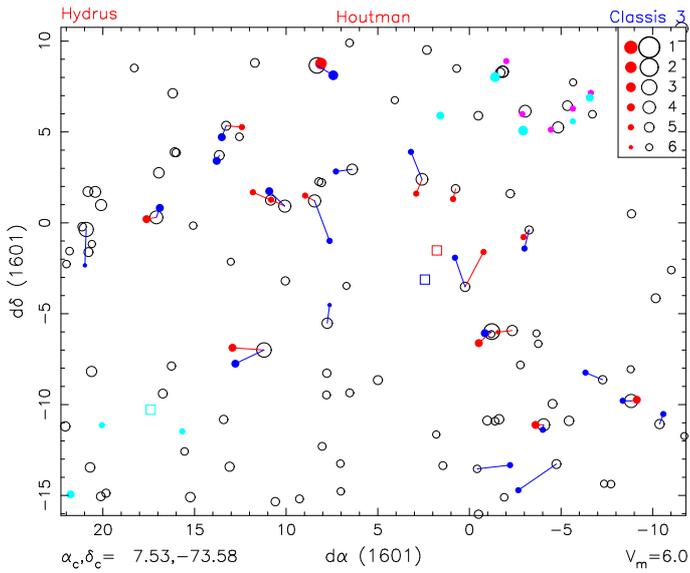}}
\caption{Hydrus, with the Small Magellanic Cloud. The nebulous object
  near 1,$-$3 is the Small Magellanic Cloud, present both in \houtman\
  and \tertia.
 \label{f:hydrus}}
\end{figure}

\noindent F\,38, near 12.4,5.3 in Fig.\,\ref{f:hydrus}, is identified by
Knobel with HIP\,13884, near 13.7,3.7, brighter but further 
($V$=5.0, $d$=2\degr) than our preferred counterpart.

\noindent F\,42,43 are identified by us with the nearest counterparts
HIP\,5896 and HIP\,4293, respectively, near 2,1.5 in
Fig.\,\ref{f:hydrus}; Knobel identifies F\,42 with HIP\,4293 and F\,43
with HIP\,1647 near $-$2.2,1.6

\noindent F\,44, near $-$0.8,$-$1.6 in Fig.\,\ref{f:hydrus}, is
identfied by Knobel with HIP\,865, a $V$=6.7 star, too faint in our view.

\noindent F\,46, near 1.8,$-$1.5 in Fig.\,\ref{f:hydrus}, is the Small Magellanic Cloud

\noindent F\,53, F\,54, near $-$7.6,5.2 and $-$3.8,2.6 in
Fig.\,\ref{f:doradus} are identified by us with HIP\,19893 and
HIP\,21281, respectively. Knobel identifies F\,53 with HIP\,21281
and F\,54 with HIP\,23693, near 0.0,1.2. 

\noindent F\,59, near $-$3.4,5.2 in Fig.\,\ref{f:colomba}, is identified by Knobel with 
HIP\,26862 ($V$=6.2), one magnitude fainter than the counterpart we prefer.

\begin{figure}
\centerline{\includegraphics[angle=270,width=\columnwidth]{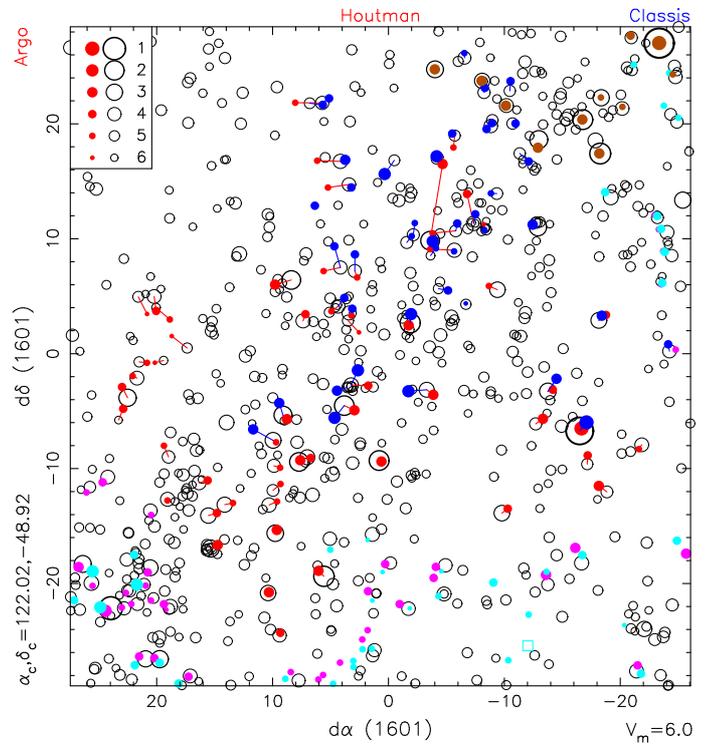}}
\caption{Argo. The crowded region near $-$6,10 is hard to
  interpret. We have chosen to identify the brightest stars in
  \houtman\ with the brightest stars in that area, even if this leads
  to large positional shifts. Stars in \keplere\ are indicated in brown.
 \label{f:argo}}
\end{figure}

\noindent F\,77, near $-$8.2,11.3 in Fig.\,\ref{f:argo} is identified
by Knobel with HIP\,37664, near $-$7.4,8.4, both fainter and further
($V$=5.1) than our counterpart.

\noindent F\,78, F\,79 near $-$6.8,13.9 and $-$5.6,17.9 in 
Fig.\,\ref{f:argo} are identified by Knobel with our counterpart for 
F\,83 and F\,82, respectively, near $-$5.7,9.0. and $-$5.8,10.7.

\noindent F\,82, F\,83, near $-$3.8,10.5 and $-$3.6,9.1 in 
Fig.\,\ref{f:argo} are identified by Knobel with HIP\,40091 and
HIP\,40326, near $-$2.1,10.4 and $-$1.6,9.7, respectively.

\noindent F\,115, F\,116, near 18.9,3.0 and 18.7,1.5 in 
Fig.\,\ref{f:argo}. F\,115 is unidentified by Knobel, and F\,116
with our counterpart for F\,115.

\noindent F\,117, F\,118, the close pair near 20.5,$-$0.8
in Fig.\,\ref{f:argo} are identified by Knobel with the pair
HIP\,51561, HIP\,51610 near 22.1,1.3.

\noindent F\,119, F\,120, F\,121 three stars near near 22.1,$-$1.9 to
$-$4.8 in Fig.\,\ref{f:argo} are identified by Knobel with our
counterpart for F\,117 near 20.8,$-$0.8, and F\,119, F\,120,
respectively, i.e\ shifted by one counterpart upwards.

\noindent F\,142-F\,146, five stars near 0,0 in
Fig.\,\ref{f:centaurus} form a pattern with four stars in a vertical row
and one star to the left. We follow Knobel in his identification
which implies a shift to the south, taking further stars as 
identifications for four stars. The alternative would leave the
top star, F\,145, unidentified.

\begin{figure}
\centerline{\includegraphics[angle=270,width=0.9\columnwidth]{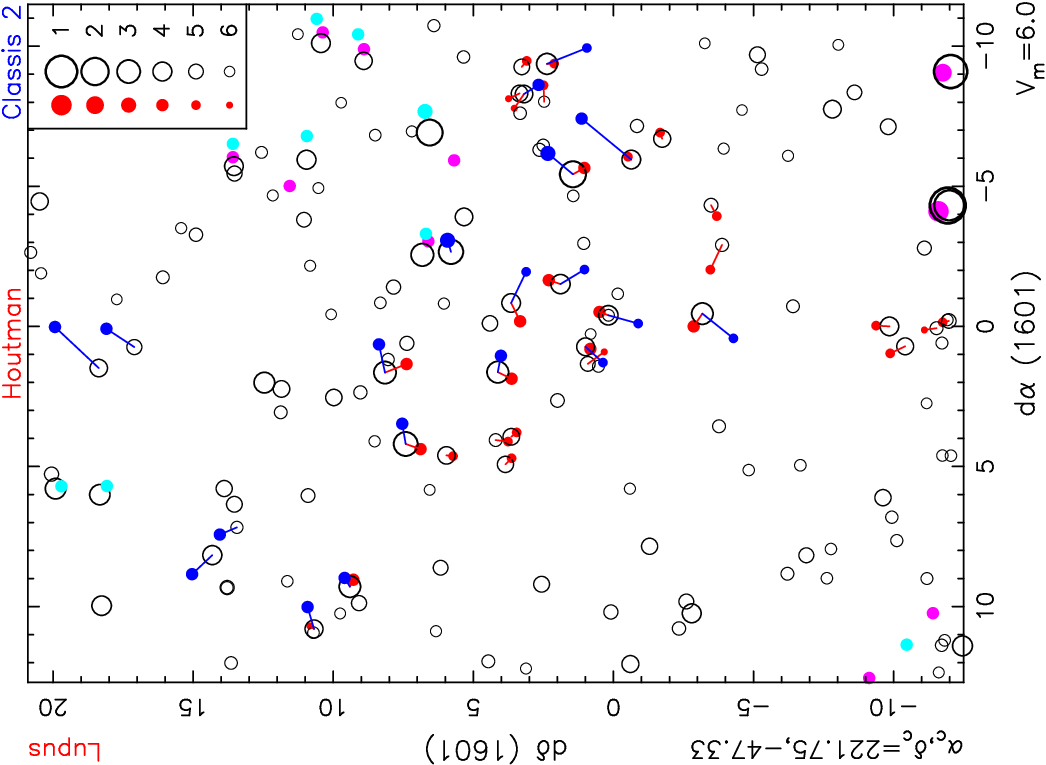}}
\caption{Lupus
 \label{f:lupus}}
\end{figure}

\noindent F\,195-197, near $-$9,3 in Fig.\,\ref{f:lupus}, are 
closely matched with our counterparts, and less so by those
proposed by Knobel. He identifies F\,195 with HIP\,69671,
too faint at $V$=6.3; F\,195 with HIP\,70054 at $-$9.6,5.4,
more than 2\degr\ distant, and F\,196 with our counterpart 
for F\,195.

\noindent F\,199, near $-$7.8,3.5 in Fig.\,\ref{f:lupus}, is closest
to HIP\,70915 ($V$=5.5, $d$=16\farcm7, but we consider the brighter
HIP\,70576 the more likely, albeit further counterpart.

\noindent F\,203, near $-$2.0,$-$3.5 in Fig.\,\ref{f:lupus} is
identified by Knobel with HIP\,73345 ($V$=6.8, $d$=55\farcm9),
much fainter and only marginally closer than our preferred 
counterpart.

\noindent F\,214, near 0.9,0.3 in Fig.\,\ref{f:lupus}, is closest 
to HIP\,75181 ($V$=5.7, $d$=33\farcm4), but
we slightly prefer the somewhat brighter but further HIP\,75206.

\begin{figure}
\centerline{\includegraphics[angle=270,width=0.9\columnwidth]{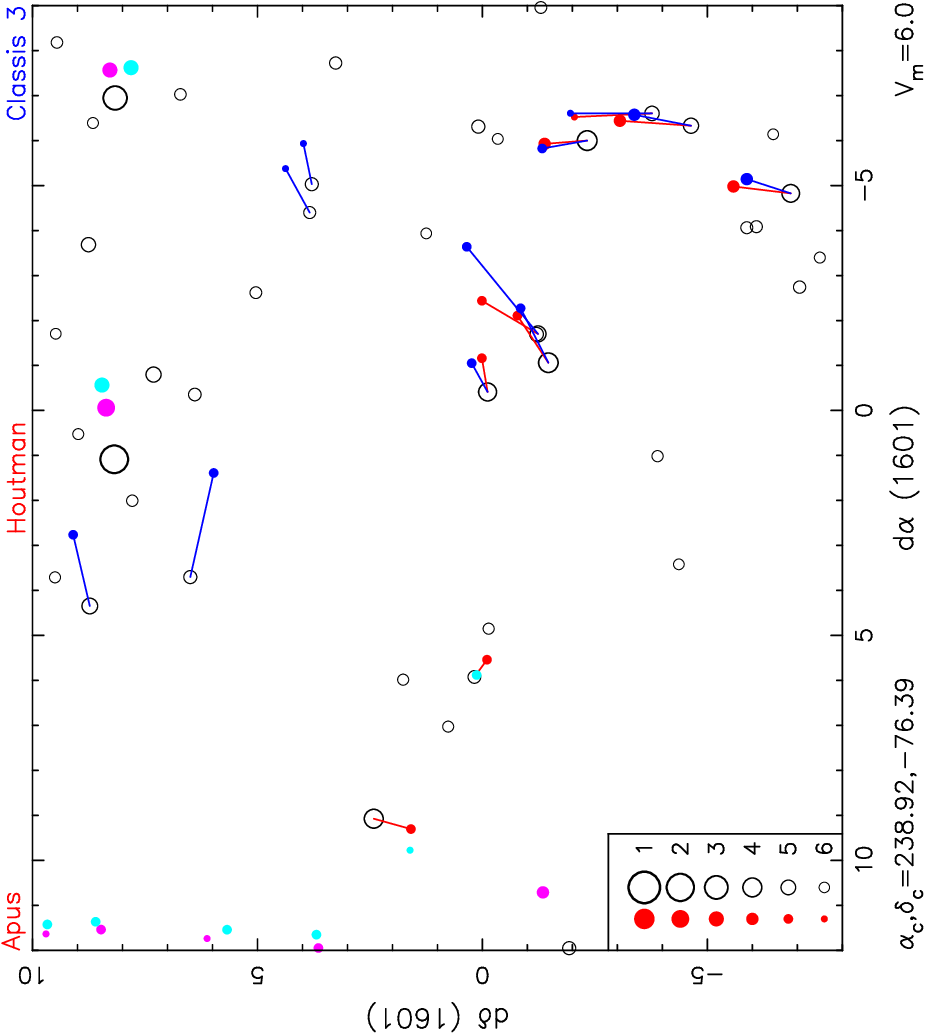}}
\caption{Apus
 \label{f:apus}}
\end{figure}

\noindent F\,228-231, from $-$5,$-$6 to $-$6,$-$1 in
Fig.\,\ref{f:apus}. We shift these stars south by about 1\fdg5,
after which each is identified with the then nearest star, thereby
identifying these four stars with the four brightest nearby
Hipparcos stars.

\noindent F\,232-233, near $-$2.5,$-$0.5 in Fig.\,\ref{f:apus},
are both closest to HIP\,80047/80057. (These Hipparcos stars are
separated by 1\farcm7 and cannot be separated by the naked eye.)
We identify F\,232 with the Hipparcos pair, and F\,233, to the south
of F\,232, with the star east and south of the Hipparcos pair, viz.\
HIP\,81065.

\noindent F\,236, near 9.3,1.6 in Fig.\,\ref{f:apus}, is identified by
Knobel with HIP\,92394, much fainter and somewhat further
($V$=6.0, $d$=64\farcm5) than our preferred counterpart.

\noindent F\,249, near $-$5.4,2.5 in Fig.\,\ref{f:scorpius}, is
closest to HIP\,82545 ($\mu^2$\,Sco, $V$=3.6, $d$=7\farcm7),
but we take the brighter HIP\,82514 ($\mu^1$\,Sco) as more plausible
counterpart. The two Hipparcos stars are separated by 5\farcm8 only.

\noindent F\,250, F\,251, near $-$4.7,$-$1.4 and $-$4.5,$-$1.9
 in Fig.\,\ref{f:scorpius}.
Knobel identifies these with the close (7\farcm5) pair
HIP\,82671, HIP\,82729 ($\zeta^1$ and $\zeta^2$\,Sco); we 
identify F\,250 with the nearer, but fainter Hipparcos star.

\noindent F\,258 is almost equidistant from HIP\,86929 and HIP\,88866.
Because HIP\,86929 is already identified with F\,257, we identify
F\,258 with HIP\,88866

\begin{figure}
\centerline{\includegraphics[angle=270,width=0.9\columnwidth]{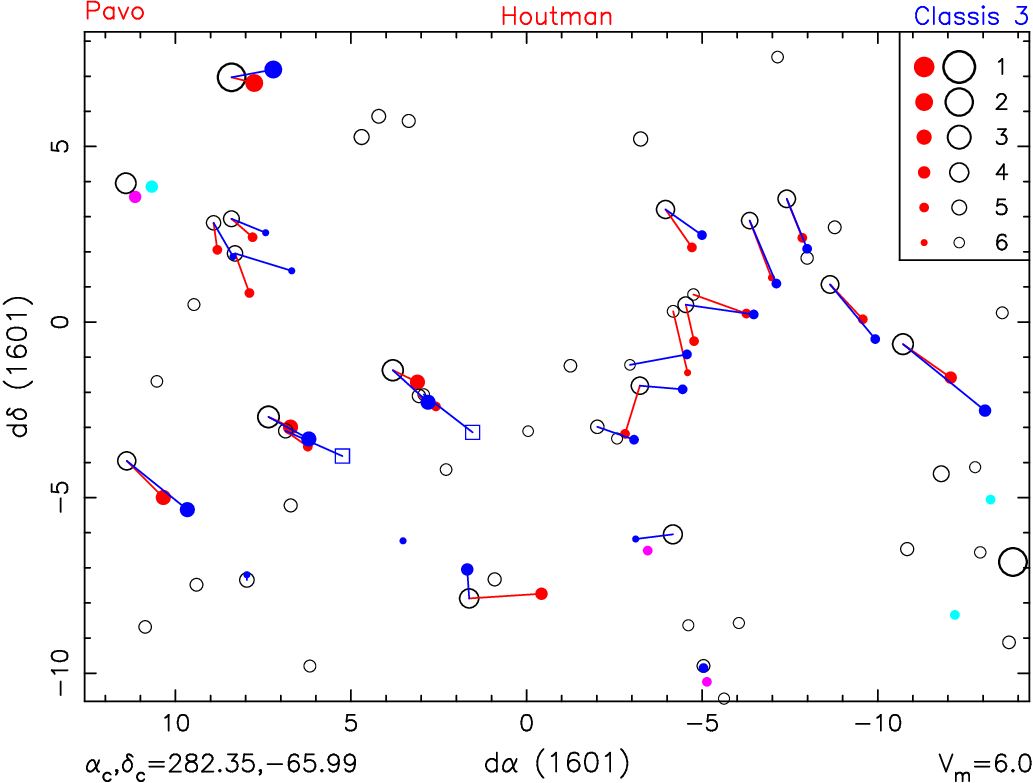}}
\caption{Pavo
 \label{f:pavo}}
\end{figure}

\noindent F\,257-F\,275, i.e. the whole constellation Pavo
(Fig.\,\ref{f:pavo}), form a pattern which appears to match the
Hipparcos stars best after a shift of about 1\fdg5 to the
north-east. This explains several differences in our identifications
with those by Knobel. Our emendations to F\,273 and F\,274, where
Knobel has no and a different identification, respectively, explain our
different identifications for these stars.  F\,268, near 2.6,$-$2.4,
is close to the pair HIP\,98478 - HIP\,98624 (which is separated by
8\farcm7), we choose the brighter of the two as counterpart.

\begin{figure}
\centerline{\includegraphics[angle=270,width=\columnwidth]{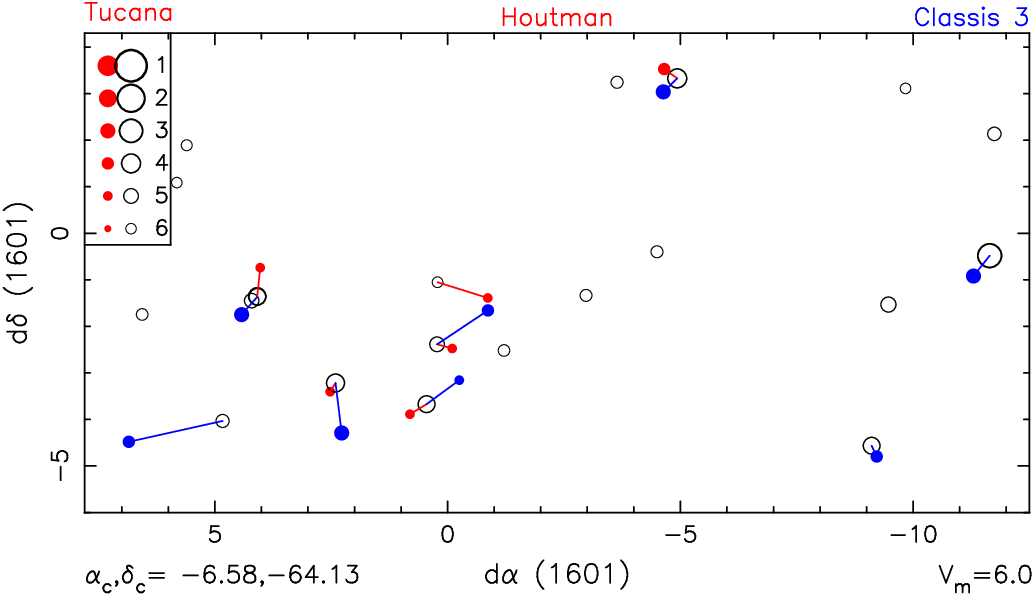}}
\caption{Tucana
 \label{f:tucana}}
\end{figure}

\noindent F\,304, near 4.0,$-$0.7 in Fig.\,\ref{f:tucana},
is the counterpart of the close ($d$=30\farcs0,
i.e.\ inseparable by the naked eye) pair HIP\,2484/HIP\,2487 ($\beta$\,Tuc).

\subsection{Secunda Classis and Tertia Classis}

In this section, we abbreviate reference to Figures C.n in Paper\,I with PI-C.n.

\noindent K\,1006, near 10.6,3.6 in Fig.\,PI-C.7, is closest to
HIP\,74596, but that star is already matched with K\,142; we identify
K\,1006 with the next nearest star, near 12.1,4.2.

\noindent K\,1009, near $-$1.4,3.2 in Fig.\,PI-C.9 is M\,13

\noindent K\,1010, near 8.9,14.9 in Fig.\,PI-C.11, is closest to
HIP\,101243 ($\omega^2$\,Cyg, $V$=5.4, $d$=67\farcm9), but we choose
HIP\,101138 ($\omega^1$\,Cyg) as a brighter, slightly further
counterpart.  The angle between HIP\,101138 and HIP\,101243 is
20\farcm3, and perhaps K\,1010 -- indicated nebulous in the catalogue,
is the combined light of these two stars.

\noindent K\,1011, near 4.8,$-$14.4 in Fig.\,PI-C.15, is nearest to
HIP\,19335 ($V$=5.5, $d$=68\farcm7) but we identify it with the further
but brighter HIP\,19811, near 6.1,$-$13.3.

\noindent K\,1012-K\,1042: Ophiuchus and Serpens. Ophiuchus in
particular has many bad and/or uncertain position in
\keplere. Many stars in \secunda\ are repeats of stars in \keplere,
sometimes with rather better positions (see Table\,\ref{t:doubles}.

\noindent K\,1036 is the supernova of 1604. For the modern position 
we use $\alpha(2000)$=17h30m36s, $\delta(2000)$=-21\degr28\arcmin56\arcsec.

\noindent K\,1054, near 4.5,7.3 in Fig.\,PI-C.30, repeat entry for K\,504

\noindent K\,1055, near 16.6,11.7 in Fig.\,PI-C.45, repeat entry for K\,823

\noindent K\,1072, near $-$10.5,5.8 in Fig.\,PI-C.31, repeat entry for K\,323

\begin{figure}
\centerline{\includegraphics[angle=270,width=\columnwidth]{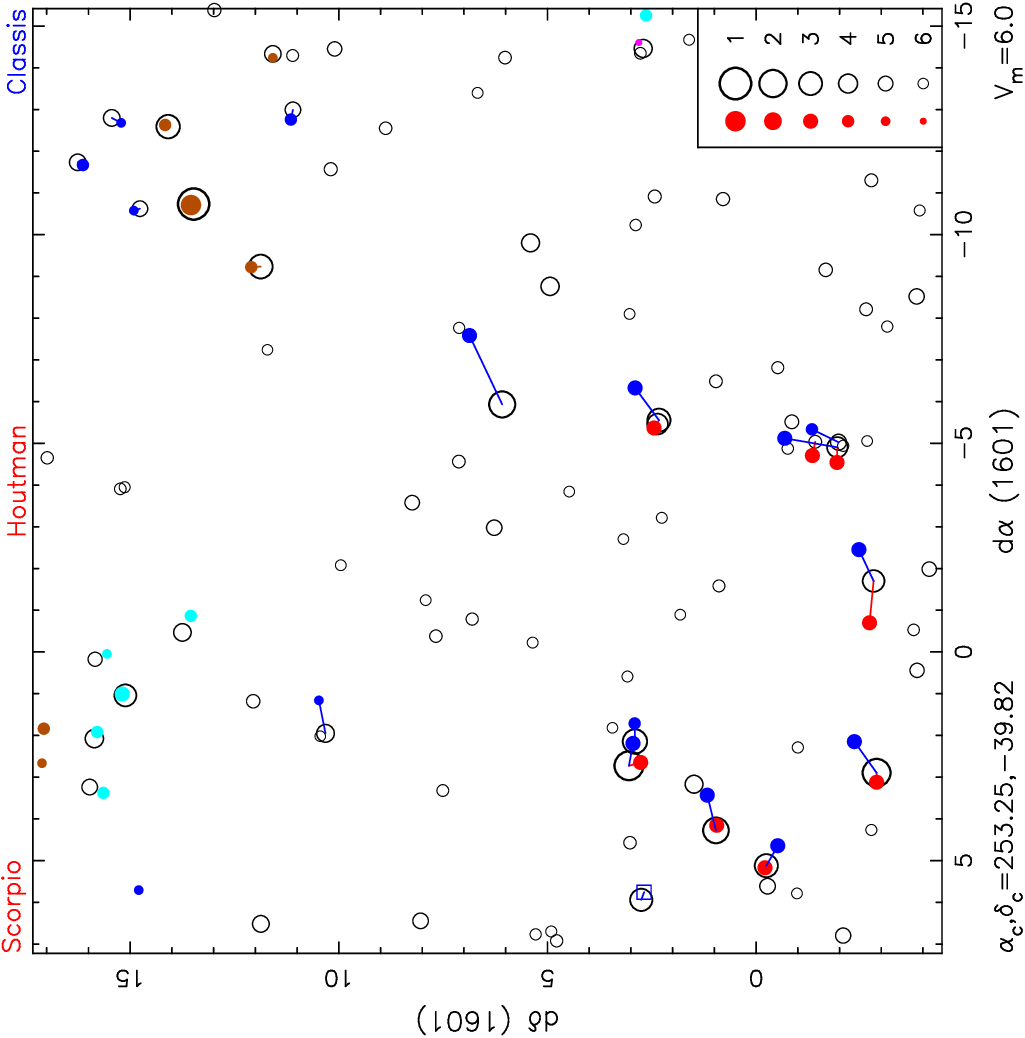}}
\caption{Scorpius. Stars in \keplere\ are indicated in brown.
 \label{f:scorpius}}
\end{figure}

\noindent K\,1084, near $-$6.3,2.9 in Fig.\,\ref{f:scorpius}, is close
to the pair HIP\,82514/HIP\,82545 (separation 5\farcm8) and perhaps represents
their combined light

\noindent K\,1095, near 5.7,14.8 in Fig.\,\ref{f:scorpius}. Kepler
mentions a latitude {\em aliter} (alternatively) for this star,
4\,30\,A, three degrees further south, which would lead to
identification with HIP\,87072, near 6.5,11.9.
 
\begin{figure}
\centerline{\includegraphics[angle=270,width=0.9\columnwidth]{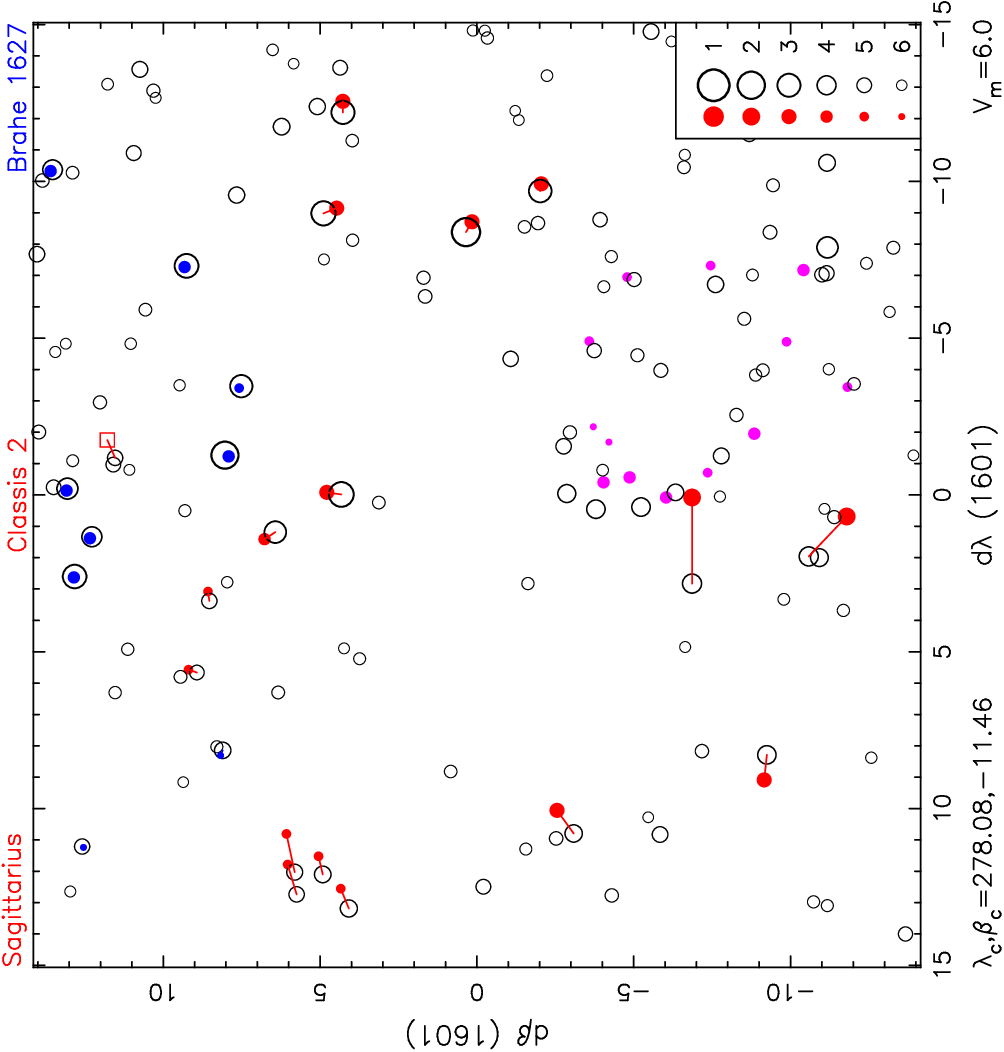}}
\caption{Sagittarius.
 \label{f:sagittarius}}
\end{figure}

\noindent K\,1099, near $-$1.8,11.8 in Fig.\,\ref{f:sagittarius}, is
described {\em In oculo nebulosa duplex} (a double nebulous [star] in
the eye), corresponding to the close (13\farcm9) pair HIP\.92761/92845
= $\nu_1$/$\nu_2$\,Sgr.

\noindent K\,1104 and K\,1105, near 0.7,$-$11.8 and 0.1,$-$6.9 in 
Fig.\,\ref{f:sagittarius} are identified by us with the two bright
stars further East, $\beta$ and $\alpha$\,Sgr, near 2.0,$-$10.6 and
2.8,$-$6.9, respectively. 

\begin{figure}
\centerline{\includegraphics[angle=270,width=0.75\columnwidth]{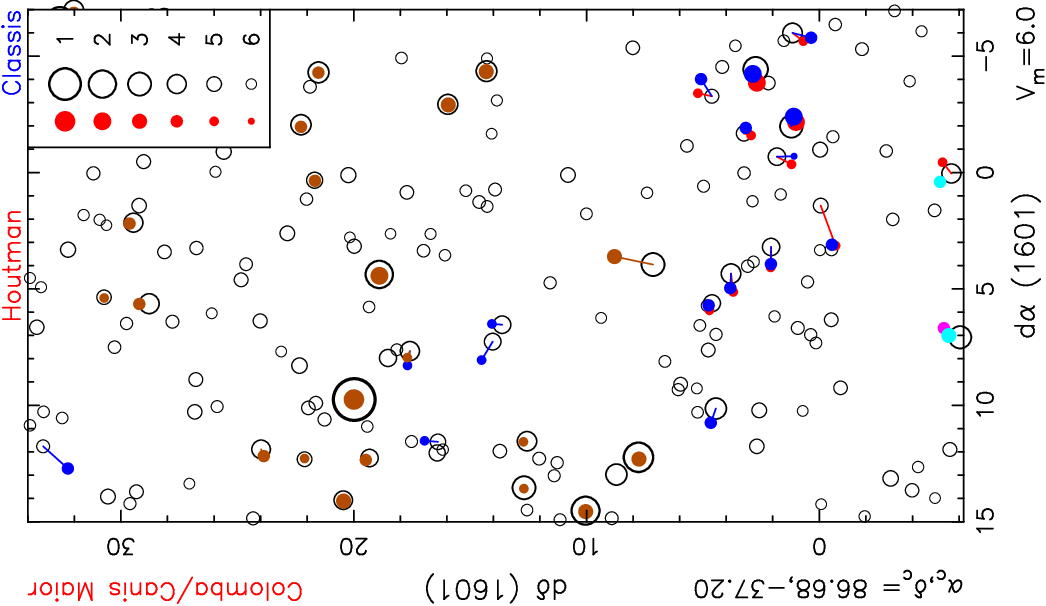}}
\caption{Colomba in \houtman\ is part of Canis Maior in
  \secunda. Stars in \keplere\ are indicated in brown.
 \label{f:colomba}}
\end{figure}

\noindent K\,1147, near 11.5,17.0 in Fig.\,\ref{f:colomba}, lies
almost exactly between two stars of different
brightness, HIP\,33077 to the North and HIP\,33092 to the South;
we take the brighter star as the counterpart.

\noindent K\,1148 repeat of K\,938

\noindent K\,1152, near 12.7,32.3 in Fig.\,\ref{f:colomba}. We
identify this star with HIP\,33184, northwest of it, but perhaps
HIP\,33971, to the southeast, is also possible, being brighter but
further ($V$=5.0, $d$=135.5\arcmin).

\noindent K\,1170, near $-$12.1,16.7 in Fig.\,\ref{f:argo} is surrounded
by faint ($V$$<$5.0 stars, but we choose a brighter star near
$-$11.2,17.4 as counterpart.

\noindent K\,1171, K\,1173 and K\,1174, respectively near
$-$12.4,11.2, $-$8.3,10.8 and $-$7.5,12.1 in Fig.\,\ref{f:argo} are
identical to the  brightest stars from the {\em Hipparcos Catalogue} 
near each of them, even though fainter stars are closer.

\noindent K\,1178, close to the much brighter K\,1176, both near
$-$3.9,9.5 in Fig.\,\ref{f:argo} is 
the faint star close to the bright counterpart of K\,1176, $\zeta$\,Pup.

\begin{figure}
\centerline{\includegraphics[angle=270,width=\columnwidth]{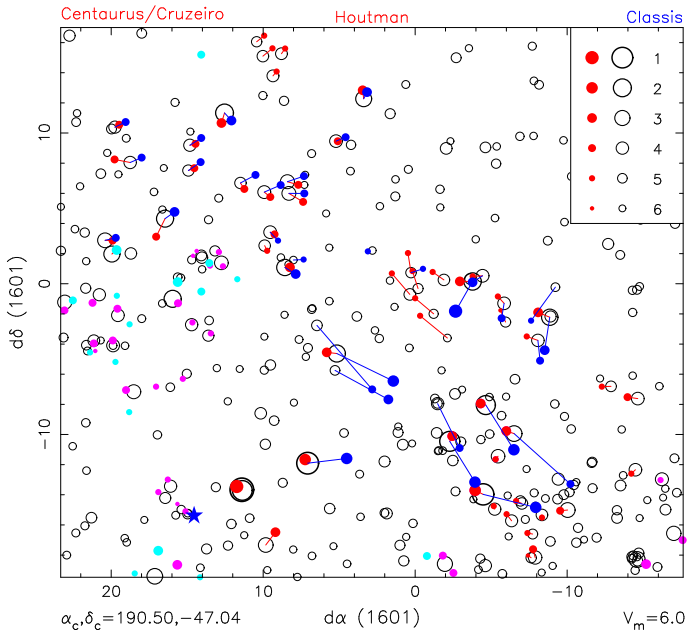}}
\caption{Centaurus and Crux. The bright star K\,1244, near
  $-$2.7,$-$1.8 is tentatively identified by us with HIP\,61932
  ($\gamma$\,Cen); an alternative position is indicated with a blue
  star, and leads to identfication with HIP\,71683 ($\alpha$\,Cen).
 \label{f:centaurus}}
\end{figure}

\noindent K\,1230, near $-$2.4,2.0 in Fig.\,\ref{f:centaurus} is the
globular cluster $\omega$\,Cen. We use the magnitude and position 
given by Harris (1996, version of February 2003). 

\noindent K\,1237-K\,1243, K\,1246. These southern stars in Centaurus
are badly matched. Comparing their pattern with the bright stars from
the {\em Hipparcos Catalogue} we suggest that they are
located about 4\degr\ too far south and varying degrees too far west
in \secunda, and we have chosen counterparts 
accordingly. See also Fig.\,\ref{f:annotcrux}.

\noindent K\,1244, near $-$2.7,$-$1.8 in Fig.\,\ref{f:centaurus}, has
no obvious counterpart. Even though HIP\,61932 is very close
(3\arcmin) to K\,1232, we prefer to identify K\,1244 with it and
K\,1232 with HIP\,61622 just above it. An alternative is to emend the
zodiacal sign of its longitude from \libra (7) to \scorpio\ (8), which brings it into
agreement with the position in the catalogue of Ptolemaios and leads
to identification with HIP\,71683 ($\alpha$\,Cen).

\noindent K\,1245, near 4.5,$-$11.6 in Fig.\,\ref{f:centaurus}, is
identified with HIP\,68702 ($\beta$\,Cen).  

\begin{figure}
\centerline{\includegraphics[angle=270,width=0.9\columnwidth]{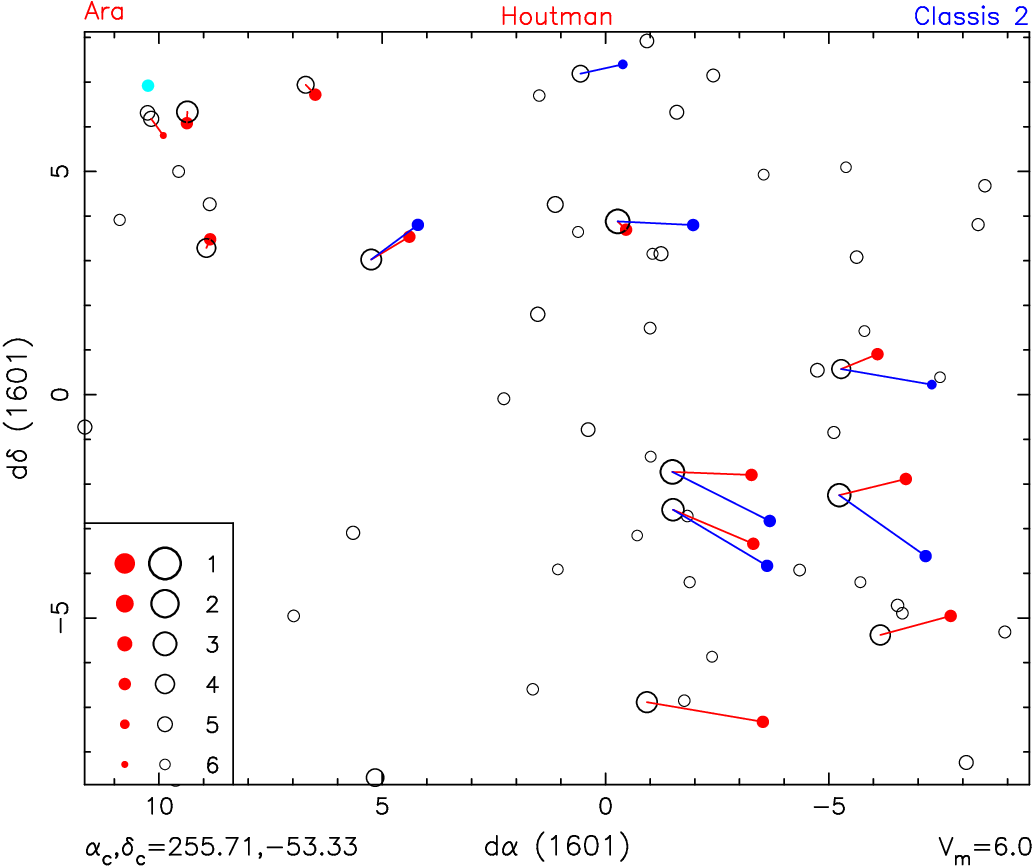}}
\caption{Ara
 \label{f:ara}}
\end{figure}

\noindent K\,1266-K\,1272, Ara (Fig.\,\ref{f:ara}). There is a general
shift of the catalogue positions of this constellation to the East,
leading to many cases where the obvious and correct Hipparcos
counterpart is not the nearest Hipparrcos star.

\begin{figure}
\centerline{\includegraphics[angle=270,width=\columnwidth]{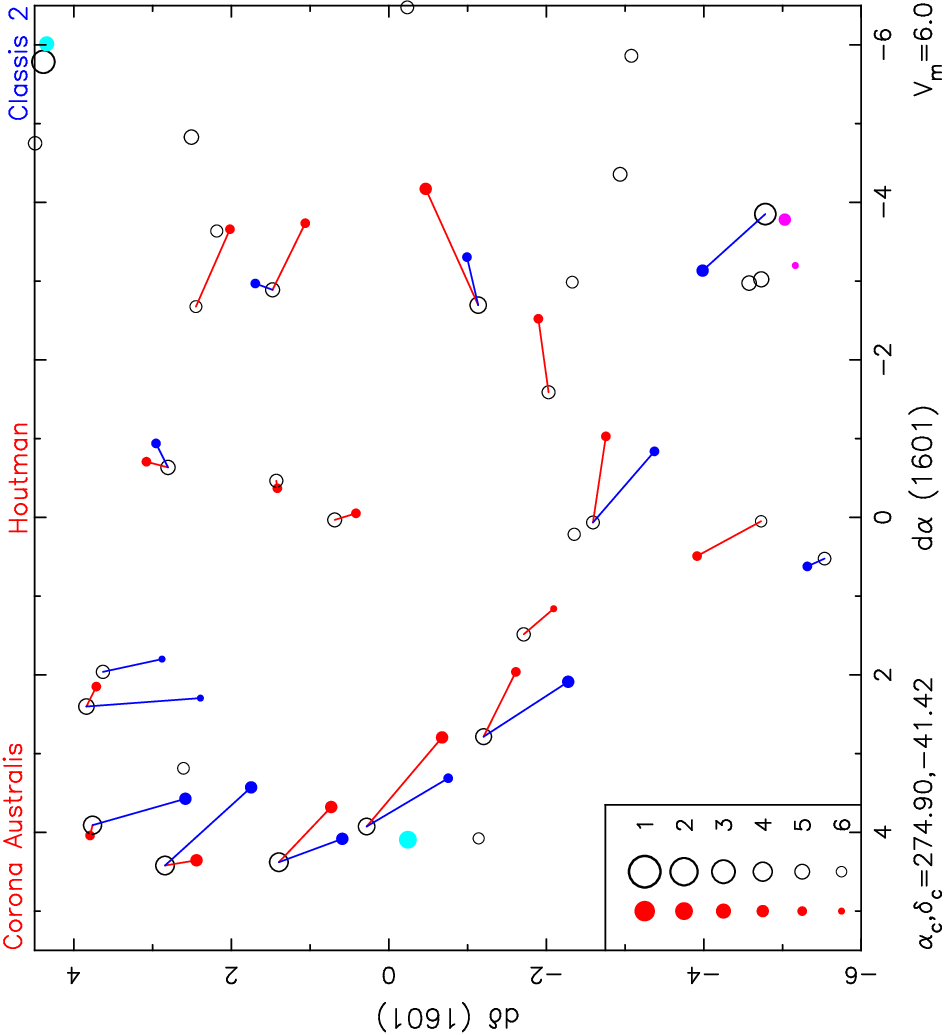}}
\caption{Corona Australis
 \label{f:coraust}}
\end{figure}

\noindent K\,1273-1285, Corona Australis (Fig.\,\ref{f:coraust}). 
There appears to be a general shift in the coordinates of the stars in
this constellation, leading us to accept Hipparcos counterparts
at larger angles in many cases where nearer Hipparcos stars are
present.

\begin{figure}
\centerline{\includegraphics[angle=270,width=\columnwidth]{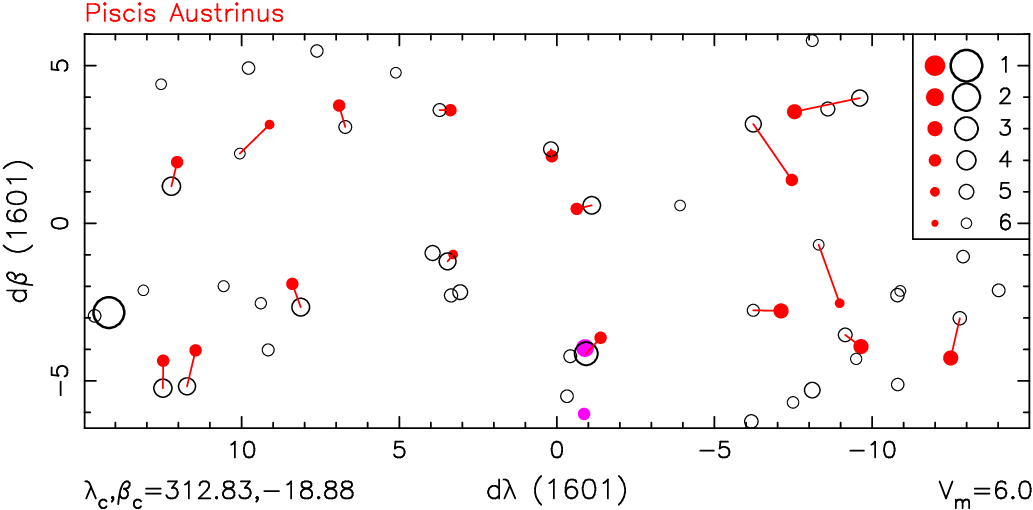}}
\caption{Piscis Austrinus.
 \label{f:pisaustr}}
\end{figure}

\noindent K\,1296, near $-$1.4,$-$3.6 in Fig.\,\ref{f:pisaustr}, has
the same counterpart as a star from Grus in \tertia, K\,1303.

\noindent K\,1300, near $-$9.0,$-$2.5 in Fig.\,\ref{f:pisaustr}, is
closest to HIP\,104738, which however is already taken by K\,1298,
hence we suggest a faint star to the North as counterpart, HIP\,104752
near $-$8.3,$-$0.7.

\noindent K\,1302, near $-$7.5,3.5  in Fig.\,\ref{f:pisaustr}, is
tentatively identified not with the nearest star to the west, but with
the brighter one beyond it, near $-$9.6,4.0.

\begin{figure}
\centerline{\includegraphics[angle=270,width=\columnwidth]{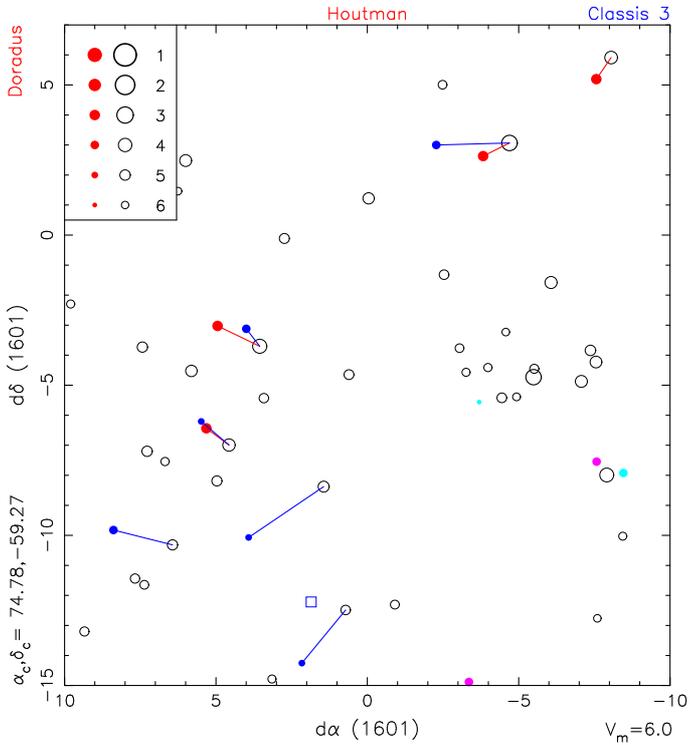}}
\caption{Doradus. The nebulous object near 2,$-$12.5 is the Large
  Magellanic Cloud, present in \tertia, but not in \houtman.
 \label{f:doradus}}
\end{figure}

\noindent K\,1408 in Fig.\,\ref{f:doradus} is the Large Magellanic Cloud

\noindent K\,1433, near 2.4,$-$3.1  in Fig.\,\ref{f:hydrus} is the Small Magellanic Cloud

\subsection{Halley}

\begin{figure}
\centerline{\includegraphics[angle=270,width=\columnwidth]{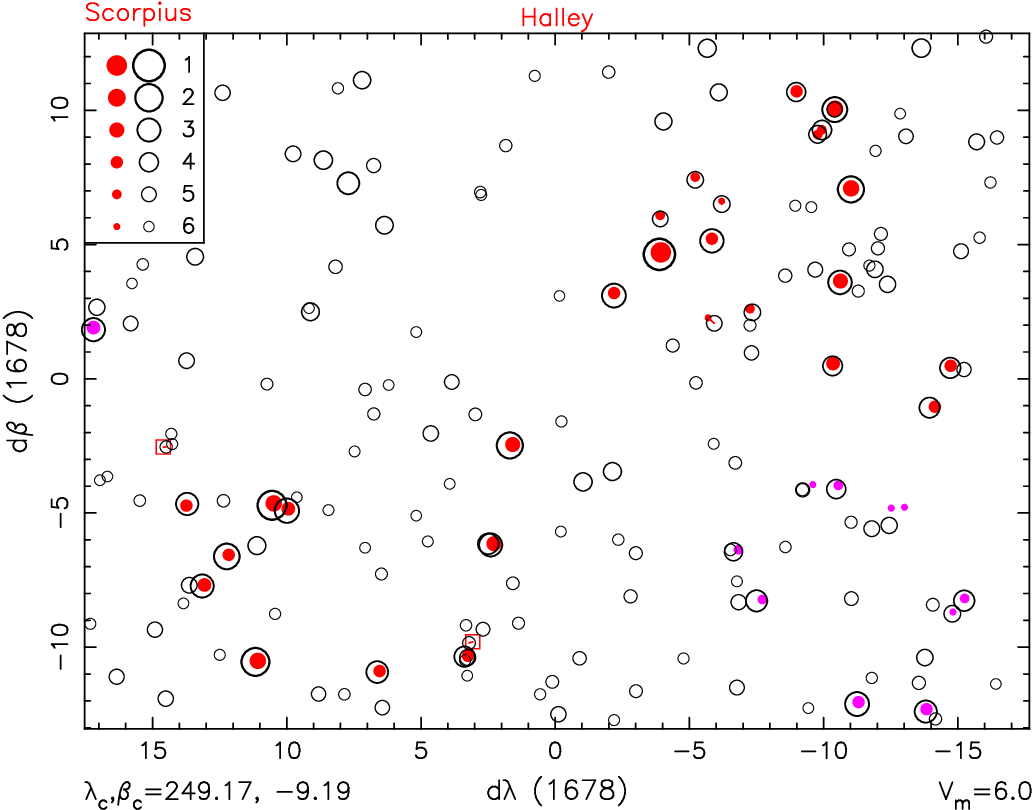}}
\caption{Scorpius.
 \label{e:scorpius}}
\end{figure}

\noindent E\,1, near $-$10.4,10.1 in Fig.\,\ref{e:scorpius}, is the
combined light of the close (15\arcsec) pair
HIP\,78820/HIP78821 ($\beta^1$\,Sco/$\beta^2$\,Sco).

\noindent E\,20, near 3.1,$-$9.8 in Fig.\,\ref{e:scorpius} is near the open
cluster NGC\,6231, which is probably why Halley calls it nebulous.

\noindent E\,21, near 3.3,$-$10.3 in Fig.\,\ref{e:scorpius} is closer
to HIP\,82671 ($V$=4.7, $d$=4.0\arcmin) than to the brighter counterpart
we prefer.

\noindent E\,29, near 14.6,$-$2.6 in Fig.\,\ref{e:scorpius}. The
Hipparcos object that we identify this with is a member of the open
cluster M\,7.

\begin{figure}
\centerline{\includegraphics[angle=270,width=\columnwidth]{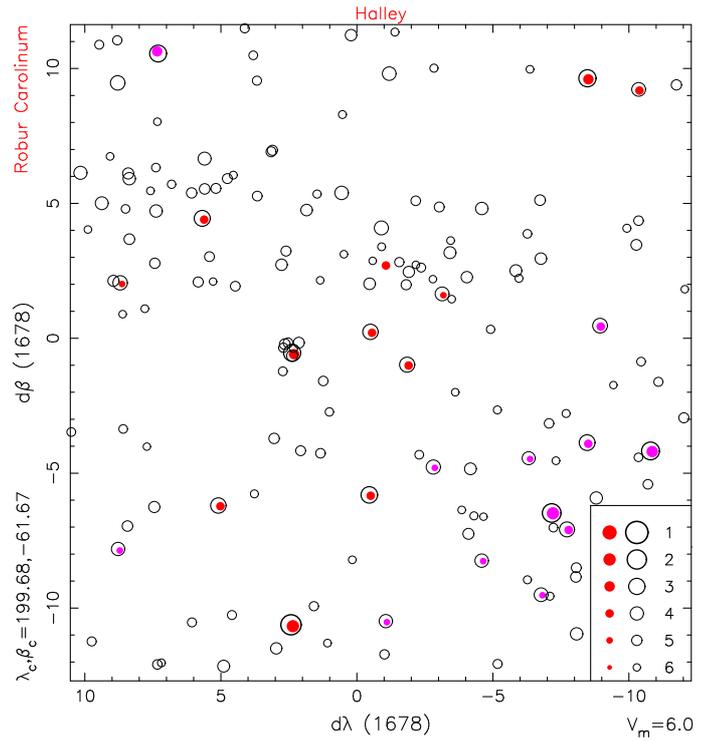}}
\caption{Robur Carolinum (the oak of Charles). The star near $-$1.1,2.7,
  not matched in the {\em Hipparcos Catalogue}, is $\eta$\,Car.
 \label{e:robur}}
\end{figure}

\noindent E\,146, near $-$1.1,2.7 in Fig.\,\ref{e:robur}, is
$\eta$\,Car. This highly variable star is not in the {\em Hipparcos}
catalogue; it is HR\,4210. Halley's catalogue is the first known
reference to this star.

\begin{figure}
\centerline{\includegraphics[angle=270,width=0.9\columnwidth]{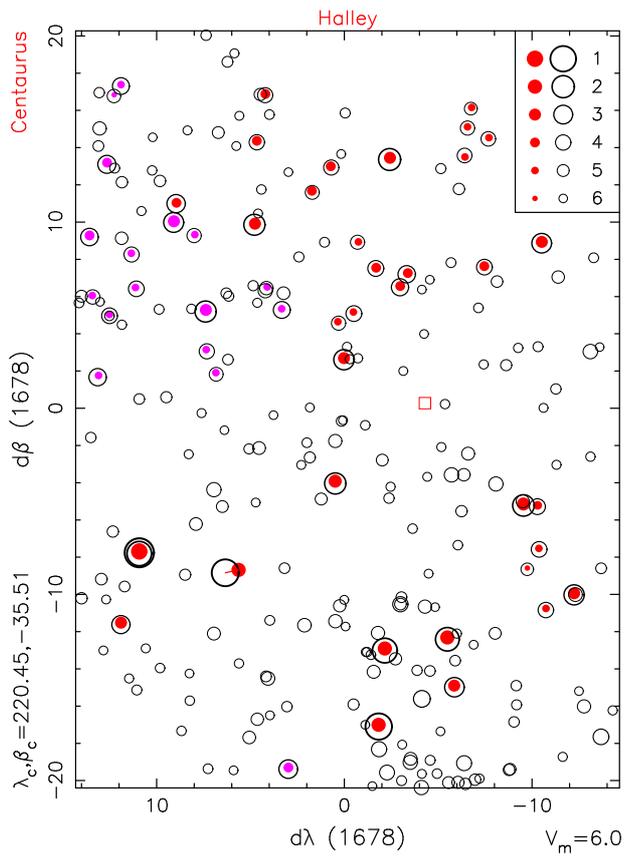}}
\caption{Centaurus. The nebulous object is the globular cluster $\omega$\,Centauri.
 \label{e:centaurus}}
\end{figure}

\noindent E\,180, near $-$4.3,0.3 in Fig.\,\ref{e:centaurus}, is the
globular cluster $\omega$\,Centauri.

\begin{figure}
\centerline{\includegraphics[angle=270,width=0.9\columnwidth]{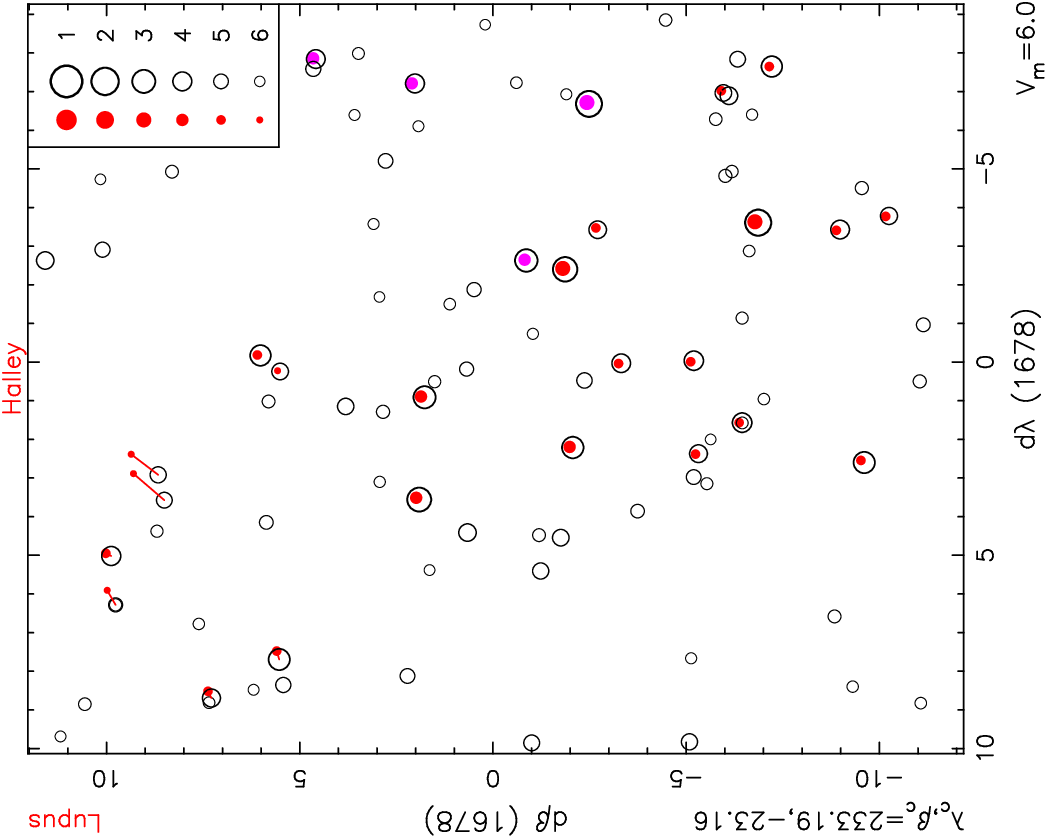}}
\caption{Lupus.
 \label{e:lupus}}
\end{figure}

\noindent E\,212-E\,217. Halley notes that these last five stars of
Lupus, in the upper left corner of Fig.\,\ref{e:lupus}, were observed
while sailing ({\em inter navigandum}), and thus less accurate but
sufficiently accurate for use on a globe. The Figure shows that these
positions are indeed less accurate than the others.

\begin{figure}
\centerline{\includegraphics[angle=270,width=0.65\columnwidth]{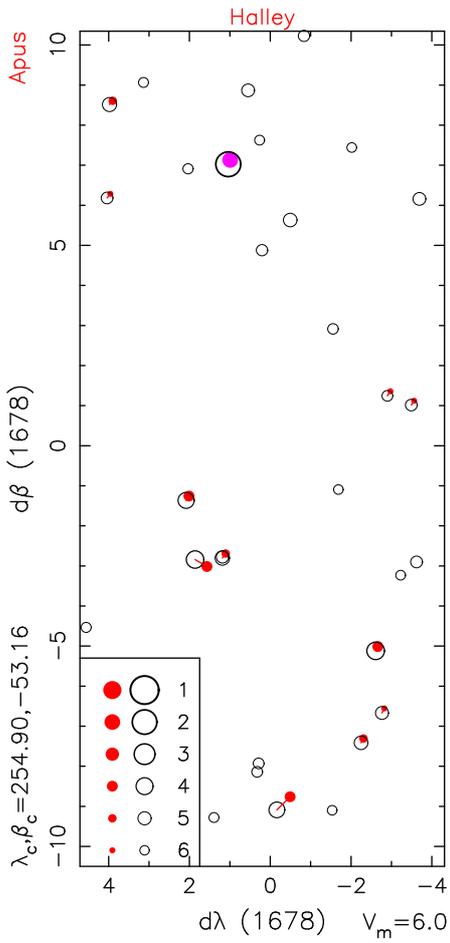}}
\caption{Apus Avis, Indica (Indian bird of paradise).
 \label{e:apus}}
\end{figure}

\noindent E\,282, near 1.1,$-$2.7 in Fig.\,\ref{e:apus}, corresponds
to the close (1.7\arcmin) pair HIP\,80047/HIP80057
($\delta^1$/$\delta^2$\,Aps)

\begin{figure}
\centerline{\includegraphics[angle=270,width=\columnwidth]{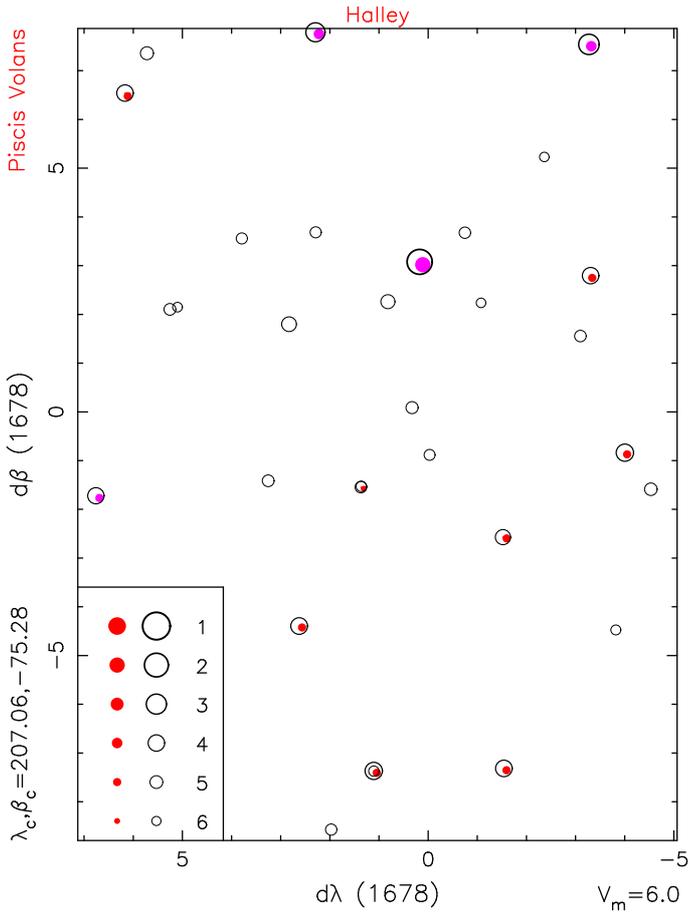}}
\caption{Piscis Volans (flying fish).
 \label{e:volans}}
\end{figure}

\noindent E\,311, near 1.1,$-$7.4 in Fig.\,\ref{e:volans}, corresponds
to the combined light of the close (13\arcsec) pair
HIP\,34473/HIP\,34481.

\noindent E\,314, near 1.3,$-$1.6 in Fig.\,\ref{e:volans}, corresponds
to the combined light of the close (1\farcm7) pair
HIP\,40817/HIP\,40834.

\end{appendix}

\clearpage

\begin{appendix}
\section{Constellations in \houtman\ and \classis\label{s:figures}}

To illustrate and clarify our identifications we provide figures for
each constellation, starting with the constellations in \houtman.  To
minimize deformation we rotate the approximate center of the
constellation to point $\aries$, as explained in Appendix\,C of
Paper\,I, but now for equatorial rather than ecliptic coordinates.
The values used for this center $\alpha_\mathrm{c}$,
$\delta_\mathrm{c}$ are indicated with each Figure, as is the
magnitude limit to which we also show all Hipparcos stars
$V_\mathrm{m}$ (usually $V_\mathrm{m}=6.0$).
The magnitudes of both catalogue and Hipparcos stars are indicated
with symbol size.

In illustrating the entries from \secunda\ and \tertia\ we wish to
facilitate comparison with the catalogues \keplere\ on one hand, and
\houtman\ on the other hand.  When \secunda\ only adds a small number
of stars to a constellation in \keplere, we have shown these added
stars in yellow in the figures for \keplere\ in Paper\,I, as indicated
in Table\,\ref{t:classis2}.  When the constellation in \secunda\ shows
a large overlap with a constellation in \houtman, and for all
constellations in \tertia, we show the stars from \secunda\ or
\tertia\ in the figure for the corresponding constellation in
\houtman, as indicated in Tables\,\ref{t:classis2} and
\ref{t:classis3}. To do so we convert the ecliptic positions as given
in \secunda\ and \tertia\ to equatorial positions, using the modern
value for the obliquity in 1601, $\epsilon=23\fdg491$.  (Kepler and
Brahe used $\epsilon=23\fdg525$.)
Constellations in \secunda\ which show little or no overlap with
\houtman\ are shown together with stars from \keplere, in rotated
ecliptic coordinates.

In these figures the stars plotted from the first catalogue are shown
in red when member of the illustrated constellation, and purple when not a member.
Stars from a second catalogue in the same figure and member of the
constellation are shown in blue, and other stars in light-blue.

\begin{figure}
\centerline{\includegraphics[angle=270,width=\columnwidth]{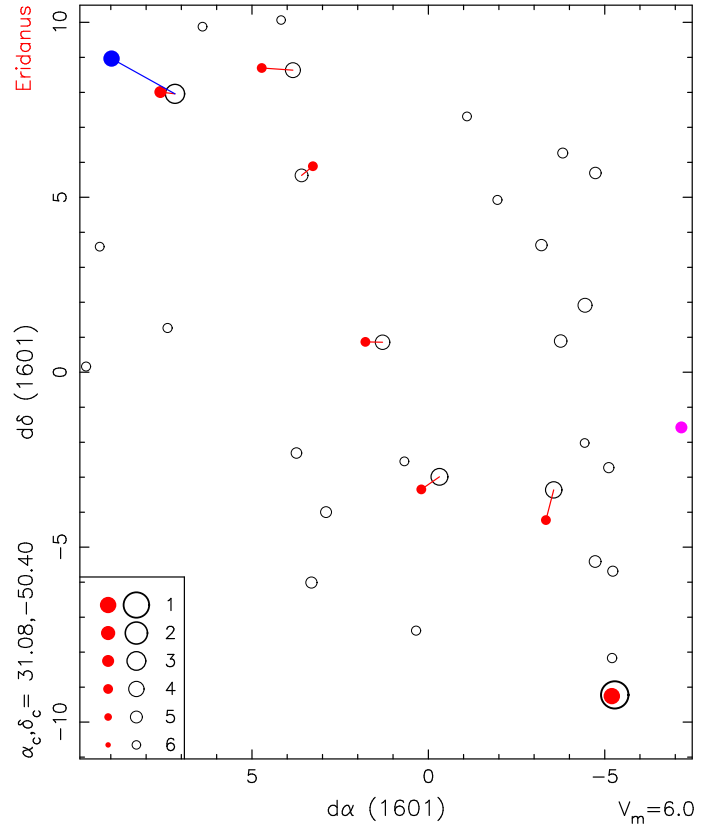}}
\caption{Eridanus in \houtman. K\,1146, the only star in \classis\ in common
  with \houtman, is indicated in blue. See Fig.\,\ref{f:eridanusc},
  where it is at $-$10.6,$-$16.0.
 \label{f:eridanus}}
\end{figure}

\begin{figure}
\centerline{\includegraphics[angle=270,width=\columnwidth]{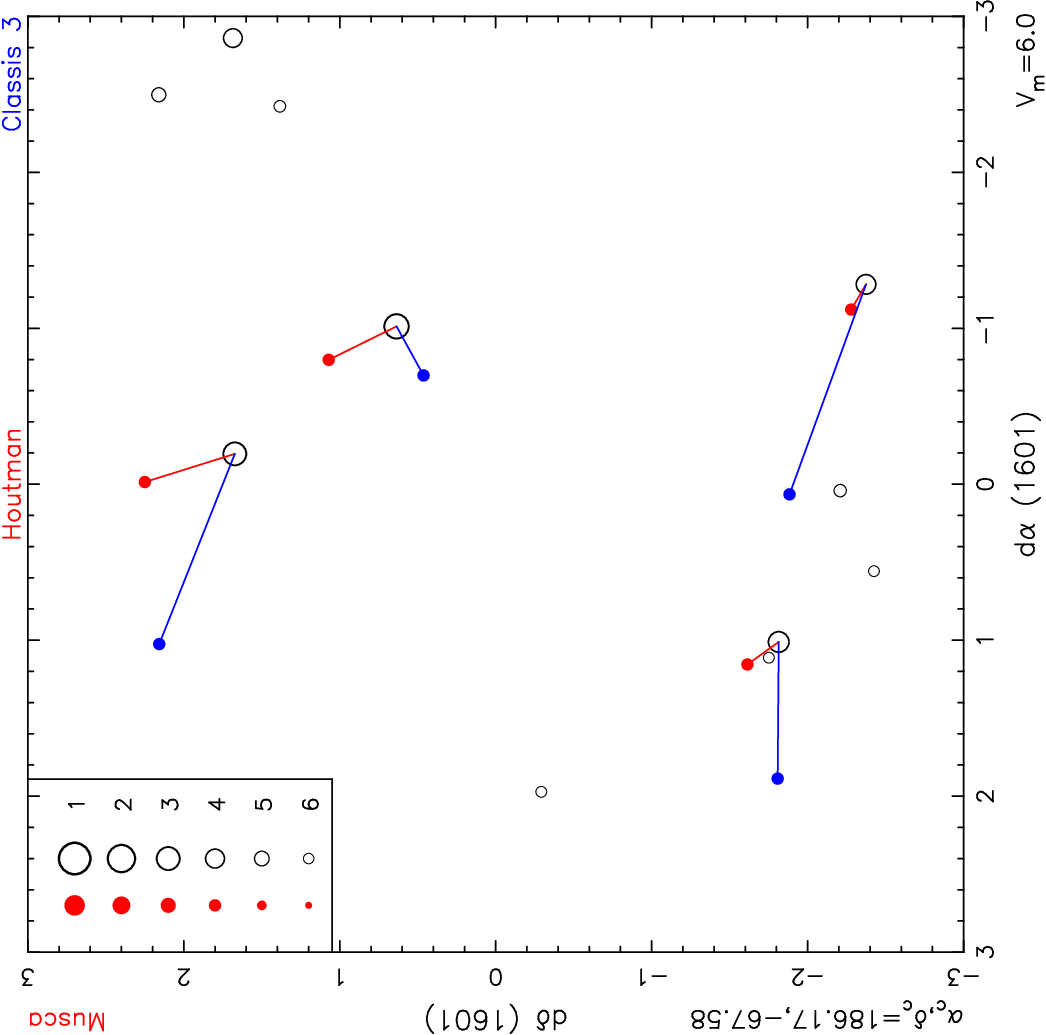}}
\caption{Musca
 \label{f:musca}}
\end{figure}

\begin{figure}
\centerline{\includegraphics[angle=270,width=0.9\columnwidth]{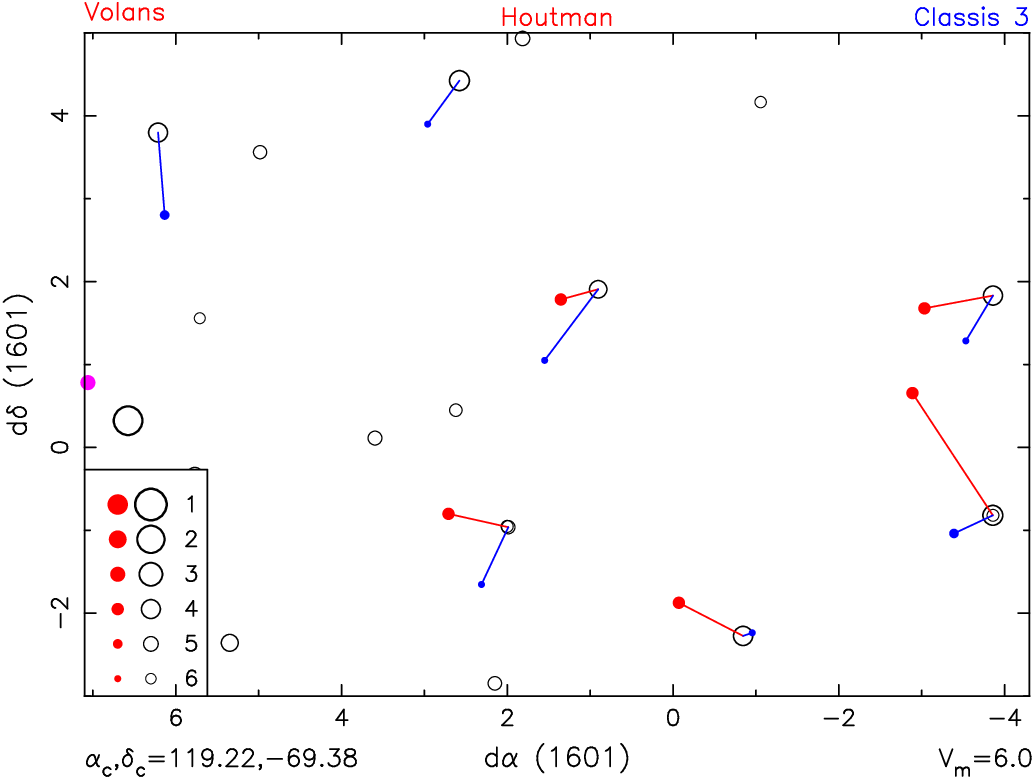}}
\caption{Volans
 \label{f:volans}}
\end{figure}

\begin{figure}
\centerline{\includegraphics[angle=270,width=\columnwidth]{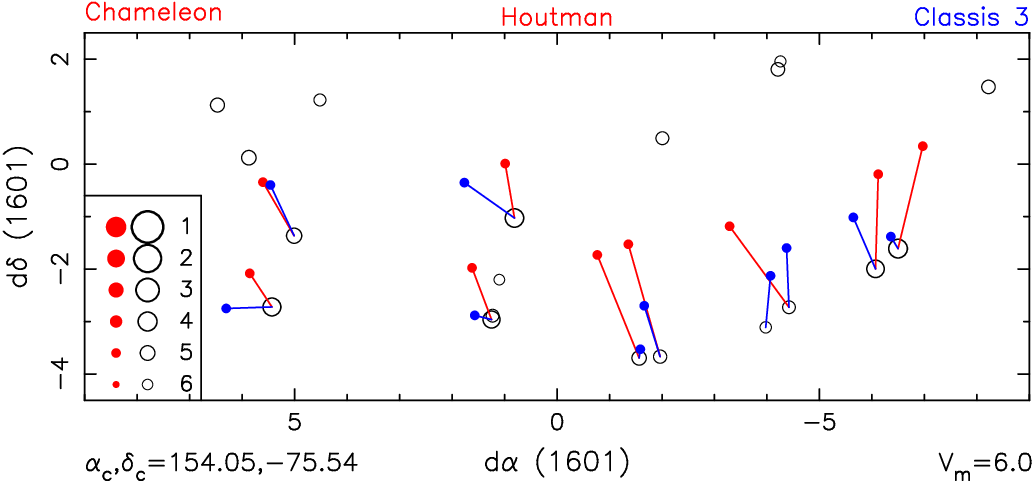}}
\caption{Chamaeleon
 \label{f:chamaeleon}}
\end{figure}

\begin{figure}
\centerline{\includegraphics[angle=270,width=0.9\columnwidth]{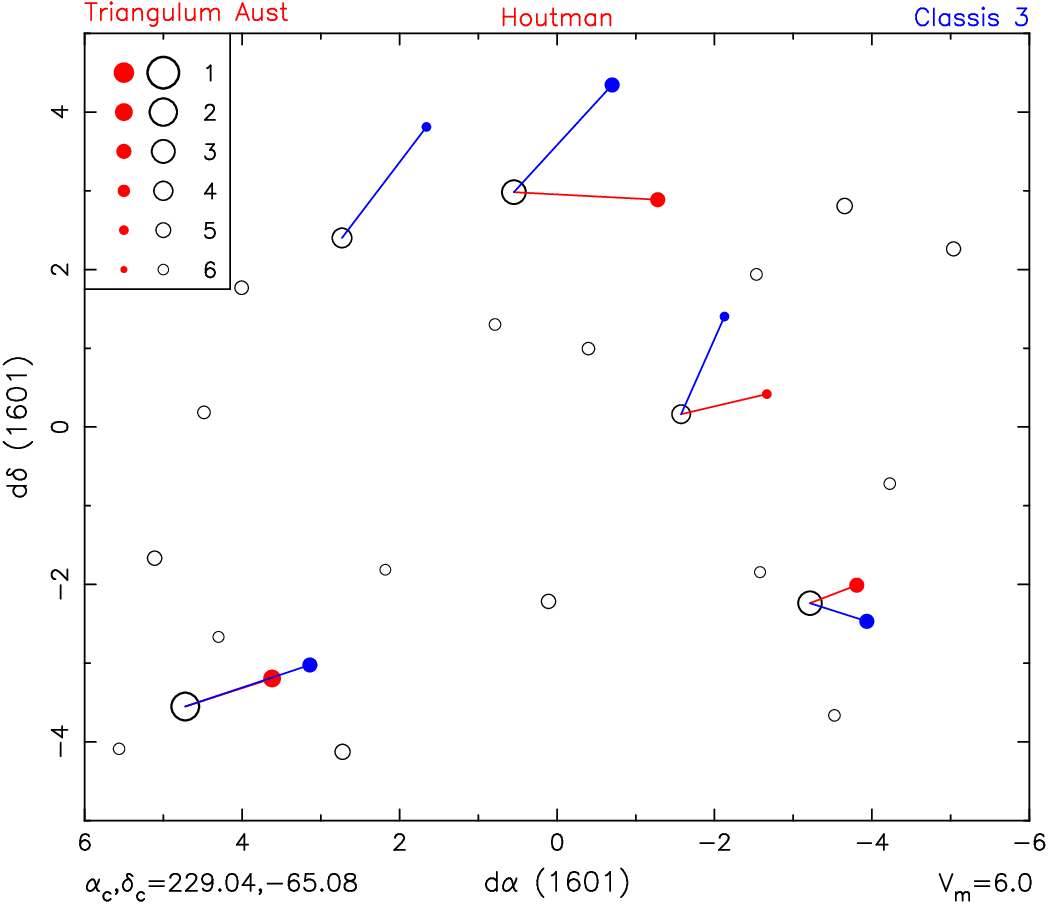}}
\caption{Triangulum Australe
 \label{f:triangaus}}
\end{figure}

\begin{figure}
\centerline{\includegraphics[angle=270,width=0.9\columnwidth]{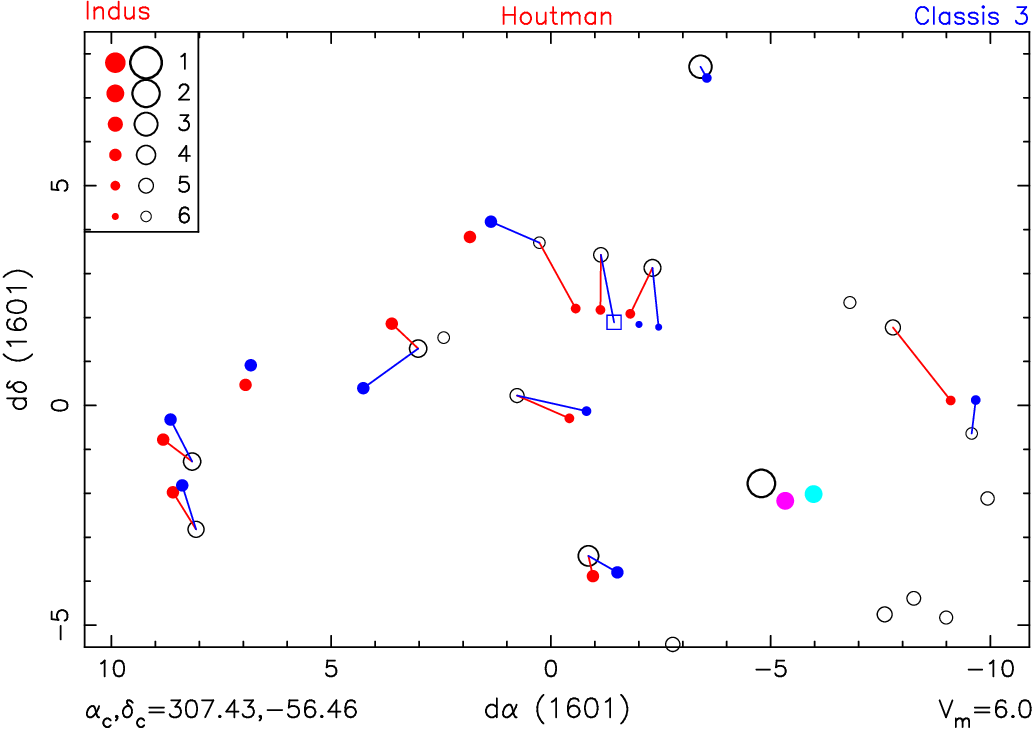}}
\caption{Indus. See also Fig.\,\ref{f:annotindus}.
 \label{f:indus}}
\end{figure}

\clearpage

\begin{figure}
\centerline{\includegraphics[angle=270,width=\columnwidth]{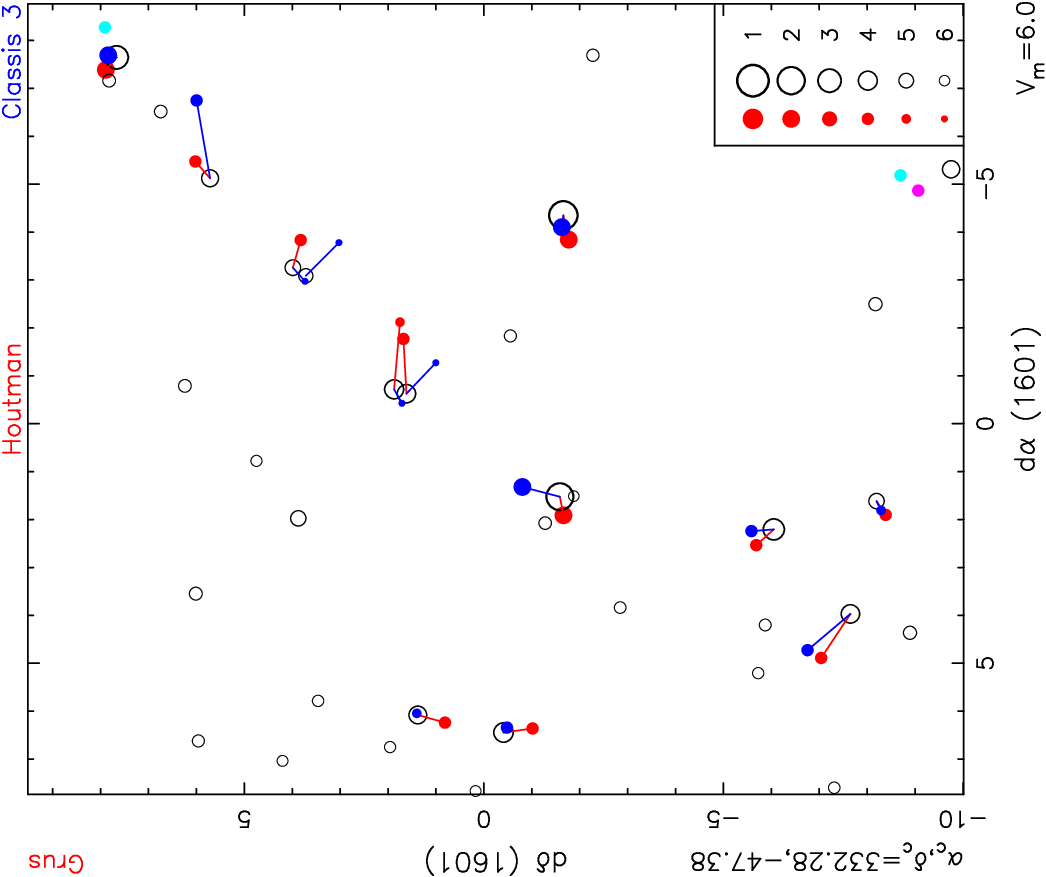}}
\caption{Grus
 \label{f:grus}}
\end{figure}

\begin{figure}
\centerline{\includegraphics[angle=270,width=\columnwidth]{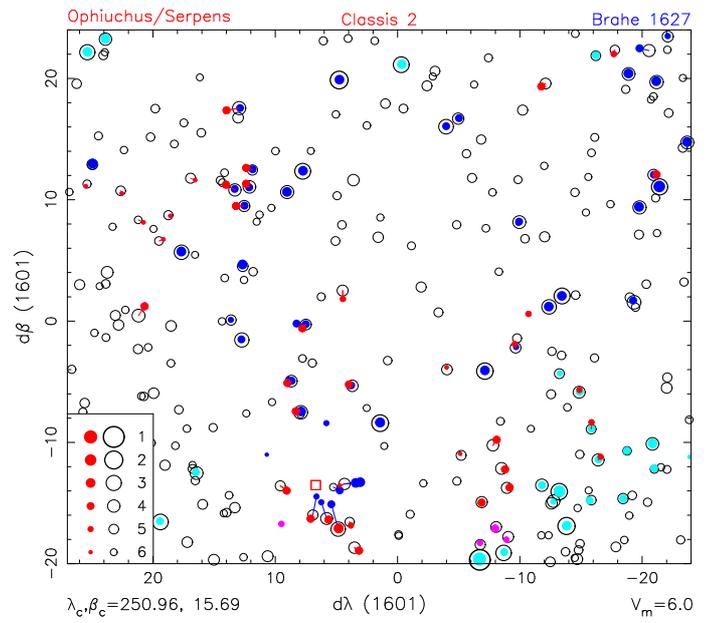}}
\caption{Ophiuchus and Serpens,  in \secunda\ and \keplere. 
K\,1036 = SN\,1604 is indicated with a red square, near
6.7,$-$13.5.  Note that for the stars near this `new star' the positions
in {\em Secunda Classis} are much better than those in \keplere.
\label{f:ophiuchus}}
\end{figure}

\begin{figure}
\centerline{\includegraphics[angle=270,width=\columnwidth]{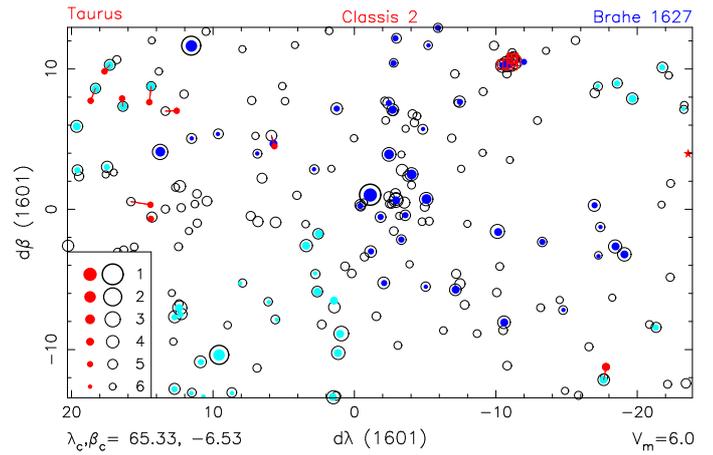}}
\caption{Taurus, in \secunda\ and in \keplere. The red star at the
  right gives the position of K\,1064 before emendation.
  For Pleiades, see Fig.\,\ref{f:pleiades}.
 \label{f:taurus}}
\end{figure}

\begin{figure}
\centerline{\includegraphics[angle=270,width=\columnwidth]{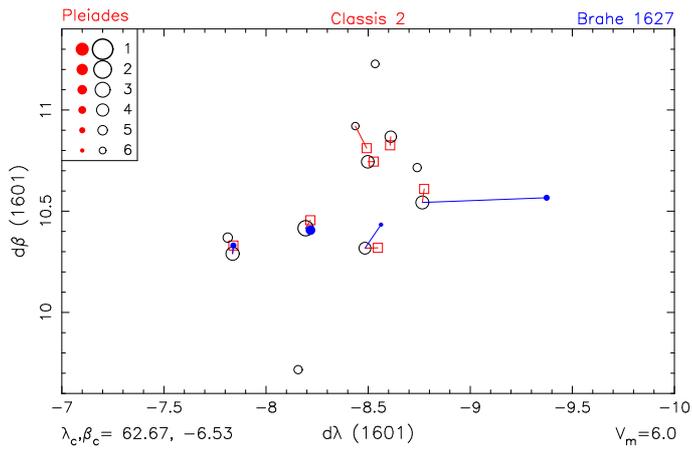}}
\caption{Pleiades. Stars from the Pleiades in \secunda\ are plotted 
 as squares since \secunda\ does not give magnitudes for these stars.
 \label{f:pleiades}}
\end{figure}
 
\begin{figure}
\centerline{\includegraphics[angle=270,width=0.9\columnwidth]{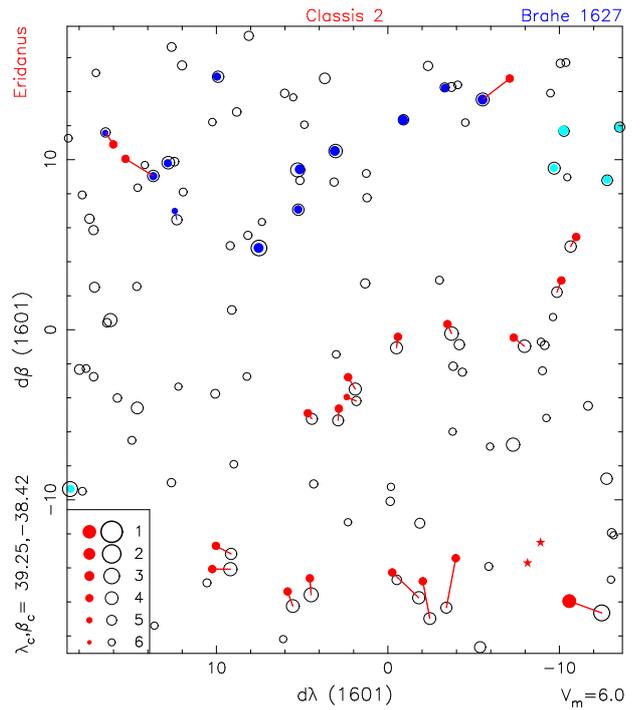}}
\caption{Eridanus. The two red stars right below give the positions of
  K\,1139 and K\,1140 before emendation.
 \label{f:eridanusc}}
\end{figure}

\begin{figure*}
\centerline{\includegraphics[angle=270,width=18.3cm]{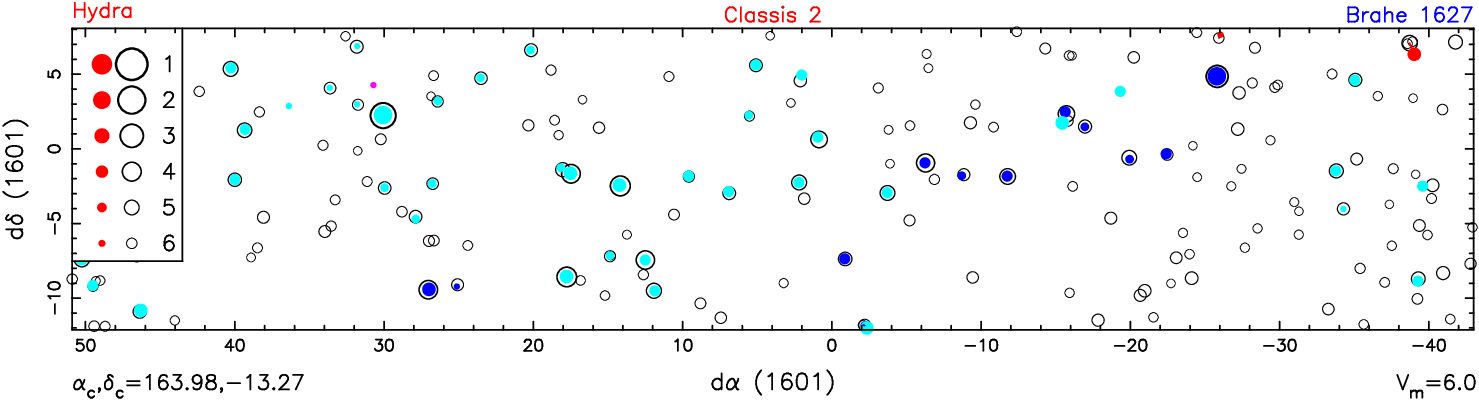}}
\caption{Hydra.
 \label{f:hydra}}
\end{figure*}

\end{appendix}

\clearpage

\begin{appendix}
\section{Constellations in \halley\label{s:fighal}}

The Figures illustrating the identifications by Halley are shown in
the rotated ecliptic system. 

\begin{figure}
\centerline{\includegraphics[angle=270,width=\columnwidth]{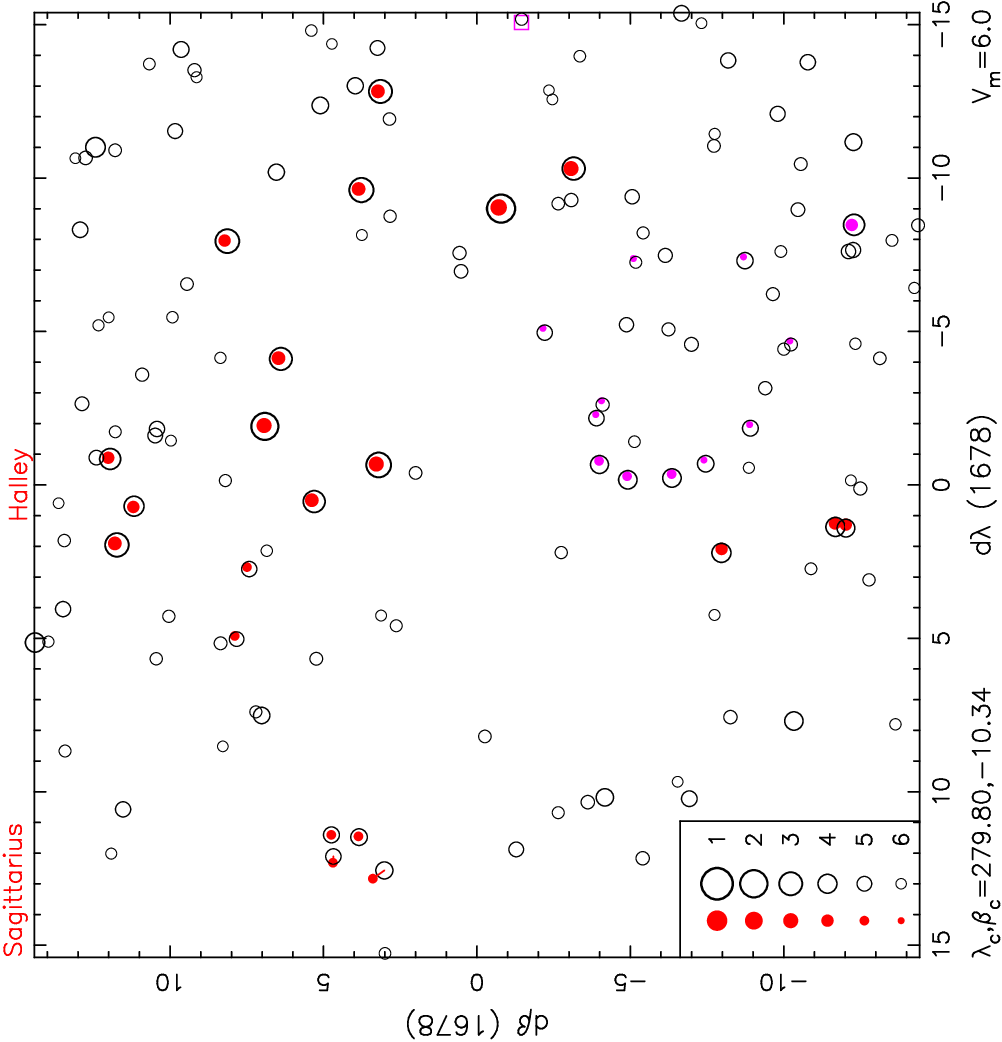}}
\caption{Sagittarius.
 \label{e:sagittarius}}
\end{figure}

\begin{figure}
\centerline{\includegraphics[angle=270,width=\columnwidth]{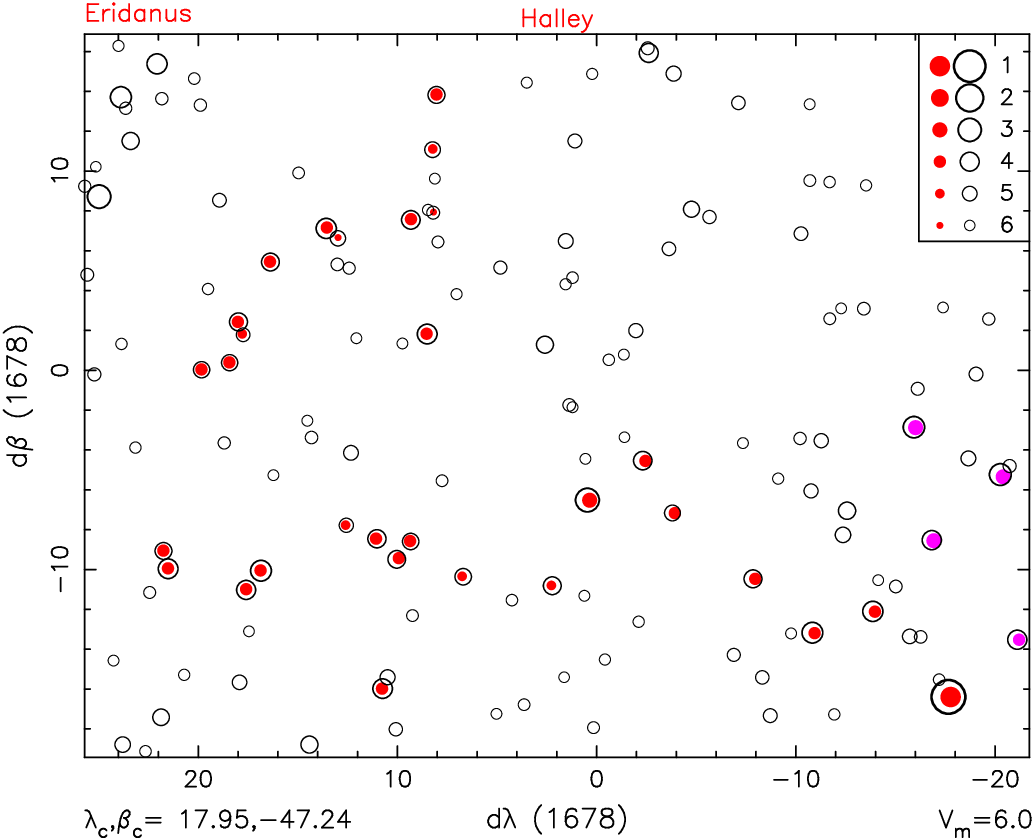}}
\caption{Eridanus.
 \label{e:eridanus}}
\end{figure}

\begin{figure}
\centerline{\includegraphics[angle=270,width=\columnwidth]{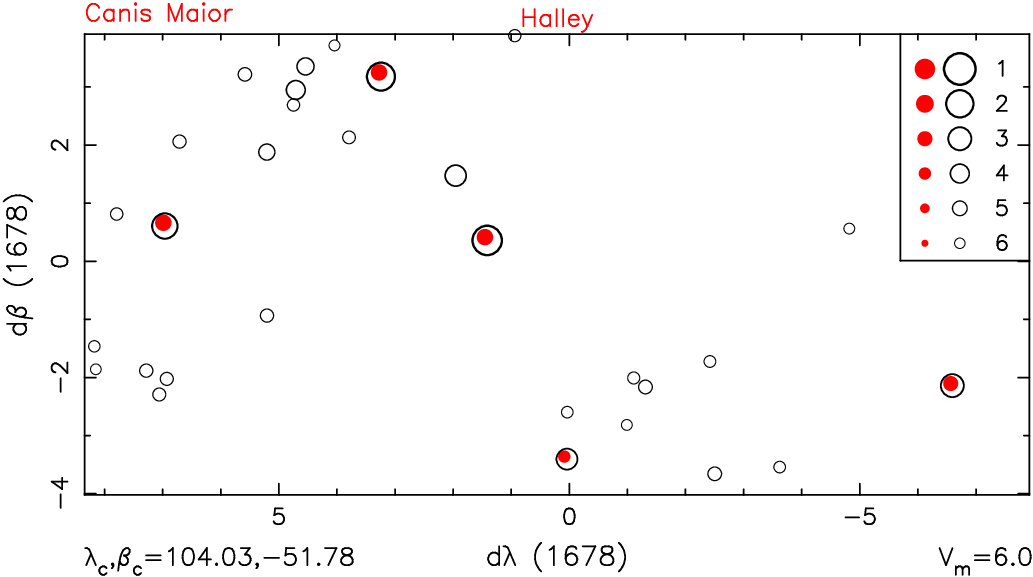}}
\caption{Canis Maior in \halley.
 \label{e:canismaior}}
\end{figure}

\begin{figure}
\centerline{\includegraphics[angle=270,width=\columnwidth]{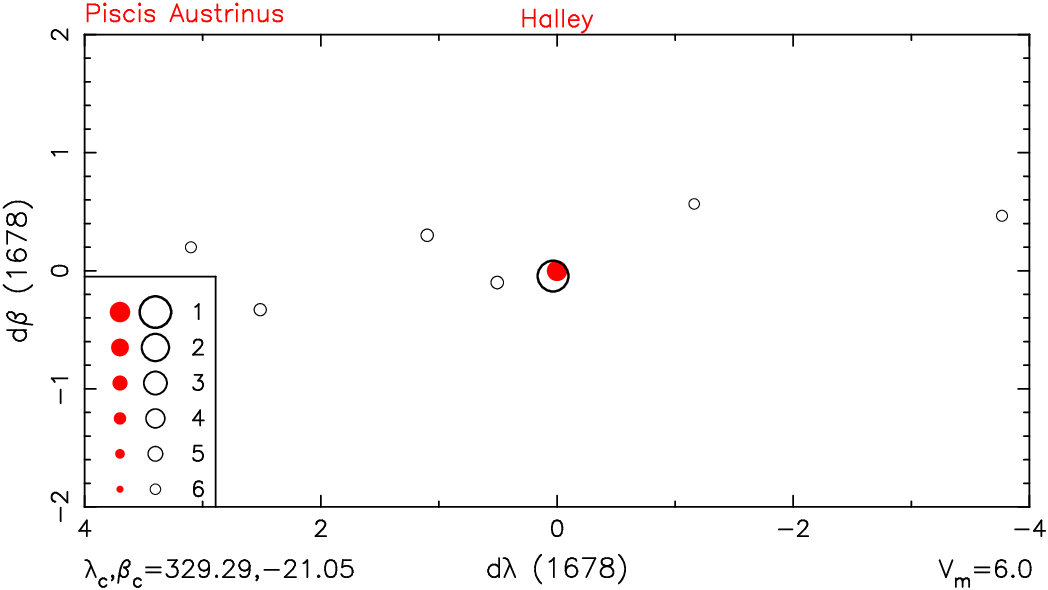}}
\caption{Piscis Austrinus.
 \label{e:pisaustr}}
\end{figure}

\begin{figure}
\centerline{\includegraphics[angle=270,width=\columnwidth]{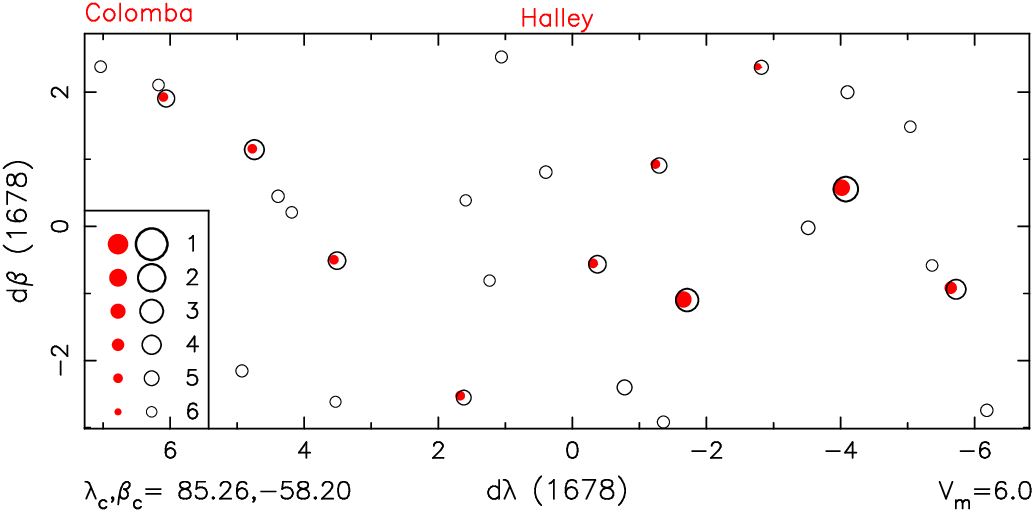}}
\caption{Colomba (the dove of Noach).
 \label{e:colomba}}
\end{figure}

\begin{figure}
\centerline{\includegraphics[angle=270,width=\columnwidth]{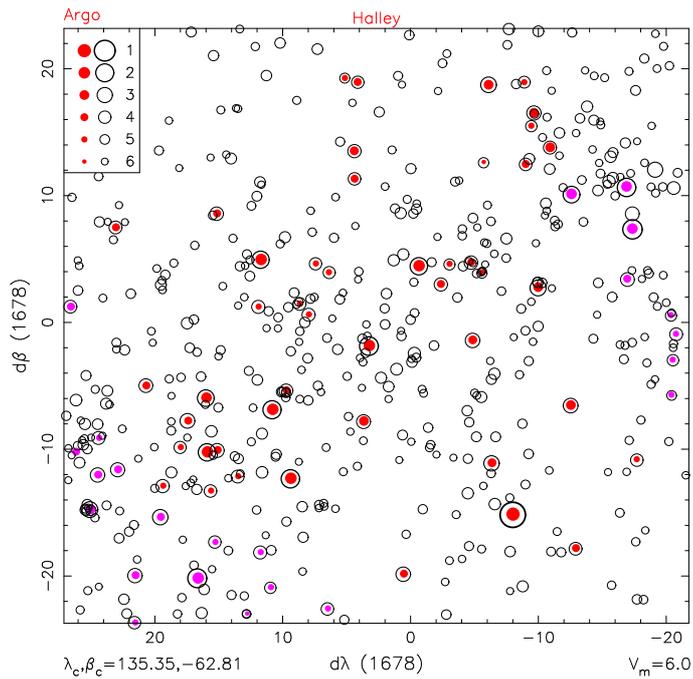}}
\caption{Argo Navis (the ship Argo).
 \label{e:argo}}
\end{figure}

\begin{figure*}
\centerline{\includegraphics[angle=270,width=18cm]{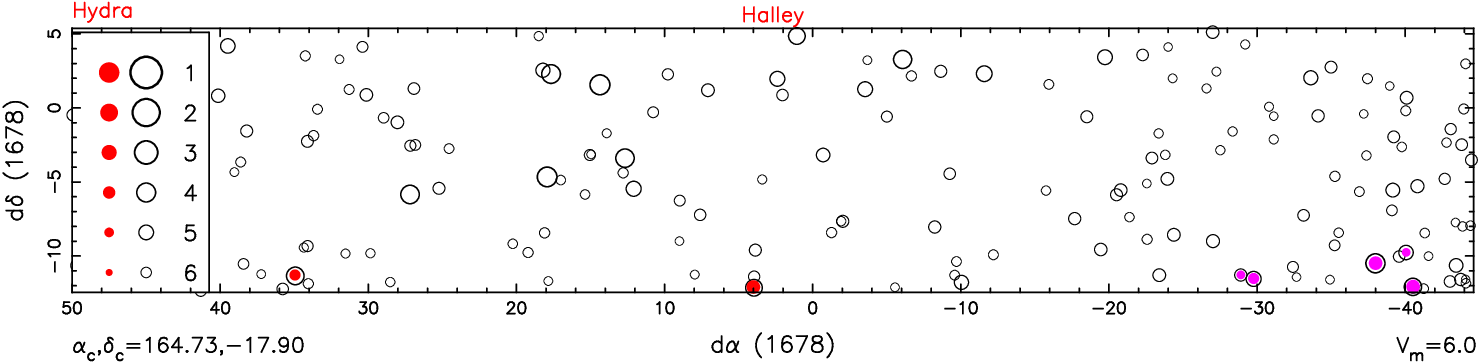}}
\caption{Hydra.
 \label{e:hydra}}
\end{figure*}

\clearpage 

\begin{figure}
\centerline{\includegraphics[angle=270,width=\columnwidth]{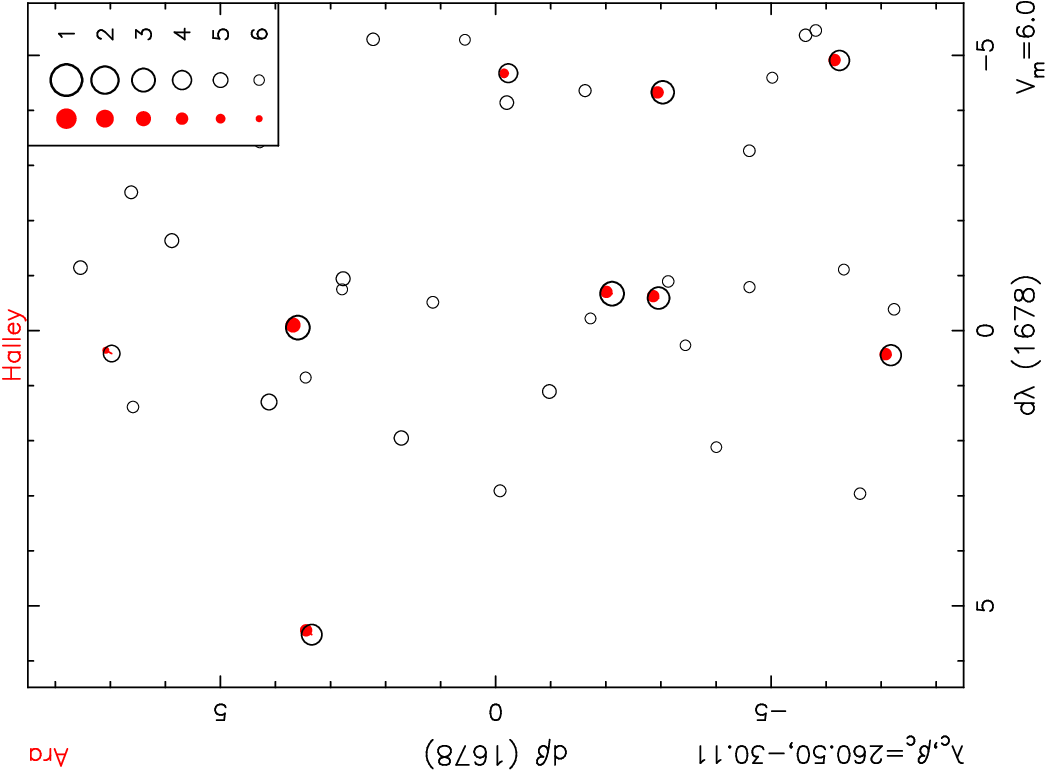}}
\caption{Ara, Thuribulum.
 \label{e:ara}}
\end{figure}

\begin{figure}
\centerline{\includegraphics[angle=270,width=\columnwidth]{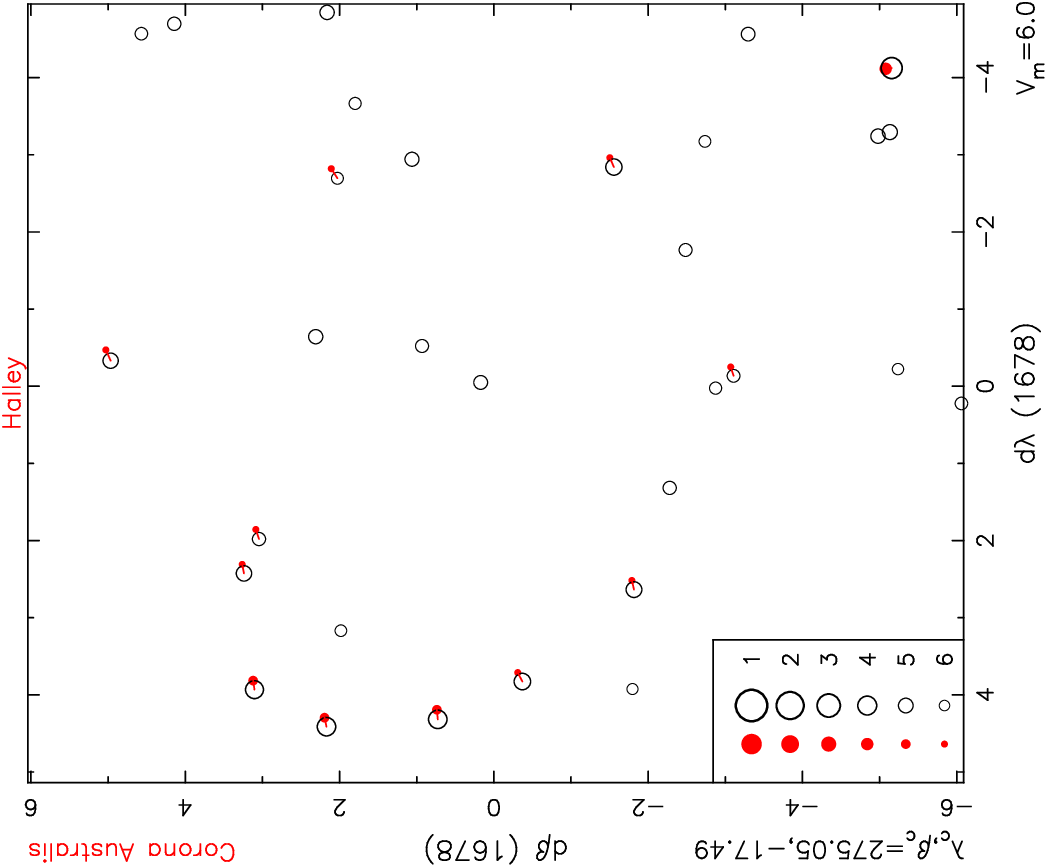}}
\caption{Corona Australis.
 \label{e:coraustr}}
\end{figure}

\begin{figure}
\centerline{\includegraphics[angle=270,width=0.9\columnwidth]{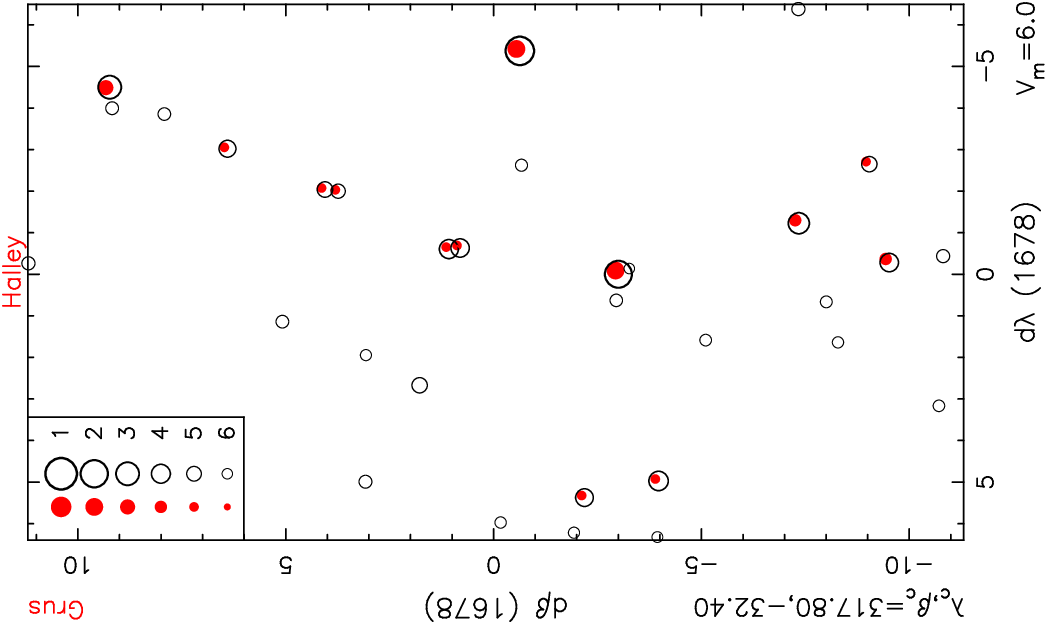}}
\caption{Grus.
 \label{e:grus}}
\end{figure}

\begin{figure}
\centerline{\includegraphics[angle=270,width=\columnwidth]{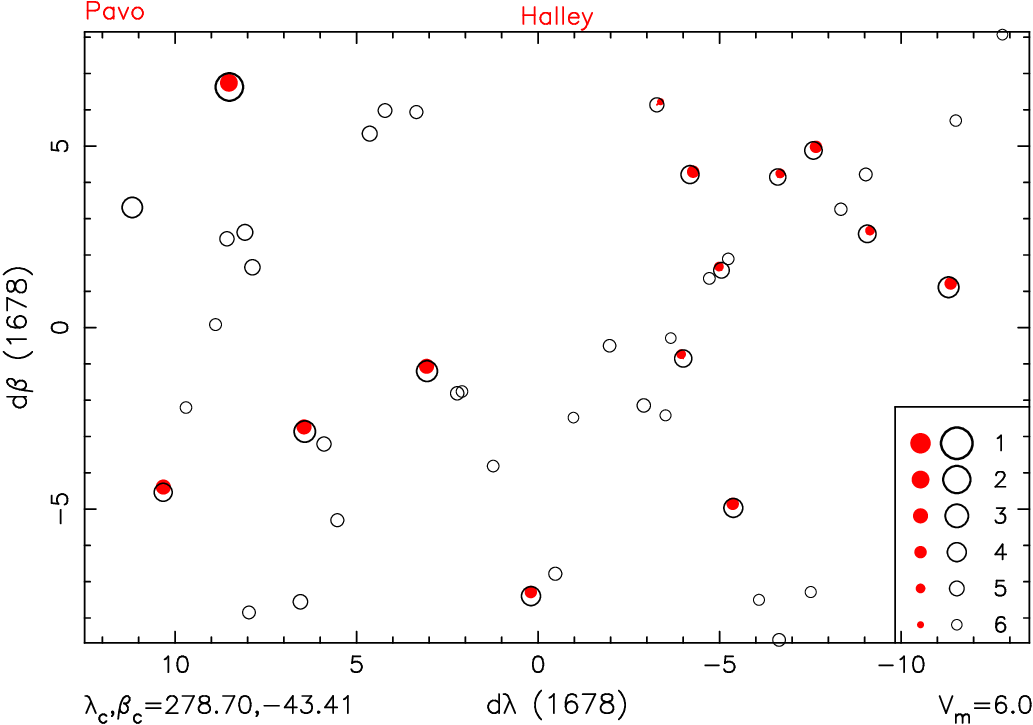}}
\caption{Pavo
 \label{e:pavo}}
\end{figure}

\begin{figure}
\centerline{\includegraphics[angle=270,width=\columnwidth]{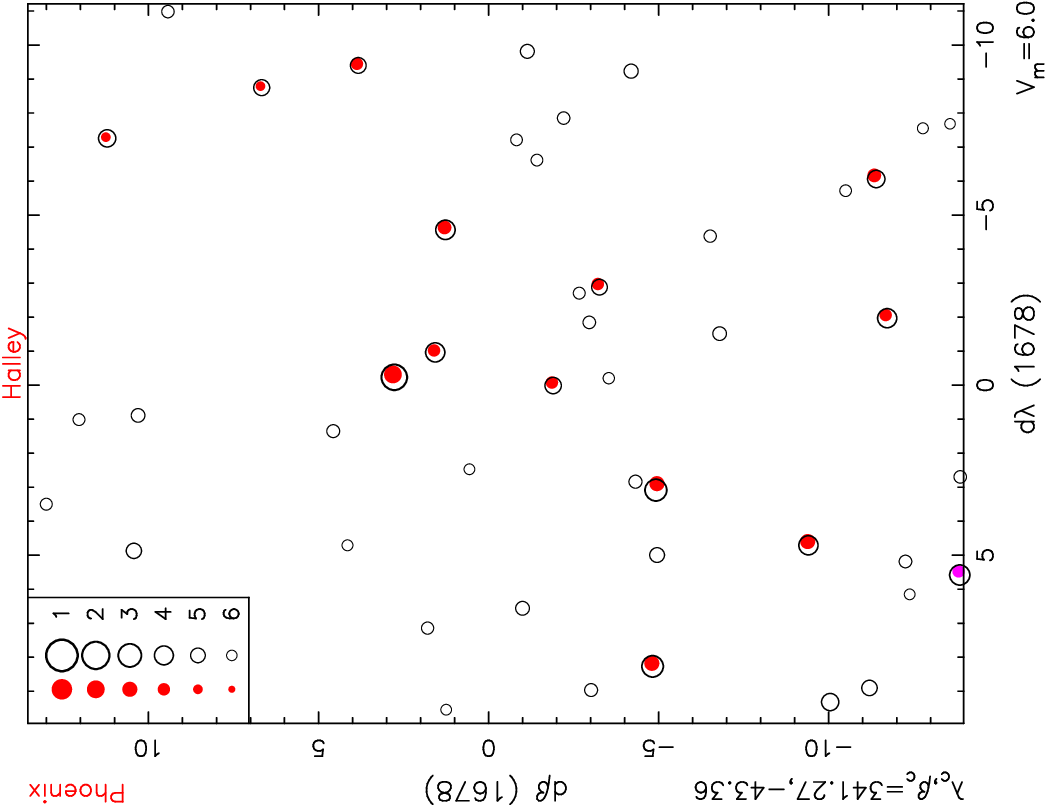}}
\caption{Phoenix.
 \label{e:phoenix}}
\end{figure}

\begin{figure}
\centerline{\includegraphics[angle=270,width=\columnwidth]{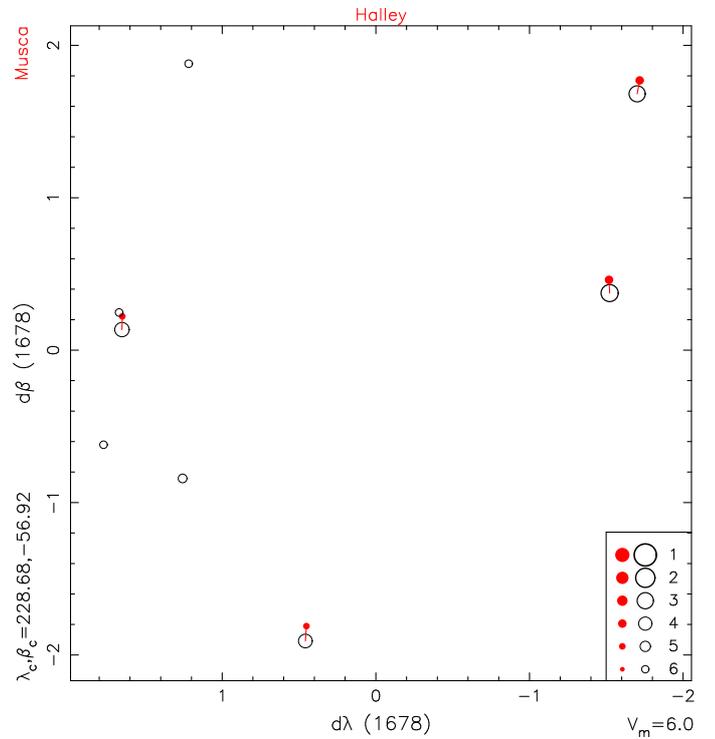}}
\caption{Musca Apis (fly, bee).
 \label{e:musca}}
\end{figure}

\begin{figure}
\centerline{\includegraphics[angle=270,width=0.75\columnwidth]{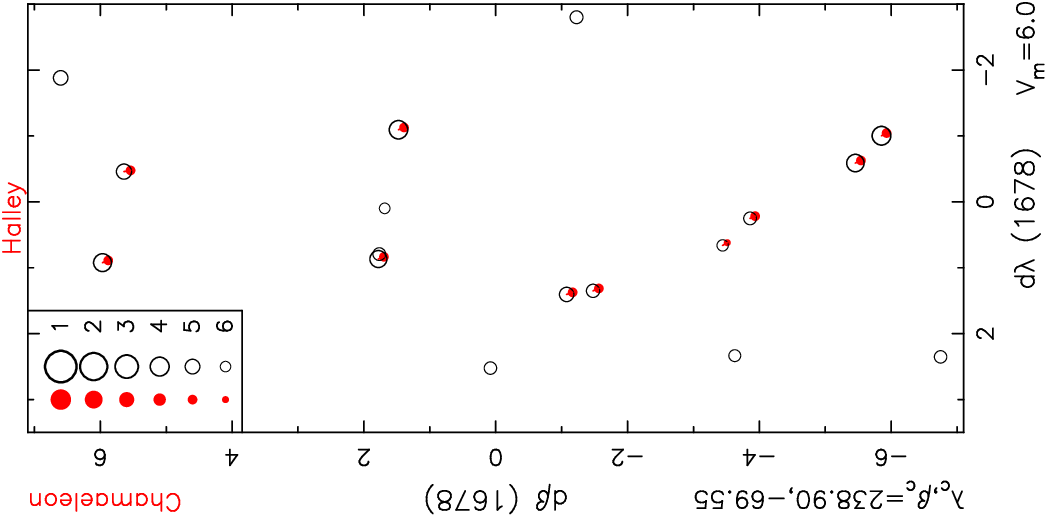}}
\caption{Chamaeleon.
 \label{e:chamaeleon}}
\end{figure}

\begin{figure}
\centerline{\includegraphics[angle=270,width=\columnwidth]{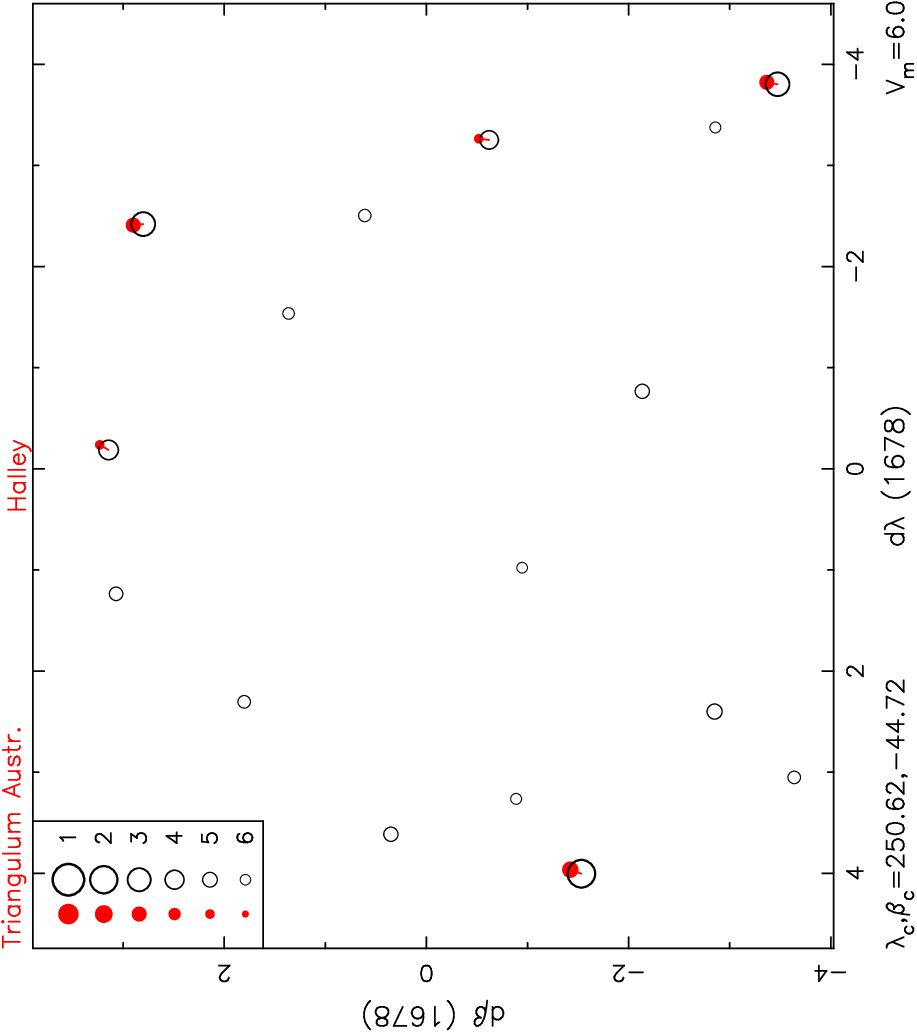}}
\caption{Triangulum Australe.
 \label{e:triaustr}}
\end{figure}

\begin{figure}
\centerline{\includegraphics[angle=270,width=\columnwidth]{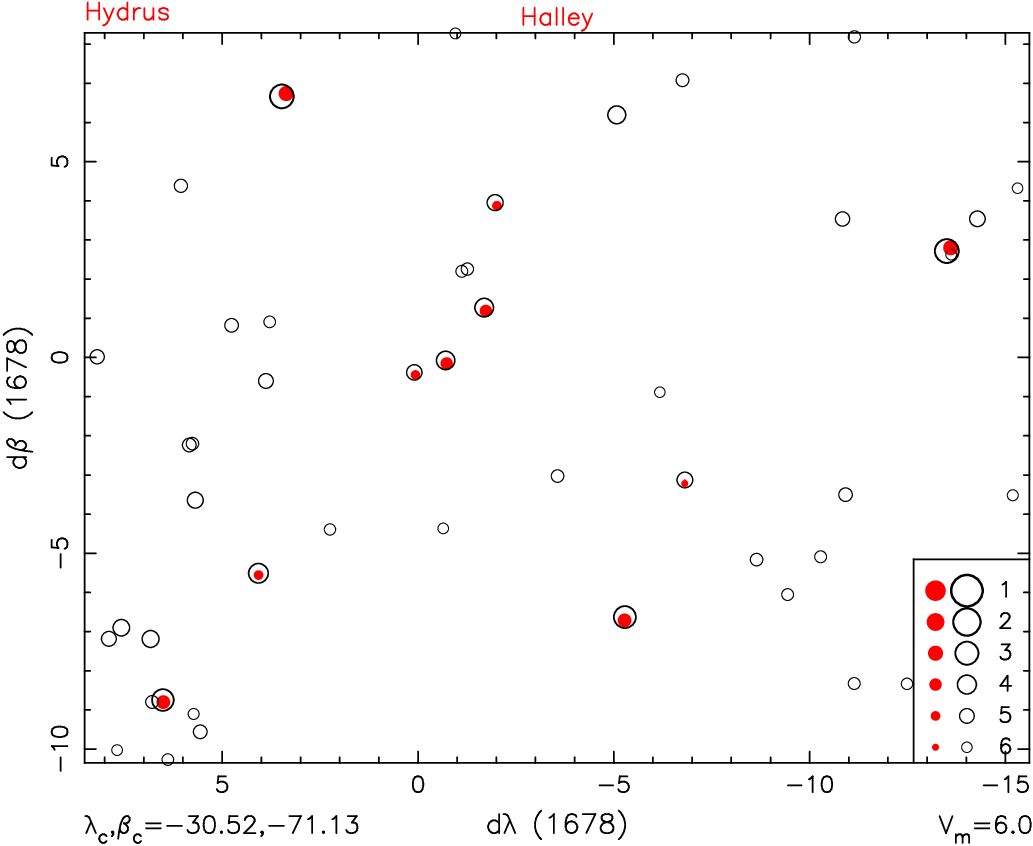}}
\caption{Hydrus.
 \label{e:hydrus}}
\end{figure}

\begin{figure}
\centerline{\includegraphics[angle=270,width=\columnwidth]{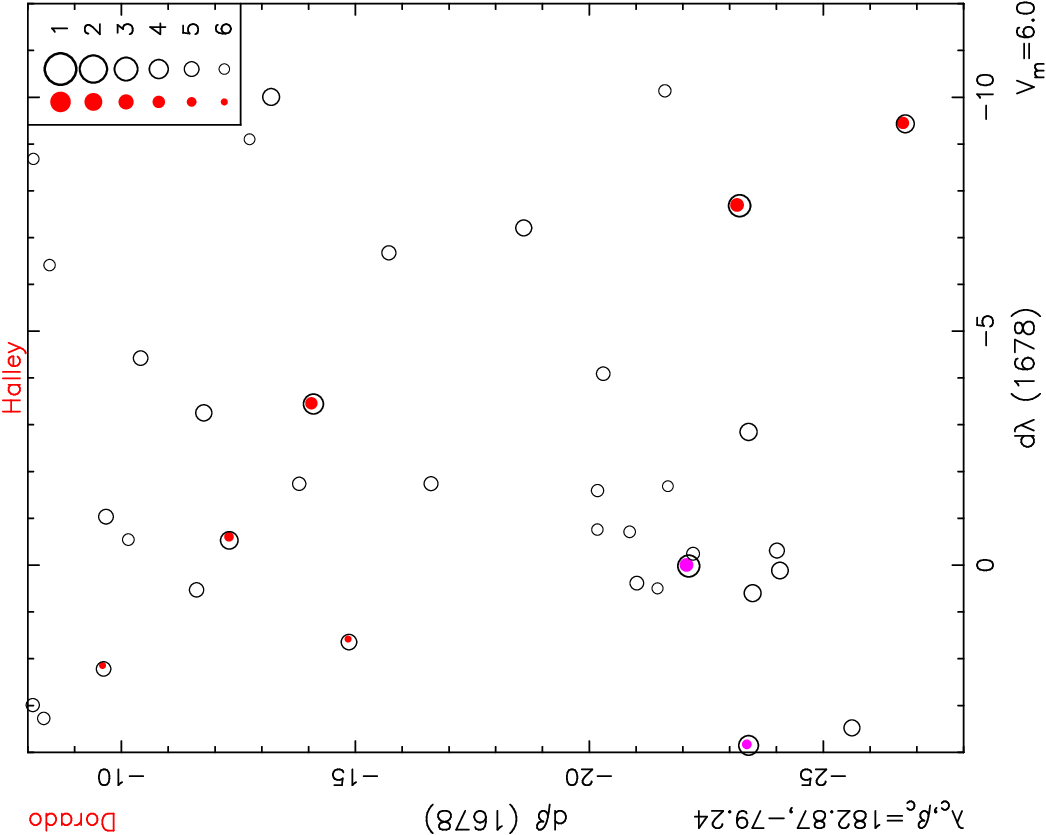}}
\caption{Dorado, Xiphias.
 \label{e:dorado}}
\end{figure}

\begin{figure}
\centerline{\includegraphics[angle=270,width=\columnwidth]{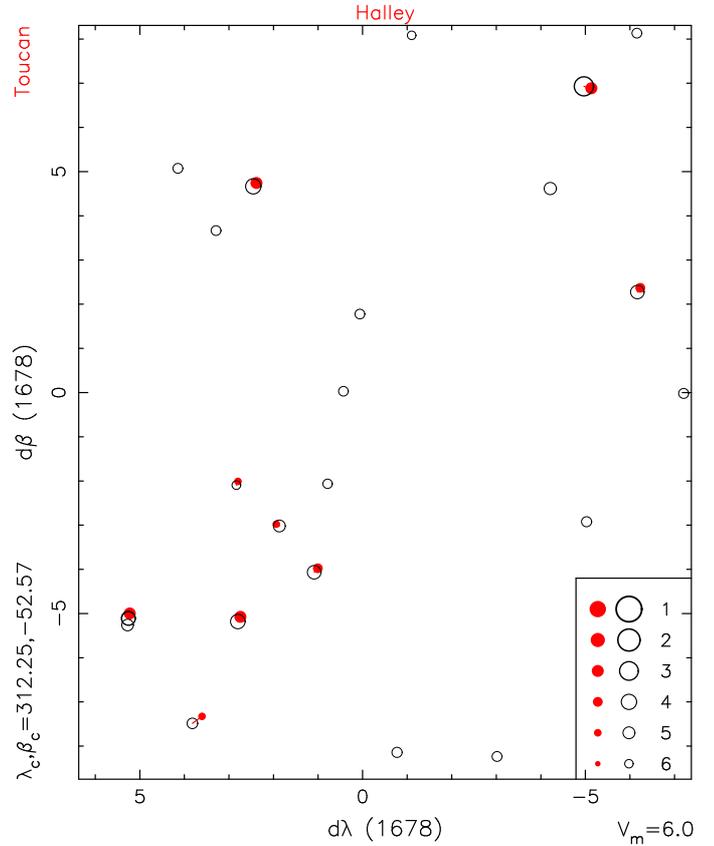}}
\caption{Tucana, Anser Americanus (American goose).
 \label{e:tucana}}
\end{figure}

\end{appendix}

\end{document}